# Two-dimensional metal halide perovskites and their heterostructures: from synthesis to applications

## Athanasia Kostopoulou,[*] Ioannis Konidakis,[*] Emmanuel Stratakis[*]

Foundation for Research & Technology – Hellas (FORTH), Institute of Electronic Structure & Laser (IESL), Vassilika Vouton, Heraklion 700 13, Greece,

E-mail: akosto@iesl.forth.gr (A. Kostopoulou), ikonid@iesl.forth.gr (I. Konidakis), stratak@iesl.forth.gr (E. Stratakis)

Size- and shape- dependent unique properties of the metal halide perovskite nanocrystals make them promising building blocks for constructing various electronic and optoelectronic devices. These unique properties together with their easy colloidal synthesis render them efficient nanoscale functional components for multiple applications ranging from light emission devices to energy conversion and storage devices. Recently, two-dimensional (2D) metal halide perovskites in the form of nanosheets (NSs) or nanoplatelets (NPls) are being intensively studied due to their promising 2D geometry which is more compatible with the conventional electronic and optoelectronic device structures where film-like components are employed. In particular, 2D perovskites exhibit unique thickness-dependent properties due to the strong quantum confinement effect, while enabling the bandgap tuning in a wide spectral range. In this review the synthesis procedures of 2D perovskite nanostructures will be summarized, while the application-related properties together with the corresponding applications will be extensively discussed. In addition, perovskite nanocrystals/2D material heterostructures will be reviewed in detail. Finally, the wide application range of the 2D perovskite-based structures developed to date, including pure perovskites and their heterostructures, will be presented while the improved synergetic properties of the multifunctional materials will be discussed in a comprehensive way.

## 1. Introduction

Perovskites were first found in the Ural Mountains and named after Lev Perovski (who was the founder of the Russian Geographical Society). Perovskite materials family has similar crystal structure with $CaTiO_3$ mineral and formula $ABX_3$. In their structure, the cation 'A' occupies the corner positions of the unit cell and the cation 'B' is located at the center of the cell, while the anion 'X' is at the unit cell faces (Figure 1a). [1–3] This family comprises oxides and metal halide perovskite compounds. Colloidal metal halide nanocrystals have gained intense attention from 2014 where the first synthesis of nanometer sized organolead $CH_3NH_3PbBr_3$ perovskites was reported,[4] due to their unique optoelectronic properties/features such as defect tolerance (for bromide and iodine-based structures), their extremely high PLQYs and photoluminescence near to unity, their capability to emit in the whole visible wavelength range by tuning their halide ratio and their thickness. [5–10]

Two-dimensional (2D) metal halide perovskites in the form of nanosheets (NSs) or nanoplatelets (NPls) are being intensively studied due to their promising 2D geometry which is more compatible with the conventional electronic and optoelectronic device structures where film-like components are employed but also due to strong confinement effects by tuning the thickness of the structures (Figure 1).[1] The 2D morphology is an ideal architecture from fundamental point of view to investigate quantum confinement effects but also to study the 2D-lateral growth and how these affect to the final properties of the final structures. In particular, 2D perovskites exhibit unique thickness-dependent properties due to the strong quantum confinement effect, while enabling the bandgap tuning in a wide spectral range (Figure 1).[1,8,11,12] Strong confinement effects were appeared when the size/thickness of the one dimension of the nanocrystals is below the Bohr diameter which is 7 nm for the cubic structure of the $CsPbBr_3$. The strong two-dimensional confinement of the carriers at such small vertical sizes is responsible for a narrow PL, strong excitonic absorption, and a blue shift of the optical band gap by more than 0.47 eV compared to that of bulk counterparts (case of CsPbBr₃). [12] While the magnitude of the band gap is clearly dependent on the extent of 2D confinement, the excitonic dynamics in lead halide perovskites is virtually independent of it (it was essentially the same in both NPls and bulk samples) and is instead sensitive only to the presence of trap states.[12]

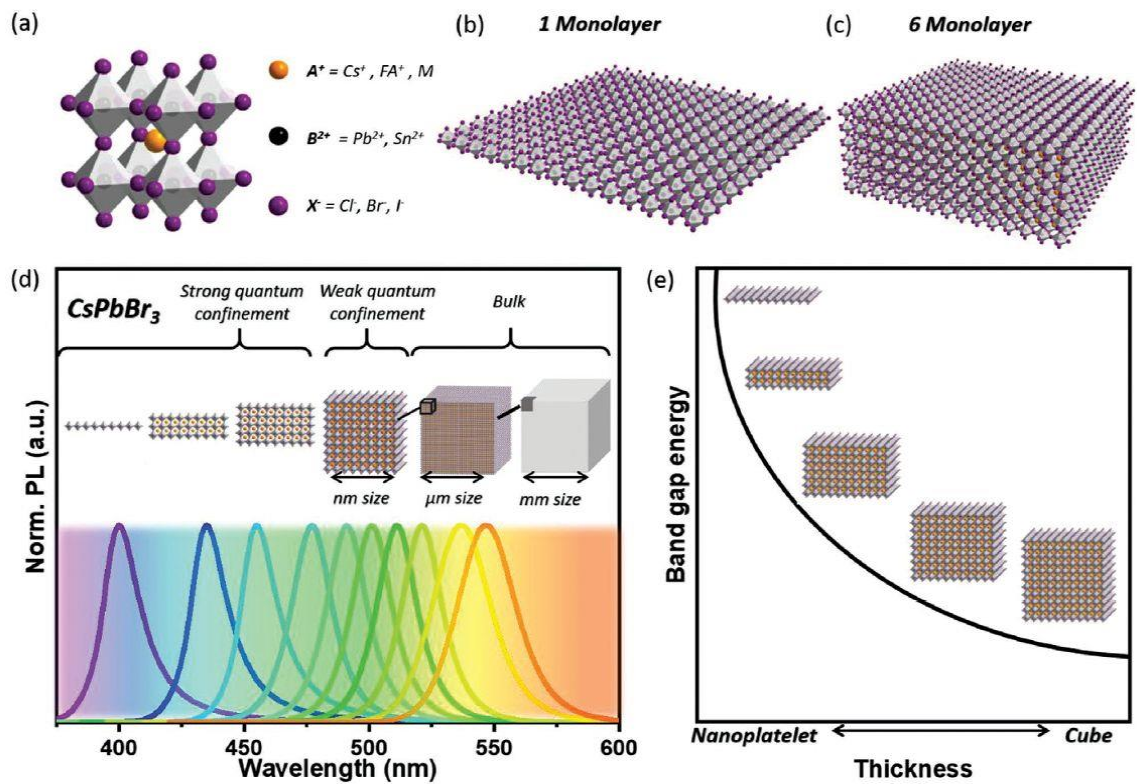

**Figure 1:** Schematic illustration of the crystal structure of the metal halide perovskites ABX₃ and the formation of the 2D platelets-like morphologies composed of one and six monolayers (a- c). PL spectra and band gap energy for the modification of the bulk structures, to weakly confined nanocubes to strongly confined 2D nanoplatelets/nanosheets (d-e). Adapted with permission from ref. [1]. Copyright 2021, Wiley.

Various synthesis methods have been introduced for the synthesis of 2D metal halide perovskite nanostructures such as wet chemistry approaches,[1] vapor phase deposition[13], mechanical exfoliation[14] and solvent evaporation method[15]. But over the years, the ligand-assisted colloidal methods, the hot injection method and room-temperature reprecipitation, have been adopted and developed the most. In addition, 2D materials in the form of nanoplatelets and nanosheets can be formed by the exfoliation of bulker materials of the same chemical phase to thinner structures in colloidal solution using ligands as spacer[16] but also exfoliation can be taken place upon irradiation with a femtosecond (fs) laser or using an ultasonication-tip device. [17,18] Specifically,  the synthesis protocols for the fabrication of the metal halide perovskites NPls and NSs can be classified into six categories: i) ligand-assisted hot injection and precipitation, [11, 19, 20, 12, 21, 22, 23, 24, 25, 26, 27, 28] ii) spontaneous crystallization in nonpolar organic media, [29] iii) cation-mediated colloidal method, [30] iv) solvothermal method, [31] v) ligand- and laser-assisted exfoliation [16, 17] and vi) external stimuli-triggered process[9,18].

The unique morphologies which combined the excellent optoelectronic properties of the perovskite nanomaterials with large lateral dimensions make them ideal building blocks for optoelectronic devices such as photodetectors and LEDs. Rigid or flexible photodetectors with high on/off ratios and fast response times have been reported and also a thickness dependent photoresponse was also revealed. [32, 33] The external quantum efficiency (EQE) of the most sufficient green NPls-based LEDs has been enhanced drastically from 0.48 % in 2016 to up to 23.6 % in 2020 discovering at the same time the factors affecting the LEDs performance. [34] Identification of the dominant type of surface traps and passivation with appropriate ligands, charge injection and charge balance by improving the quality of the perovskite film and maximizing the light outcoupling efficiency were among these factors and how these were improved.[34] Moreover, NPls/NSs were lead in a remarkable improvement of the power conversion efficiency (PCE) and stability of perovskite solar cells, when compared to the corresponding devices with typical films fabricated by conventional spin-coating methods. [35] Also, the 2D structures except of the enhanced photocatalytic activity due to their large proportion of low-coordinated metal atoms together with the short carrier diffusion distance revealed a higher stability upon irradiation time compared to the small nanocrystals of the same chemical phase. [36] In their application as photocatalysts for the reduction of the $CO_2$, it was found that the generation rate of the CO was thickness and stoichiometry dependent. [37] In addition, the strong conductivity together with the fast and reproducible the PL response of these structures in the presence of moisture or gas have been used for the fabrication of sensors with high responsiveness. [38, 39]

Despite great progress in the synthesis protocols for metal halide perovskite 2D morphologies, it is still challenging to achieve monodisperse NPls with a single PL peak, while being stable in their colloidal solutions. Furthermore, the cleaning and purification processes after the synthesis is non-trivial, and the washing techniques used till now, many times affect the stability and the PL of the 2D nanocrystals. [40] This sensitivity of the NPls/NSs affects also the performance of the devices using these materials. Challenges still remain in further improve the device performance taking into account stability issues of the 2D materials. In this direction, metal halide perovskite nanocrystals/2D material heterostructures have been proposed that combine the properties of the perovskite nanomaterials with their unique properties of the 2D materials and the large lateral dimensions. The large number of different perovskite nanocrystals and together with the plethora of 2D materials (graphene-based materials, hexagonal boron nitride or Transition Metal Dichalcogenides) can result to heterostructures with designed functionalities.[41] Additionally, new physics and synergetic effects can emerge from the coupling between the two different materials and new or improved functionalities have been arisen due to the interfacial phenomena.

Metal halide perovskite of lead-based or lead free different compositions such as $CsPbX_3$ [42, 43, 44, 45, 46, 47, 48, 49,50,51,52,53,54,55,56], $CsPbBr_{3-x}I_x$ [42, 57, 58, 59], $Cs_4PbBr_6$ [60], $CH_3NH_3PbX_3$[61, 62, 59, 63, 64], $Cs_2AgBiBr_6$ [65] and morphologies have been conjugated on 2D materials such as sheets of graphene-based [42, 43, 44, 62, 45, 63, 64, 50, 65, 53, 60] materials, Transition Metal Dichalcogenides [57, 58, 54, 59, 55, 66], graphitic carbon nitride [61, 46, 49], Hexagonal Boron Nitride, [52] and black phosphorous. [47, 48] These heterostructures have been fabricated by: i) spin coating or drop-casting of as-synthesized perovskite nanocrystals on the film of the 2D materials with a well-defined contact,[45,55,58] ii) in situ growth of the nanocrystals on the 2D materials, when the later are presented through the nanocrystals' synthesis,[43,44,51,52,61,64] iii) simply mixing of the two materials and chemical or non-covalent interactions,[46,48,49,54,66] iv) irradiation with laser[60] and v) solid state reactions[56].

Such heterostructures have used effectively as photocatalysts for $CO_2$[43,49,64, 46,67, 68] or for hydrogen splitting[69,62,59,65,70] due their increased number of catalytic sites together with the sufficient charge separation compared with the perovskite nanocrystals and 2D materials independently. Moreover, the synergetic effects in these heterostructures were the reason behind the enhanced sensing capability using these materials as sensing elements. The presence of the perovskite nanocrystals found that enhanced the reactivity of the 2D materials that usually performed a poor sensing performance. The 2D materials play mainly the role of the efficient charge conductor, ensuring that the charges generated upon the interaction between gas molecules and perovskite nanocrystals reach the device electrodes. [71] They used for the detection of toxic gases such as ammonia ($NH_3$) and nitrogen dioxide ($NO_2$) [63] or volatile organic compounds (VOCs) [71] without requiring high working temperature or UV irradiation to activate the sensing process. Furthermore, the superior optical features of the metal halide perovskite nanocrystals together with the carrier mobility of the 2D materials provides the ideal combination for using this materials in photodetectors. [72] Metal halide nanoislands grown on graphene film showed the extremely high responsivity $6.0 \times 10^5$ A $W^{-1}$ and a photoconductive gain of $\approx 10^9$ electrons per photon because of effective photogating effect applied on graphene along with increased lifetime of trapped photocarriers in separate perovskite islands. [72] In addition, the humidity and thermal stability of the perovskite nanocrystals has been enhanced with their conjugation with the 2D materials displaying better performance and higher color purity compared to the single perovskite nanocrystals in LEDs.[51,52] Finally, the instability originating from the thermal/moisture-induced perovskite nanocrystal agglomeration in nanocrystal films in the photovoltaics found that also greatly suppressed by the 2D crosslinking. [73] The films of the perovskite nanocrystals/2D materials heterostructures provides not only an effective channel for carrier transport, as witnessed by much improved conductivity but also significantly better stability against moisture, humidity, and high temperature stresses.

This review paper is organized in different sections, which will present the metal halide perovskite 2D NSs/NPls morphologies and the metal halide nanocrystals/2D materials heterostructures and their applications (Figure 2). Firstly, the synthesis protocols and the unique features/properties of organic-inorganic and all-inorganic metal halide NSs and NPls will be presented and discussed. The synthesis of the metal halide nanocrystals/2D materials heterostructures will describe as well and how the properties of the 2D materials were affected by the presence of the perovskite nanocrystals due to synergetic effects emerged from the coupling between the two different materials. Throughout the paper, focus will be given on the unique properties of the 2D perovskites and the synergistic effects of the perovskite nanocrystals/2D materials heterostructures, when possible, which endow new or improved functionalities and improved performance in applications. Each application of the two 2D single-phase or heterostructures materials will be analyzed. The application of these materials in photovoltaics and photocatalysis will be analyzed followed by the demonstration of application in photodetectors and LEDs. The use of these materials for gas and humidity sensors will finally be explained, before providing their future perspectives.

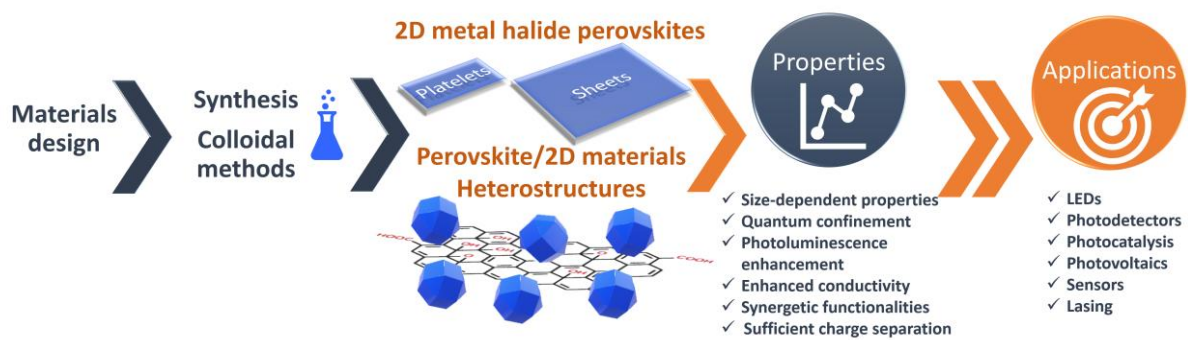

**Figure 2:** Review content in one scheme. 2D metal halide perovskites, metal halide nanocrystals/2D material heterostructures, enhanced properties and applications.

# 2. Synthesis of the 2D metal halide perovskites

## 2.1 Organic-inorganic metal halide perovskite nanosheets/nanoplatelets

In this Section, we will consider an overview of the available colloidal routes for the synthesis of organic-inorganic 2D perovskite nanocrystals (PNCs), while summarizing the key features and properties of the so-formed materials towards potential applications. In general, the key parameters for synthesis optimization and properties control of 2D PNCs are the reaction temperature, the precursors concentration, the ligands ratio, the acid-base features of ligands, and the chain length of the ligands [28,34,74,75]. Over the years, two main colloidal synthesis methods have been adopted and developed the most, namely the hot injection (HI), and the ligand-assisted reprecipitation (LARP) [34,75]. Notably, the former requires high temperatures and inert atmosphere, and thus, it is not convenient for large scale fabrication. Alternatively, the LARP approach offers a more cost-effective solution as it can provide high quality PNCs at room temperature under ambient air. In fact, the vast majority of organic-inorganic PNCs are synthesized by means of LARP (Table 1). Herein, we will present examples of HI synthesis first, followed by the evolution of the LARP routes over the years. Finally, a brief summary of the alternative synthesis strategies will be provided.

A HI synthesis method without the use of polar solvents for the development of organic-inorganic PNCs was introduced by Vybornyi et al. in 2016 [23]. Namely, the reaction between methylamine cation and $PbX_2$ salts was conducted in 1-octadecene, i.e. a nonpolar solvent, in the presence of oleylamine and oleic acid as coordinating ligands. The so-formed $CH_3NH_3PbX_3$ PNCs exhibited PL quantum efficiencies of 15-50% and good phase purity, however multiple types of shapes were obtained including nanocubes, nanowires, and NPIs [23]. Another example of organic-inorganic PNCs synthesis by means of HI was reported by Protesescu et al. a year later [24]. In particular, the authors developed highly luminescent formamidinium-based $FA_{0.1}Cs_{0.9}PbI_3$ and $FAPbI_3$ PNCs, exhibiting stable red (680 nm) and near-IR (780 nm) emissions, respectively. For the development of phase pure perovskite NSs, the same outcome is achieved upon following two synthesis methods. In the first, a two-precursor approach was followed where lead halide was set to react with FA-oleate. The formation of $FAPbI_3$ NSs was successful at injection temperatures below 50 °C, with the size of the so-formed NSs ranging from 0.2 to 0.5 μm [24]. In the second, a three-precursor approach was adopted where a solution mixture of FA-oleate and Pb-oleate was formed upon reaction with FA-acetate. Figure 3a depicts the photoluminescence (PL) and absorbance spectra of the of the $FAPbI_3$ NSs, along with indicative transmission electron microscopy (TEM) images [24]. Imran et al. advanced further the HI route by means of employing benzoyl halides as the perovskite halide precursor [76]. The great advantage of this approach is the ability to feasibly regulate the composition of the final PNCs, on the contrary to HI routes where $PbX_2$ salts are employed as both lead and halide precursors. Both $MAPbX_3$ and $FAPbX_3$ PNCs with good size distribution control and phase purity were synthesized. Figure 3b shows a schematic animation of the HI colloidal synthesis, whereas Figures 3c-e depict indicative TEM photos of the so-formed $MAPbCl_3$, $MAPbBr_3$, and $MAPbI_3$ PNCs [76]. Similar TEM profiles were captured for the $FAPbX_3$ compounds (not shown here). Moreover, Figures 3f-h present the corresponding PL and optical absorbance spectra. Inspections of the latter figures reveals sharp PL features at 404 nm, 527 nm, and 730 nm, for the Cl, Br, and I PNCs, respectively, while the obtained quantum yields were determined equal to 5%, 92%, and 45% [76].

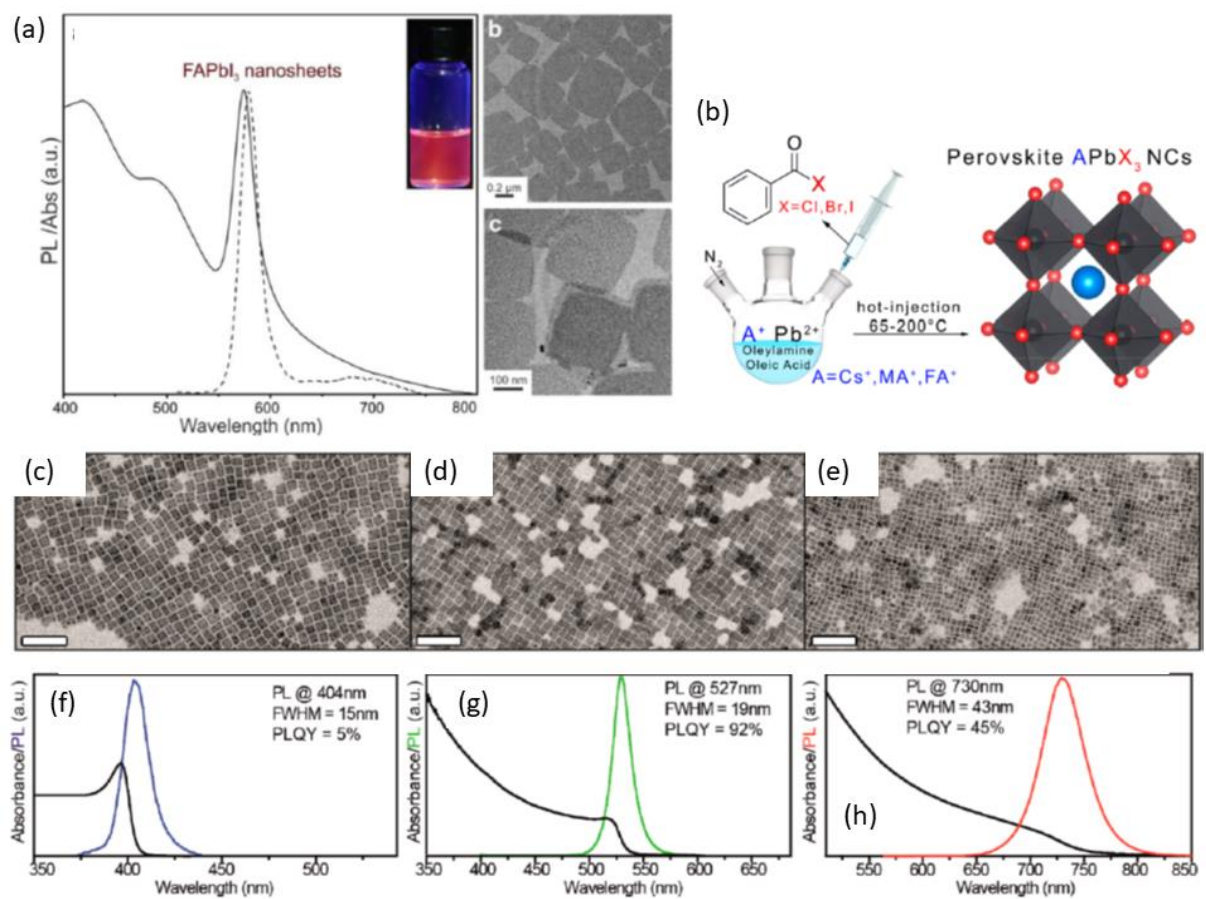

**Figure 3:** (a) Photoluminescence (PL) and absorbance spectra of the of the $FAPbI_3$ NSs, along with indicative transmission electron microscopy (TEM) images. (a has been reproduced from ref. [24] with permission from American Chemical Society,

copyright 2017). (b) Schematic animation of the HI colloidal synthesis with benzoyl halides. Transmission electron microscopy (TEM) photos of the so-formed $MAPbCl_3$ (c), $MAPbBr_3$ (d), and $MAPbl_3$ PNCs (e). Absorbance and photoluminescence (PL) spectra of $MAPbCl_3$ (f), $MAPbBr_3$ (g), and $MAPbl_3$ PNCs (h). (b-h have been reproduced from ref. [76] with permission from American Chemical Society, copyright 2018).

We move on now to present the progress of LARP method for the synthesis of organic-inorganic PNCs. In this widely used approach, the perovskite precursors are dissolved in a good polar solvent such as N,N-dimethylformamide (DFM) or dimethylsulfoxide (DMSO), and a poor nonpolar solvent such as toluene or hexane, is added in the presence of ligands to induce the formation of PNCs by means of supersaturation. Figure 4a depicts the method schematically [28]. Typically, for organic-inorganic perovskites this approach produces nanoplatelets (NPls), in which the thickness is controlled by varying the ratio between long-chain and short-chain alkylammonium ligands [28,34]. Alternatively, in some LARP cases the addition of acetone (polar aprotic solvent) to a nonpolar solvent containing the precursors and ligands induces also the growth of NPls at room temperature.

Let us first consider various synthesis cases of $MAPbBr_3$-based PNCs over the years by means of LARP route. Sichert et al. in 2015 demonstrated the synthesis of $MAPbBr_3$ NPls by means of LARP [77]. Namely, by altering the ratio of the employed organic cations the thickness of the so-formed NPls was controlled. As a result, NPls with tunable emission ranging from blue to green were fabricated (Table 1). A year later, Yuan et al. synthesized $MAPbBr_3$ NPls with quasi-2D layered structures [78]. Different ligand lengths and precursors ratios were employed in order to plausibly alter the number of layers, and, consequently, the optical properties of the NPls. PL emissions from deep-blue to intense green were achieved (Table 1) [78]. Similarly, upon varying the cation, metal, and halide composition, Weidman et al. demonstrated the feasible tuning of PL emission of $MAPbBr_3$ NPls within the 398-437 nm range (Table 1) [79], whereas for the same aim Cho et al. employed various ligand chain lengths and ratios [80], and Bhaumik et al. altered the ligand and precursor ratio (Table 1) [81]. In addition, Kumar et al. developed ultrathin $MAPbBr_3$ PNCs with efficient blue electroluminescence upon employing low-k organic hosts at various ratios [82]. Another set of $MAPbX_3$ NPls with X=$Br^-$ or $I^-$ were synthesized by Levchuk et al. [83]. The authors performed ligand-assisted tailoring for controlling the thickness and properties of the PNCs. In particular, the oleylamine and oleic acid ratio was varied. As a result, broadly tunable emission wavelengths in the 450-730 nm range were achieved, by means of the quantum size effect, i.e. without anion-halide mixing [83]. In another study, Ahmed et al. highlighted the role of pyridine and temperature during LARP synthesis of $MAPbBr_3$ NPls [25]. The authors found that the presence of pyridine causes the transformation of three-dimensional (3D) cubes into 2D nanostructures, as depicted schematically in Figure 4(b). The so-formed PNCs exhibit high photoluminescence quantum yields (PLQY) in the visible, rendering themselves suitable for optoelectronic applications.

Apart from $MAPbBr_3$, the synthesis of $MAPbl_3$ PNCs also has attracted significant attention over the recent years (Table 1) [79,83–87]. Weidman et al. developed $MAPbl_3$ NPls by means of LARP, while tuning the PL properties in the 512-573 nm range upon altering the composition [79]. Similarly, Stoumpos et al. formed $MAPbl_3$ NPls with tunable emission in the 527-652 nm range [84], and Blancon et al. in the wider 516-677 nm range [85], upon changing the precursors ratio (Table 1). Hautzinger et al. focused on the A-cation cavities of the developed PNCs, namely attributing the observed band gap shifting to the variation of chemical pressure inside the cavity and the Pb-I framework [87].

Similar LARP approaches have been followed over the years when the longer chain formamidinium cation is employed for the development of $FAPbX_3$-based PNCs (Table 1) [26,79,87–91]. As was the case for methylammonium-based perovskites, the main tuning tools for the composition, morphology, and optical properties of the $FAPbX_3$ crystals rely on employing various types of ligands, as well as different ligands and precursors ratio. Indicatively, Levchuk et al. demonstrated the facile room temperature synthesis of a $FAPbX_3$ (with X=Cl⁻, Br⁻ and I⁻ or mixed) PNCs by means of LARP [26]. Figure 4c shows a photograph of typical dispersion solutions of the so-formed PNCs under UV-light, along with the PL spectra that cover a wide range of the visible, i.e. from 415 to 760 nm. Moreover, the authors showed a nice thickness tuning and a variation of the band gap with respect to the number of NPls layers stacked together (Figure 4c). Figure 4c depicts also a typical TEM photo of vertically stacked $FAPbBr_3$ NPls. Furthermore, it was proposed that upon appropriate surface modifications, the PNCs crystals exhibited great resistance over water, and thus, making them suitable candidates for optoelectronic and photonic applications, that demand extra stability. Finally, it is worth to note that, the LARP method can be adopted for the synthesis of lead-free organic-inorganic PNCs. As for instance, Weidman et al. reported on the synthesis of $FASnI_3$ PNCs with tunable emission within the 628-689 nm range (Table 1) [79].

In the final part of the Section, we will consider alternative synthesis approaches, summarized also in Table 1, i.e. other than typical HI and the common LARP procedure. Hintermayr et al. reported on a simple, versatile, and efficient two-step process for the synthesis of $MAPbX_3$ PNCs (with X=Cl⁻, Br⁻ and I⁻) [92]. Based on this method, methylammonium halide and lead halide powders are dry-grounded first, while the perovskite formation is monitored by the color change. After the grounding, the perovskite powder is dispersed in toluene with oleylamine and sonicated for 30 min. Following sonication, the mixture provides highly fluorescent PNCs of good crystalline quality, however the so-formed crystals are polydisperse in both size and shape. Lignos et al. prepared blue-emitting $FAPb(Cl_{1-x}Br_x)_3$ PNCs by means of an automated microfluidic platform [27]. The main principle of this method is the use of a droplet-based microfluidic reactor that permitted a thorough screening of the reaction parameters, and thus the elucidation of optimum Cl/Br ratios. After this optimization the mixtures were transferred to conventional reactor flask for upscaling. As depicted in Figure 4d, PNCs with tunable emission between 440 and 515 nm were synthesized, while indicative TEM photos are also included. Finally, it should be noted that ultrafast laser processing fabrication techniques have been employed for the development of high-resolution luminescent 2D perovskite patterns. Such methods offer the advantage of avoiding solvents and chemical processes, while allowing the controlled growth of the perovskite components.[93,94]

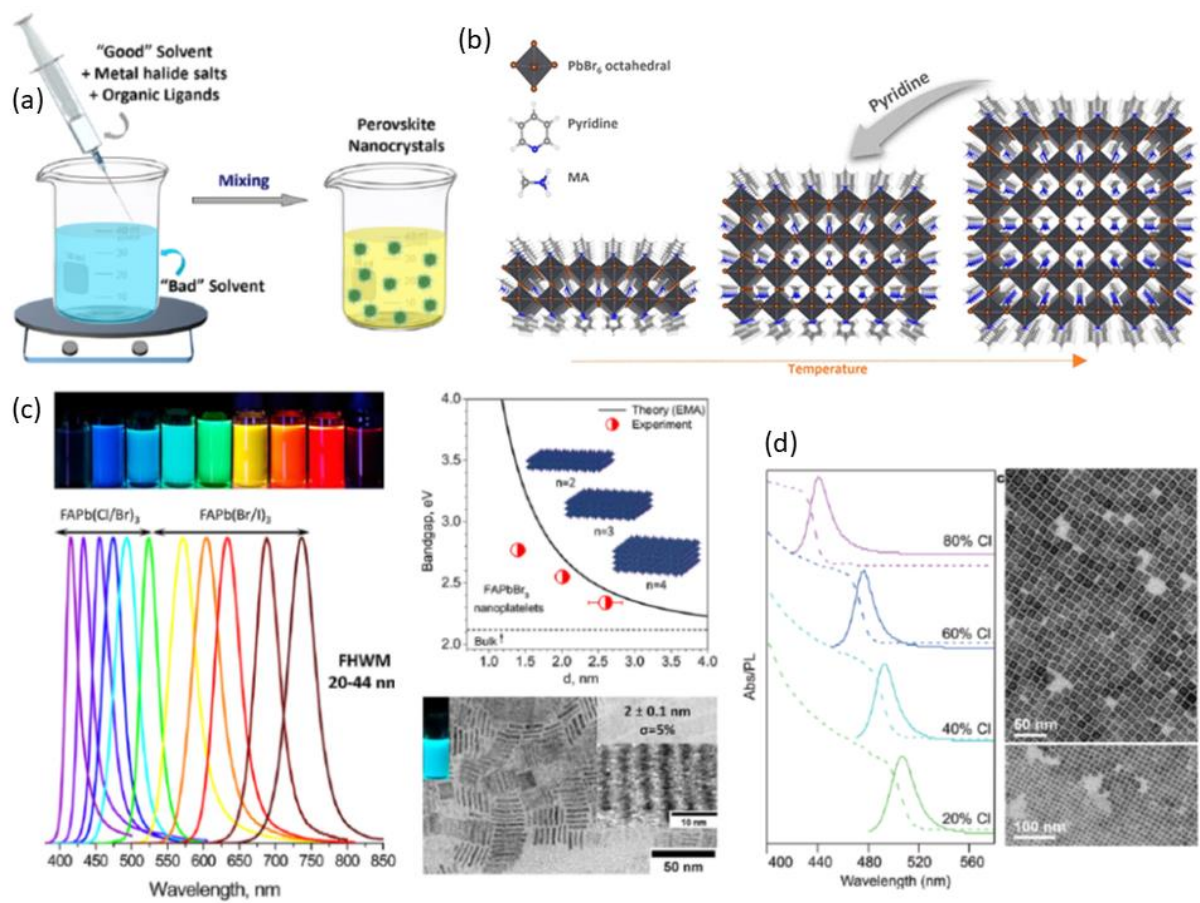

**Figure 4:** (a) Schematic representation of the ligand-assisted reprecipitation (LARP) method. (a has been reproduced from ref. [28] with permission from American Chemical Society, copyright 2019). (b) Transformation of three-dimensional (3D) cubes into 2D nanostructures in the presence of pyridine. (b has been reproduced from ref. [25] with permission from American Chemical Society, copyright 2017). (c) Photograph of typical dispersion solutions of the so-formed FAPbX$_3$ PNCs under UV-light, along with the photoluminescence (PL) spectra that cover a wide range of the visible. On the right, energy bandgap variation versus thickness and typical transmission electron microscopy (TEM) photos of the PNCs. (c has been reproduced from ref. [26] with permission from American Chemical Society, copyright 2017). (d) Absorbance and PL of FAPb(Cl$_{1-x}$Br$_x$)$_3$ PNCs prepared by means of an automated microfluidic platform. (d has been reproduced from ref. [27] with permission from American Chemical Society, copyright 2018).

**Table 1:** Summary of two-dimensional (2D) organic-inorganic perovskite nanocrystals (PNCs) synthesis routes. *NR stands for not reported.

| Study-year | Composition | Key features | Photoluminescence (PL) range |
|---|---|---|---|
| **Hot Injection (HI) synthesis** | | | |
| Vybornyi et al., 2016[23] | $CH_3NH_3PbX_3$ | Nonpolar solvent, various shapes of PNCs (nanocubes, nanowires, nanoplatelets) | 460-750 nm |
| Protesescu et al., 2017[24] | $FA_{0.1}Cs_{0.9}PbI_3$ | $PbX_2$ + FA-oleate, FA-oleate + Pb-oleate, Nanosheets | 690-780 nm |
| Imran et al., 2018[76] | $APbX_3$, A=$MA^+$ and $FA^+$, X=$Cl^-$, $Br^-$, $I^-$ | Benzoyl halide precursor, PNCs | 404-764 nm |
| **Ligand-assisted reprecipitation (LARP)** | | | |
| Sichert et al., 2015[77] | $MAPbBr_3$ | Organic cation ratio, NPls | 427-508 nm |
| Yuan et al., 2016[78] | $MAPbBr_3$ | Ligand length, precursors ration, NPls | 403-530 nm |
| Weidman et al., 2016[79] | $MAPbBr_3$ | Cation, metal, halide ratio, NPls | 398-437 nm |
| Cho et al., 2016[80] | $MAPbBr_3$ | Ligand-mediated thickness modulation, NPls | 433-506 nm |
| Bhaumik et al., 2016[81] | $MAPbBr_3$ | Ligand-precursors ratio, NPls | 440-490 nm |
| Kumar et al., 2016[82] | $MAPbBr_3$ | Ligands ratio, NPls | 436-489 nm |
| Levchuk et al., 2016[83] | $MAPbBr_3$ | Ligand-assisted tailoring, NPls | 450-730 nm |
| Ahmed et al., 2017[25] | $MAPbBr_3$ | Pyridine-temperature effect, NPls | 445-508 nm |
| Weidman et al. 2016[79] | $MAPbI_3$ | Cation, metal, halide ratio, NPls | 512-573 nm |
| Levchuk et al., 2016[83] | $MAPbI_3$ | Ligand-assisted tailoring, NPls | 549-722 nm |
| Stoumpos et al., 2016[84] | $MAPbI_3$ | Precursors ration, NPls | 527-652 nm |
| Blancon et al., 2017[85] | $MAPbI_3$ | Precursors ratio, NPls | 516-677 nm |
| Dahlman et al., 2020[86] | $MAPbI_3$ | Ligands composition and ratio, NPls | 510-650 nm |
| Hautzinger et al., 2020[87] | $MAPbI_3$ | NR | 579 nm |
| Levchuk et al., 2017[26] | $FAPbBr_3$ | Ligands ratio, NPls | 438-530 nm |
| Minh et al., 2017[88] | $FAPbBr_3$ | Ligands ratio, NPls | 438-533 nm |
| Yu et al., 2018[89] | $FAPbBr_3$ | Ion-exchange mediated self-assembly, NPls | 398-490 nm |
| Fang et al., 2019[90] | $FAPbBr_3$ | Ligands ratio, NPls | 440-532 nm |

| Peng et al., 2020[91] | FAPbBr$_3$ | NR, NPLs | 440 nm |
|---|---|---|---|
| Weidman et al., 2016[79] | FAPbI$_3$ | Cation, metal, halide ratio, NPIs | 512-575 nm |
| Levchuk et al., 2017[26] | FAPbI$_3$ | Ligands ratio, NPIs | 668-737 nm |
| Hautzinger et al., 2020[87] | FAPbI$_3$ | NR | 582 nm |
| **Alternative synthesis routes** | | | |
| Hintermayr et al., 2016[92] | MAPbX$_3$ (X=Cl$^-$, Br$^-$, I$^-$) | Two-step method, dry-grounding, sonication, Polydisperse PNCs | 400-750 nm |
| Lignos et al., 2018[27] | FAPb(Cl$_{1-x}$Br$_x$)$_3$ | Microfluidic platform, NPIs | 440-520 nm |

## 2.2 All-inorganic metal halide perovskite nanosheets/nanoplatelets

### 2.2.1 All-inorganic metal halide NPls/NSs synthesis

The synthesis methods for the fabrication of all-inorganic metal halide perovskites NPls and NSs can be classified into six categories: i) ligand-assisted hot injection and precipitation, ii) spontaneous crystallization in nonpolar organic media, iii) cation-mediated colloidal method, iv) solvothermal method, v) ligand- and laser-assisted exfoliation and vi) external stimuli-triggered process (Table 2). In the first method, perovskite NPls/NSs can be obtained by injected monovalent cation-oleic acid complex (Cs-oleate) in the reaction mixture, usually containing divalent metal salt, octadecene, Oleic acid, Oleylamine at a specific reaction temperature. [1] Furthermore, 2D nanomaterials can be fabricated by a room temperature ligand-assisted repreciptation (LARP) method. In this synthesis approach, CsX and $PbX_2$ that were selected as ion sources, were dissolved in a good solvent such as N,N-dimethylformamide (DMF), or dimethylsulfoxide (DMSO), together with the ligands. Then a small quantity of this solution was injected in a bad solvent such as toluene or hexane at room temperature producing immediately a highly supersaturated state and inducing then the rapid recrystallization of the nanocrystals.[95] In the spontaneous crystallization of the 2D nanocrystals, simple mixing of the precursor-ligand complexes in organic media at ambient atmosphere was carried out. [29] No polar solvent as DMF or DMSO was needed to dissolve the precursors; instead, ligands act as coordinating solvents. In the hot injection method, the acid–base equilibrium of the system can disturbed by $Sn^{4+}$ cations injected by $SnX_4$ salts facilitating the formation of 2D structures. [30] In solvothermal fabrication method, the reaction between Cs-oleate and $PbX_2$ precursors takes place in a Teflon-line autoclave at relatively low temperature. [31] 2D metal halide nanocrystals can be also by the exfoliation and then fusion of bulker materials through a ligand mediated process [16] or by irradiation with laser. [17] In the last fabrication process, a tip-ultrasonication device or microwave tube was used for the exfoliation of bulk perovskites to thinner 2D morphologies or for the mixing of the precursors with the ligand in a non-polar solvent.[9,18]

**All inorganic metal halide nanoplatelets.** The first synthesis of 2D $CsPbBr_3$ NPls morphology reported in 2015 by Bekenstein et al., using the ligand-assisted hot injection method Figure 5a-b.[40] This synthesis was carried out by modifying the nanocubes' synthesis protocol reported previously by Protesescu et al. which performed at 140-200 °C. [96] This study revealed that when the temperature was decreased at 90-130 °C the asymmetric growth and the formation of 2D $CsPbBr_3$ NPls was favored. Specifically, the precursor $PbBr_2$ was solubilized in octadecene (ODE) together with the capping ligands OA and Olam, and then the Cs-oleate was injected at elevated temperatures (90-130ºC) to form the NPls. The temperature at which the Cs-oleate was injected was crucial parameter for the morphology and size of the nanocrystals. Symmetrical nanocubes were formed for the reactions conducted at 150 °C, NPls of lower symmetry at 130° and very thin NPls with lateral size of 200 to 300 nm at 90 and 100 °C (Figure 5a, b). However, the problem with these too thin structures were to centrifuge and collect them. After the first centrifugation and redispersion, any additional cleaning step is non-trivial, as they dramatically affected their bright PL and degrade the NPls. Upon testing a variety of antisolvents that would destabilize the NPls dispersion, the ethyl acetate and methyl-ethyl-ketone with polarity indexes of 4.4 and 4.7 respectively were sufficient to precipitate the NPls but also prevent their degradation. These NPls showed surprising PLQYs that reach the value of 84.4% despite that these 2D morphology would be expected to be sensitive to surface defects and very low PLQY due to the inherently high surface to volume ratio. The PL emission shifts from green (512 nm 2.5 eV) to deep purple (405 nm 3 eV) for the strongly quantum confined band edge emission (Figure 5c). Such PLQYs were observed only for semiconducting nanocrystals passivated from a second layer. Their emission spectra can be tuned to cover all of the visible spectrum by introducing an anion exchange process in which the $Br^-$ anions are replaced with either $Cl^-$ or $I^-$ through which the NPls' composition can be tuned (Figure 5c).

Later Akkerman et al, synthesized $CsPbBr_3$ NPls by adding HBr during the reaction (Figure 3d-e). The reaction in this case was taken place at room temperature while the precursors and ligands were the same with the previous report. [12] In particular, the synthesis was starting from a mixture of Cs-oleate, Olam, OA, $PbBr_2$ and HBr that remained stable unless an additional component was added to it. Then, acetone was added to the precursor mixture. The role of the acetone was to destabilize the complexes of $Cs^+$ and $Pb^{2+}$ ions with the various molecules in solution and therefore to trigger the nucleation of the particles. Other polar solvents, for example, protic solvents like isopropanol and ethanol, were also tested but they were not such effective as acetone in the control of the NPls' shape. The stability of the NPls was affected by the number of layers: the 4 monolayered (ML) sample, emitting at 2.75 eV, was only stable for a day, whereas the 3 and 5 ML samples were stable for at least one month, with only the appearance of a secondary emission peak at 2.36 eV, as well as a sedimentation of a green/yellow precipitate, indicating the formation of larger nonquantum confined CsPbBr$_3$ aggregates. The optical bandgap was remarkably sensitive to the number of layers in this 3–5 ML regime, as shown in Figure 5f, with a blue shift and strengthening of the excitonic transition for thinner sheets as a result of confinement effects. Their emission spectra can be tuned by changing the halide anion (Figure 5f).

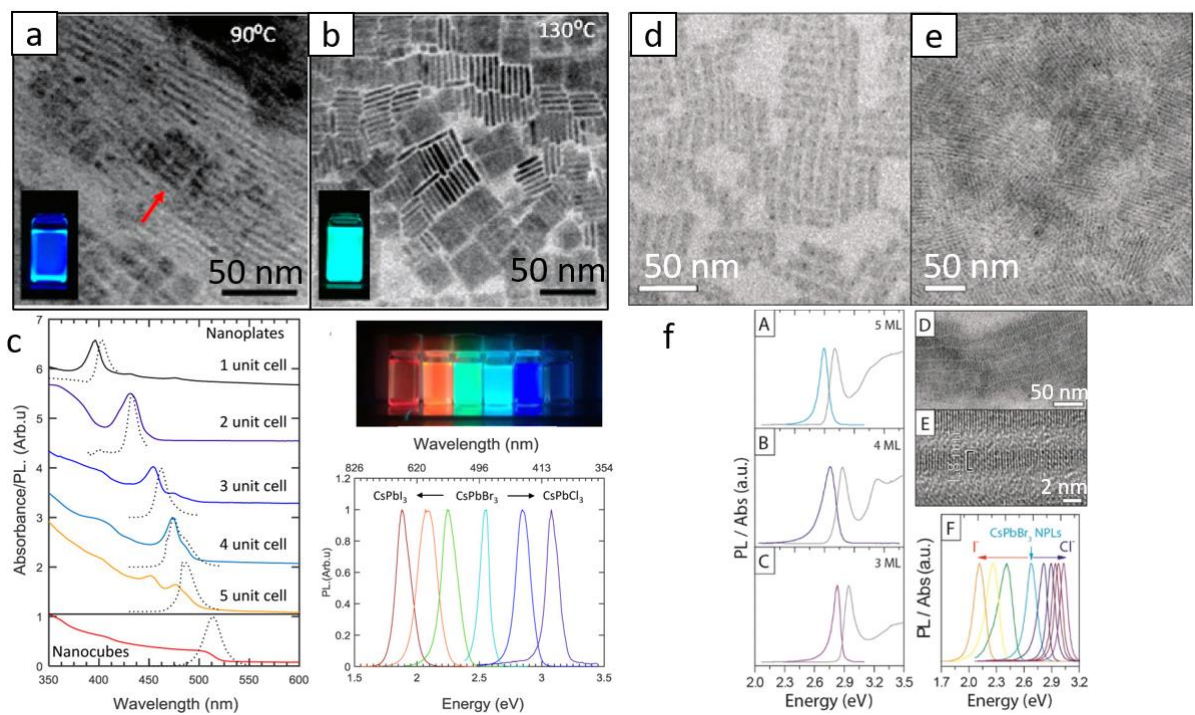

**Figure 5:** First all-inorganic metal halide perovskite NPls synthesized at high temperature (hot-injection method, reaction temperature 90 and 130 °C) (a-b) and at room temperature (LARP) (d, e, low and high concentration samples). Optical properties for NPls of different thicknesses and halide ratio (through anion exchange reactions) (c, f). Reprinted with permission from Ref.[40], copyright 2015, American Chemical Society (a-c). Reprinted with permission from Ref.[12] , copyright 2016, American Chemical Society, https://pubs.acs.org/doi/10.1021/jacs.5b12124 that further permissions related to the material excerpted should be directed to the ACS (d-f).

Almost at the same time, CsPbX$_3$ NPls have been reported by Sun et al by a mixing of the precursors in a good solvent (DMF) and injected a small amount of this into a poor solvent (toluene) at room temperature (ligand-assisted reprecipitation-LARP). [97] The color of the solution was changed indicating the formation of CsPbX$_3$ due to the co-precipitation of Cs$^+$, Pb$^{2+}$, and X$^-$ in the presence of the organic acid and amine ligands. In the precursor solution, polar solvent DMF was acted as a good solvent to dissolve the inorganic salts and molecule ligands, while the nonpolar solvent toluene as a bad solvent to facilitate the reprecipitation process. The final morphology of the nanocrystals obtained by this precipitation-based approach was determined by the type of the organic acid and the type of the amine ligands (Figure 6a). Quantum dots, nanocubes, nanorods and few-unit- cell-thick NPls have been obtained by different combination of organic ligands (Figure 6a). In particular, NPls were formed by mixing oleic acid and octylamine. These NPls had a typical edge length of 100 nm and a typical thickness of 5.2 nm corresponding to 4 monolayers of CsPbBr$_3$. Similar chemical approach has been reported also for the synthesis of Cs$_4$PbBr$_6$ NPls.[98] The anisotropic growth anisotropic 2D growth of CsPb$_2$Br$_5$ chemical phase is facilitated by the intrinsic symmetry of its tetragonal crystal structure.

The effect of the ligand type and chain length on the size and shape of the all-inorganic NPls has been studied for the hot injection approaches by Pan al al. [99] Different combination of carboxylic acids and organic amines with varying chain lengths were utilized. (Figure 6b). Both ligands were bound on the surface but the ammonium ligands were seemed to be detached easier during the centrifugation cleaning processes. From reactions carried out at 170 °C using oleylamine and carboxylic acids with decreasing carbon chain lengths, nanocube with increasing edge lengths were obtained, and from reactions carried out at 170 °C using oleic acid and amines with shorter carbon chain lengths, NPls were obtained except using oleylamine. Also, the thickness of the NPls is depended of the ligand chain-length, thinner NPls down to three-unit cells thickness were produced when shorter chain amines were introduced in the reaction.

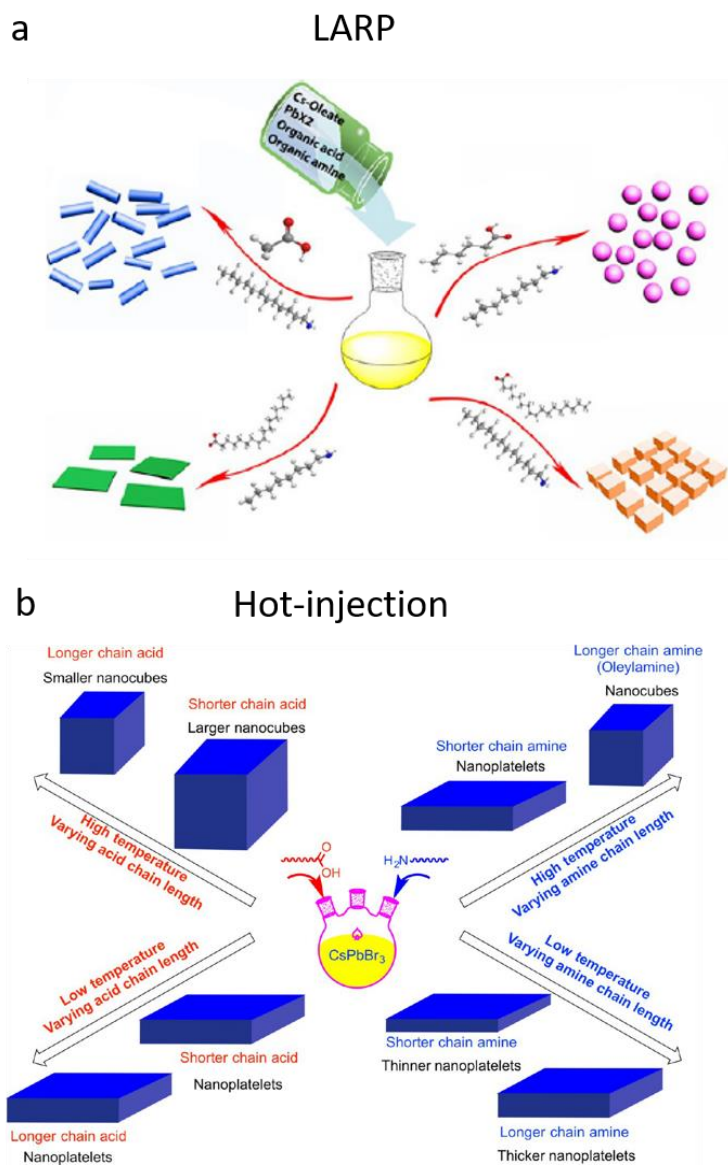

**Figure 6:** Schematic illustrating the effect of the ligand type on the morphology of the final all-inorganic metal halide perovskite nanocrystals synthesized by LARP method at room temperature (a) and hot injection (b) method. Reprinted with permission from Ref. [97] , copyright 2016, American Chemical Society. Reprinted with permission from Ref.[99] , copyright 2016, American Chemical Society.

Furthermore, Samsi et al. [21] confirmed what Bekenstein et al reported earlier[11] that the 2D NPls formation can be facilitated when the Cs-oleate dissolved in OA instead of the octadecene that resulted in nanocubes[96]. In addition, according to this protocol, the NPls were obtained in a wider temperature range (50-150 °C) compared to the previous report in which the NPls were formed in the range of 90 to 130 °C. Moreover, by tuning the time of the reaction from a few seconds to a few minutes, NPls of lateral size from 50 to 200 nm can be obtained but when shorter ligands were introduced nanosheets of up to 5 µm were taken with thickness in the strong quantum confinement effect. Octanoic acid (OctAc) and octylamine (OctAm) were added in addition of the OA and Olam. The use of both short ligands in conjunction was necessary and the optimum NSs were obtained for the temperature range of 145-155 °C and time of the reaction up to 5 min. The PLQY of these NSs found to be ~33%.

Two years later, Almeida et al., reported the role of the concentration of the ligands and acidity in the hot injection approaches of the CsPbBr₃ nanocrystals at high temperature. [19] By changing acid-base ligand pair (OLA-OA), NPls with thickness down to a single monolayer can be acquired. Almost simultaneously, Sheng et al studied the role of an additional halide in the final morphology of the CsPbBr₃ NPls. [100] Among multiple tuning factors such as component and ratio of the metal halides, temperature, time, and ligands, the additional metal halide was seemed to be the essential one for the perovskite NPls formation. The evolution of morphology by tuning the ratio [CuCl₂]: [PbBr₂] using the CuCl₂ as additional halide is illustrated in the Figure 7. Furthermore, the type of the additional halide was played a crucial role in the final NPls' size. For example, $CsPbCl_xBr_{3-x}$ NPls obtained by addition of $CuCl_2$ have the average size of ≈2.9 × 30 × 30 nm; $AlCl_3$ leads to ≈ 4.6 × 12 × 12 nm; $HAuCl_4$ yields ≈ 3.5 × 40 × 40 nm; $ZnCl_2$ results ≈2.3 × 26 × 26 nm. $NH_4Cl$ or $NH_4Br$ as additive, the products were the perovskite nanocubes in majority, while $Cu(NO_3)_2$ or Cu-oleate no perovskite nanocrystals can be obtained. Finally, by increasing the ratio of the ligands [Olam]/[OA], larger NPls were obtained.

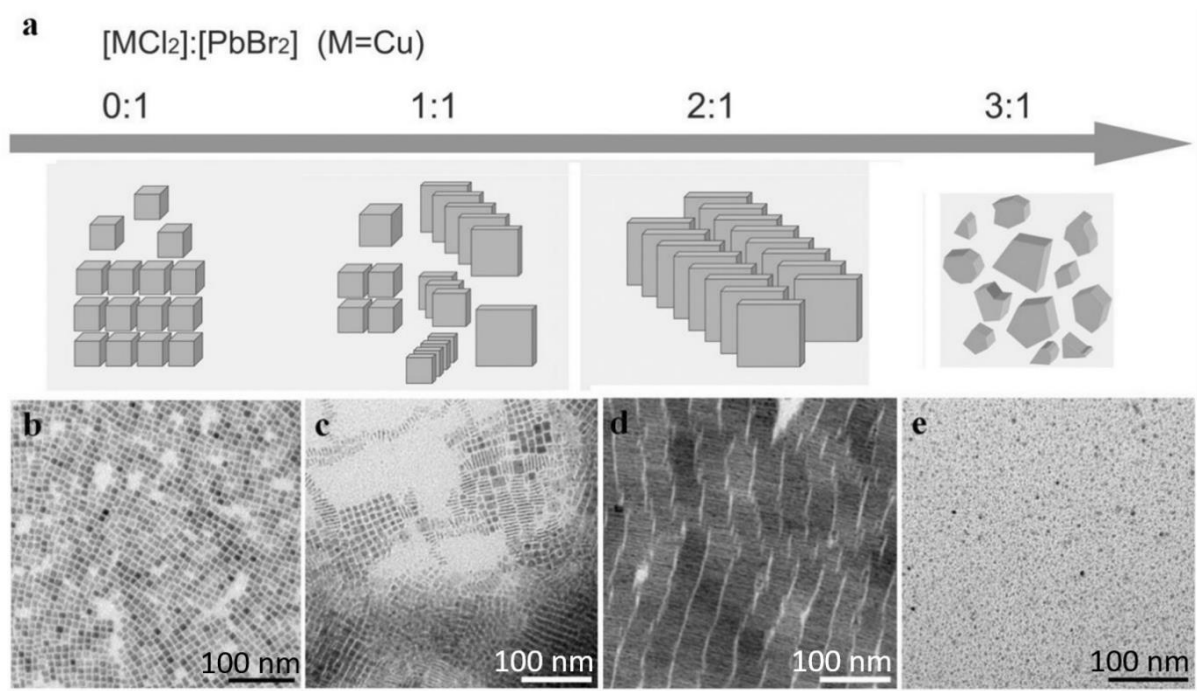

**Figure 7:** Effect of the metal halides ratio on the NPls morphology in the hot injection approach. Reprinted with permission from Ref.[100], copyright 2018, Wiley.

In the quest to find the optimum conditions for stable NPls of high PLQYs, different parameters have been tuned. For example, Bertolotti et al reported that the reduction of the reaction temperature (115 °C) and using two-fold higher concentrations of reagents compared to that for nanocubes in the hot injection methods, ultra-uniform NPls of six-monolayers-thickness have been formed [101] while by using a different Cs precursor such as cesium cholate the PLQYs of the NPls reached the value of 94 %.[20] Furthermore, the addition of a short ligand instead of Olam such as the hexylphosphonate ligand was also a good alternative ligand for the formation of ultra-thin NPls down to the 2.4 nm thickness.[102]

In a different study, the role of the aging of the oleate in LARP synthesis has been evaluated by Zeng et al.[103] In this approach, ultra-thin CsPbBr₃ NPls with PLQYs approaching the unity and high stability have been synthesized using a triple-source ligand assisted reprecipitation method (TSLARP) with an aged metal-oleate precursors (Figure 8a). Using fresh Pb-oleate precursor, hybrid CsPbBr$_3$@Cs$_4$PbBr$_6$ NPls with low PLQY (28 %) were formed while with aged one, CsPbBr$_3$ NPls with PLQYs of 97.4% were obtained. Cs-oleate, Pb-oleate, and OAmBr (octylamine, HBr, ethanol) precursor solutions were dropped into toluene solvent in sequence and then vigorously stirred for 5 min (Figure 8a), yielding NPls' solution with blue emission under UV excitation. The aging treatment of Cs-oleate and Pb-oleate precursors was performed in sealed glass vials and stored in air for different times (from 0 month to 12 months). The PL and absorbance curves of the NPls obtained with the aged oleate under different times are illustrated in the Figure 8b and the NPls with the highest PLQY are those with the aged oleate for 12 months (upper curve in Figure 8b). This different behavior found that could be originated from the transformation of the Pb-oleate in toluene from isolated molecules into clusters after the aging process (Figure 8c). The Pb oleate clusters serve as a template for the formation of the clustered [PbBr$_x$] octahedra intermediates, which thermodynamically favors the formation of the CsPbBr$_3$ NPls (Figure 8c).

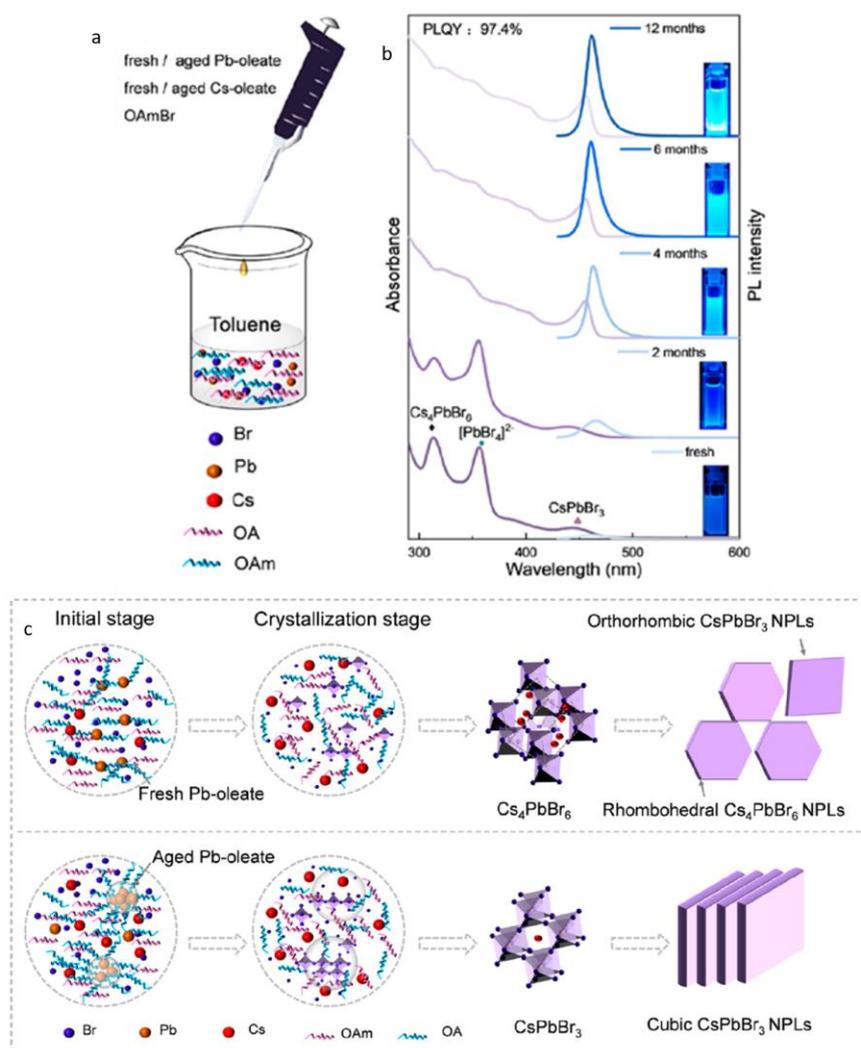

**Figure 8:** All-inorganic metal halide perovskite NPls synthesized with triple-source ligand assisted reprecipitation method (a). Absorption spectra (purple) and corresponding PL spectra (blue) of the CsPbBr$_3$ NPls' solutions synthesized using metal-oleate precursors with different aging times (b). Proposed mechanism of the NPls formation (c). Reprinted with permission from Ref.[103], copyright 2021, American Chemical Society.

A different approach than LARP but conducted also at room temperature has been reported by Huang et al. that does not require external stimuli such as heat, microwave irradiation, ultrasonication, mechanical force, or polar solvent, in which perovskite nanocrystals were easily obtained through spontaneous crystallization upon simple mixing of precursor-ligand complexes in organic media at ambient atmosphere. No polar solvent such as DMF or DMSO was needed to dissolve the precursors; instead, ligands act as coordinating solvents. This protocol could be transfer in large scale (× 50 scale up) as the precursor stock solutions remain stable upon the time.[29]

Cation mediated approaches, also found that can result in the formation of the 2D nanostructures. In a typical synthesis of CsPbBr₃ NPIs reported by Ding et al., a mixture of PbBr₂, SnBr₄ and Cs₂CO₃ was loaded in a three-neck flask containing ODE, OA and Olam and heated until 110°C. [30] The presence of foreign Sn⁴⁺ ions in the precursors led to the formation of thin perovskite NPIs with a thickness of around 3 unit-cells. Increasing the Sn⁴⁺ (or Sn²⁺ in the open-air case) amount led to an increasing fraction of thin NPIs in final solution. In the same direction, Bonato et al added SnX₄ salts during the hot injection process and the results were similar to that of previous report. [104] The acid–base equilibrium of the system was disturbed by Sn⁴⁺ cations increasing OA/Ola molar ratio, which induces the formation of the NPIs. The formation of tin–oleylamine complexes proved that was the cause of such disturbance. The NPIs formation was facilitated by the addition of the Sn⁴⁺ cations but no signal assigned to those species was confirmed from the X-ray photoelectron spectroscopy (XPS) data. It is important to notice here that two previous reports also discussed the role of the additional halide in the formation of the NPIs. [19,100]

Moreover, NPIs found to be promoted by using hydrophobic acid (Trp, Cys, His, Leu, Phe, and Trp Derivatives)[105] or by adding a controllable amount of water[106]. Different passivation strategies have been applied to increase the PLQY of the NPIs.[22,107] The addition of PbBr₂-ligand solution was found that can repair surface defects likely stemming from bromide and lead vacancies in a subensemble of weakly emissive nanoplatelets and the PLQY can be increased from 7 to 49 % for NPIs of 2ML thickness and form 42 to 73 % for those of 6ML.[107] Moreover, the addition of HBr providing additional Br⁻ was introduced to drive the ionic equilibrium and form intact Pb-Br octahedra. [22] With the second strategy NPIs of 96 % PLQY have been fabricated.

In addition to LARP and hot-injection methods to synthesize NPIs, a pre-dissolution assisted solvothermal method has been introduced to fabricate CsPbBr₃ NPIs of ~28 % PLQY. In particular, the PbX₂ precursor was dissolved in octadecene and dried OA and Olam was injected at 120 °C. After complete solubilization of the Pb- precursor, the temperature was decreased at room and a Cs-oleate stock solution was injected slowly. Then this solution was transferred in a Teflon-lined autoclave and let at 140 °C for different reaction times (Figure 9 a-e). [108] Almost simultaneously, a similar approach has been reported from Zhai et al[31] but in this protocol the reaction temperature was lower (100 °C) while the tuning parameter was the same, the reaction time. In this case the NPIs seemed to be more homogeneous with very sharp edges compared to the previous NPIs synthesized with solvothermal method and the PLQYs reached the value of 50 % (Figure 9 f-j). The bandgap of the NPIs was gradually decreased upon an increase in the reaction time in both reports (Figure 9 e, j).

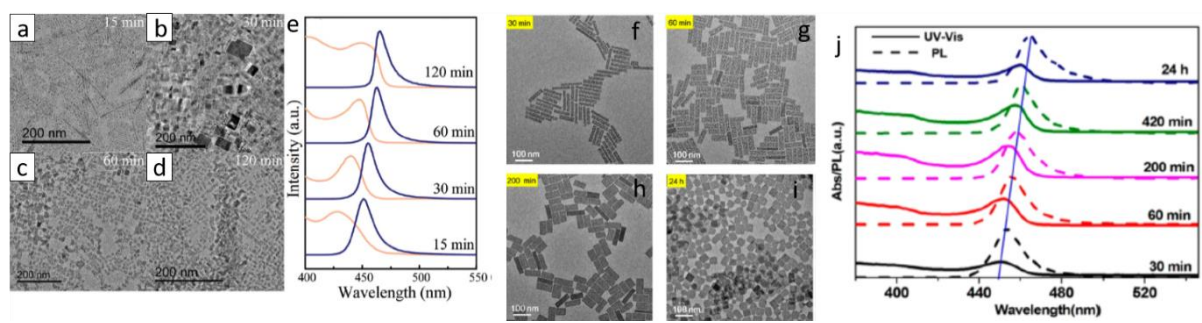

**Figure 9:** NPIs synthesized with solvothermal method at 140 (a-d) and 100 °C (f-i) with different reaction times. Optical absorption and emission spectra for the corresponding samples (e, j). (a-e) Reprinted with permission from Ref.[108], copyright 2018, The Royal Society of Chemistry. (f-j) Reprinted with permission from Ref.[31], copyright 2018, American Chemical Society.

Ligand mediated anion exchange route also used for the fabrication of the all-inorganic metal halide NPIs.[16] Treatment of the pre-synthesized CsPbX₃ nanocrystals with Dichlorodiphenyltrichloroethane (DDT) and AIX₃ was included in this process. The pre-synthesized nanocrystals first were exfoliated and then was fused in larger NPIs structures (Figure 10). The exfoliation, the removal of the ligands from the nanocrystals and the fusion were taken place almost simultaneous. The thickness of the fabricated NPIs was 4-15 monolayers and the PLQYs of the them was 50-65% for the violet-blue emitting NPIs, near unity for the green and 81% for the red emitting NPIs. Exfoliation and then fusion were also occurred when nanocrystals were irradiated with a fs laser with 513 nm wavelength (Figure 10 d-f). [109] Specifically in this case except of the shape transformation also a dimensionality transformation was taken place. $Cs_4PbBr_6$ hexagonally-shaped nanocrystals were irradiated in a liquid environment. An exfoliation of the initial nanohexagons was taken place in the first 30 s of the irradiation, then these particles started to fragment into smaller cubic particles until 4 min, and finally, these small cubes were enlarged to nanoplatelet-type morphologies until 12 min through a side-by-side-oriented attachment and fusion process. Sheets of lateral sizes of around 1–1.5 μm could also be obtained for longer irradiation times. Partial anion exchange was occurred due to the chlorinated solvent very fast and completed in the case of the NPIs.

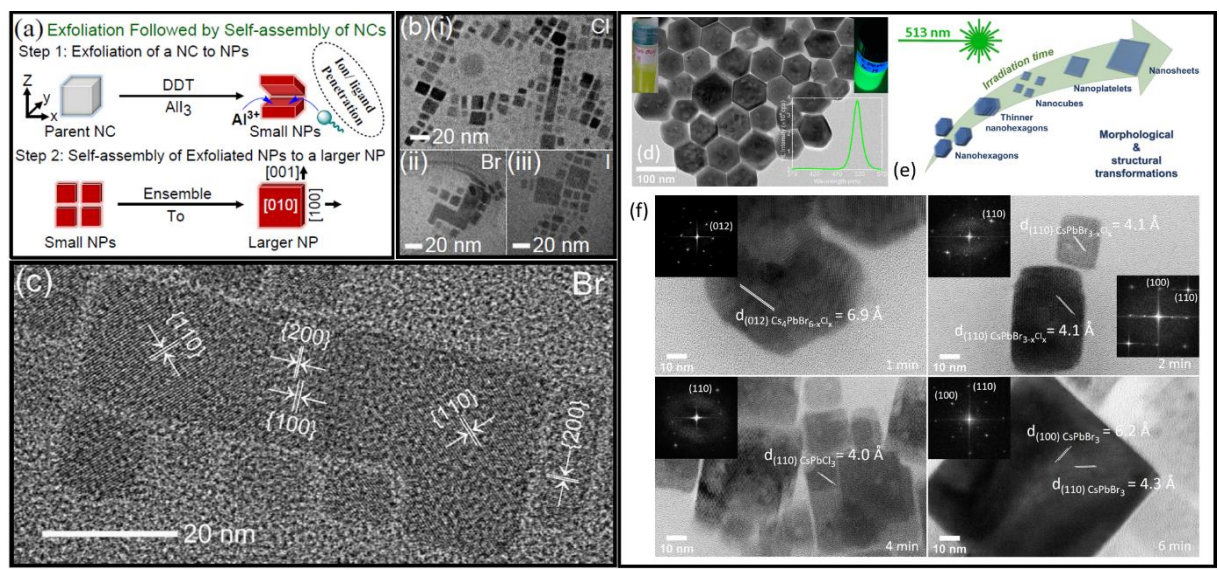

**Figure 10:** Exfoliation and fusion though a ligand-mediated (a-c) anion exchange and laser-triggered (d-f) process. (a-c) Reprinted with permission from Ref.[16], copyright 2020, American Chemical Society. (d-f) Reprinted with permission from Ref.[109], copyright 2022, MDPI.

In a different approach, Tong et al presented an one pot polar-solvent-free synthesis of $CsPbX_3$ NPIs using sonication to induce the formation of a cesium-oleate complex. Then this was reacted with the $PbX_2$ in the presence of OA and Olam. [18] This method gave well crystalline nanocubes slightly larger than those prepared with hot injection methods but lowering the cesium content, 2D nanoplatelets-like morphologies were obtained (Figure 11a-e). This is the first method that gave $CsPbI_3$ NPLs of the 75 % PLQY. Furthermore, a microwave-assisted method has been utilized to form NPIs (Figure 11f-k). [110] In this process, all the reactants were mixed together in a microwave quartz tube in air, which was then put into a microwave reactor. No other pre-treatment is needed. The reaction system was then heated to 80 °C and kept at this temperature for 5 min.

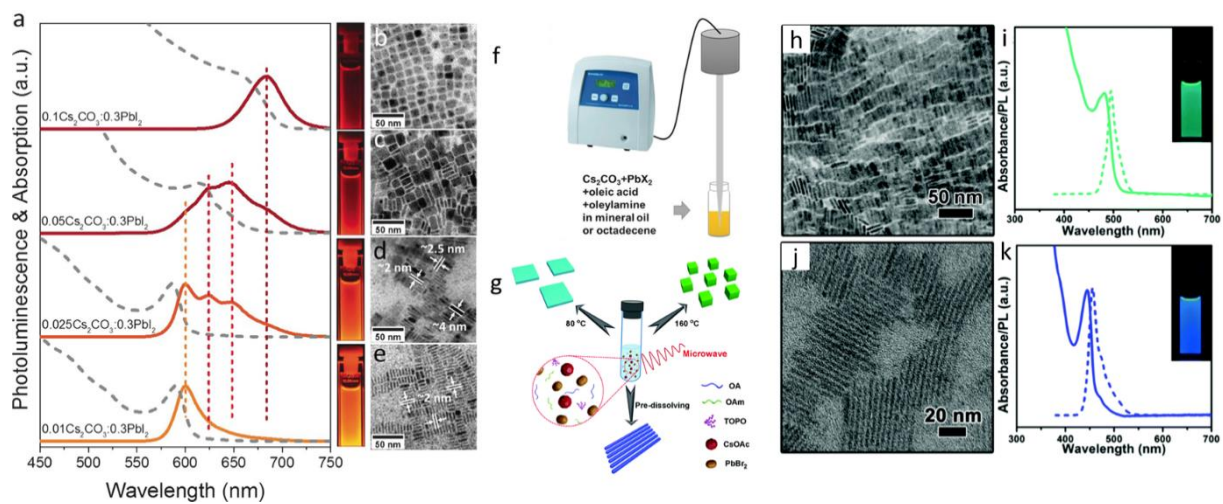

**Figure 11:** All-inorganic metal halide NPls fabricated with external stimuli-triggered processes. NPls synthesized using a sonication (a-f) and microwave reactor (g-k). (a-f) Reprinted with permission from Ref.[18], copyright 2016, Wiley. (g-k) Reprinted with permission from Ref.[110], copyright 2017, Royal Society of Chemistry.

**Lead-free metal halide nanoplatelets.** Utilizing the protocols developed for the lead-based metal halide perovskite NPls, lead-free perovskites of the same morphologies have been fabricated. Stable NPls of $Cs_2SnI_6$ have been synthesized with a hot injection method at 220 °C.[111] No toxic phosphines, such as tri-n-octylphosphine (TOP) and tributylphosphine (TBP), were used during this synthesis. Typically, cesium oleate reacted with tetravalent tin(IV) iodide in the presence of oleic acid and oleylamine in octadecene (ODE) at 220 °C to generate crystalline $Cs_2SnI_6$ nanocrystals. The NPls were formed after 60 min at high temperature after the addition of the cesium oleate. In contrast, $CsSnI_3$ NPls were synthesized with thickness less than 4 nm utilizing trioctylphosphine (TOP) both as a reducing agent and as a solvent to dissolve $SnI_2$ while minimizing oxidation and then this solution was injected into a heated mixture of $CsCO_3$ with long- and short-chain amines (oleylamine and octylamine) and a short-chain organic acid (octanoic acid).[112] The combination of long- and short-chain ligands to control particle morphology as well as the use of TOP with high Sn concentrations are the main key points for the synthesis of NPls. In addition, cesium bismuth bromide perovskite NPls have been synthesized with hot injection method at 180 °C and OA and Olam as ligands and $Cs_2AgBiX_6$ at 150 °C utilizing OA, Olam, octanoic acid, octylamine. [113,114] A different approach was used for the synthesis of $Cs_2AgBiX_6$ 2D NPls which involves a room temperature precursor injection followed by a solution heating up process.[115] The lateral size of the NPls was 180 ± 130 nm when only oleylamine is used as ligand and can be reach the size of 630 ± 380 nm by using octylamine with a ratio 8:2.

**All inorganic metal halide nanosheets.** The previous described NPls were tended to be assembled in stacked assemblies limiting their use in devices which require efficient lateral charge transport. In order to increase the charge transport, nanosheets with larger lateral sizes were fabricated. A generalized approach was published from Lv et al to synthesize all-inorganic nanosheets of $CsPbX_3$ (X: Cl, Br, I) to overcome this issue and to use them in photodetectors (Figure 12 a-d) . [116] 2D NSs were obtained with lateral size of 100 nm to micrometers by controlling the reaction conditions. For the case of $CsPbBr_3$ nanosheets, the Cs-oleate in which the Cs-oleate complex was dissolved in OA was injected at the reaction mixture ($PbBr_2$, ODE, OA, Olam) at 60-150 °C and let for 30 min. Investigation of multiple aliquots obtained at different time intervals at 120° C revealed that NPls were formed very fast (Figure 12). At 30 s, NPls of 30 × 40 nm were formed, then these were laterally grown by oriented attachment to NPls of 90× 100 nm at 5 mins and at NSs of 114 × 140 nm at 30 mins. NSs of 460 nm were obtained by increased the temperature at 150 °C. These NSs were stable in their hexane-based dispersions for months stored in the glovebox and no other morphologies were observed in them in contrast to the previous reports. Furthermore, due to their lateral dimensions, these NSs tended to lie flat which respect to the substrate regardless of the concentration. Except of the reaction time that control the size of the NSs, the lateral size can be tunable by varying the ratio of shorter ligands over longer ligands (Figure 12 e-h). The thickness was mainly unaffected by tuning the same parameter and stays practically constant at 3 nm in all the syntheses conducted at short-to-long ligands volumetric ratio below 0.67. By varying the ratio of two short ligands (octanoic acid octylamine) over that of the longer ligands (OA and Olam), the lateral size of the NSs could be tuned from 300 nm up to 5 µm (with the higher ratio of short to long ligands to yield in larger NSs).

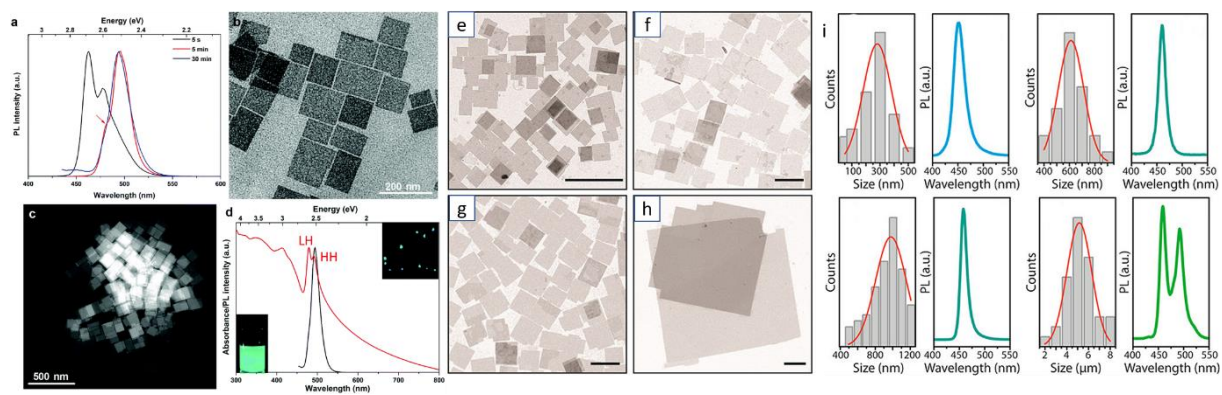

**Figure 12:** Nanosheets' size control by the reaction time (a-d) and the ratio of short to long ligands (e-i). (a-d) Reprinted with permission from Ref.[116], copyright 2016, Royal Society of Chemistry. (e-i) Reprinted with permission from Ref.[21], copyright 2016, American Chemical Society, https://pubs.acs.org/doi/10.1021/jacs.6b03166 and further permissions related to the material excerpted should be directed to the ACS.

Later, in 2019 Zhang et al. achieved to increase the PLQY of the 2D NSs morphologies caped with the same ligands with previous work and to use them in X-ray scintillators.[117] In this case, the Cs-precursor and PbBr$_2$ were prepared separately and the second solution, which is hot, was injected into the first solution swiftly under vigorous stirring at room temperature. In particular the Cs-precursor solution was prepared by dissolving CsAc in propanol followed by hexane and propanol. The PbBr$_2$ precursor solution was prepared by the dissolving the PbBr$_2$ into a mixture of propanol, octylamine and octanoic acid at 90 °C. The thickness of these NSs is down to the 2.8-4.2 nm.

All-inorganic 2D NSs with PLQY to reach the value of 128 %, which is the highest PLQL reported for perovskite NSs, have been reported very recently by doping the CsPbCl$_x$Br$_{3-x}$ with rare earth ions. [118] The Yb$^{3+}$ doped CsPbCl$_3$ NSs were synthesized by the injection of Cs-precursor into the mixed solution of OA and octadecene dissolved with lead acetate (Pb(OAc)$_2$) and ytterbium acetate (Yb(OAc)$_3$). Furthermore, short-chain n-octylamine was utilized instead of octadecene to dissolve cesium acetate (CsOAc), which can avoid the low boiling point issues originated from the short chain amines. The same ligand type was also used to replace the Olam and facilitate the formation of the NSs instead of nanocubes while the metal acetates (instead of metal halides) was easier to dissolve in organic solvents. After different reaction experiments, the optimum temperature found for the growth of the NSs was 200-240 °C while the optimum Yb$^{3+}$ doping concentration could be increased to 6.8 mol%.

**Table 2.** Synthesis strategies summary for the all-inorganic metal halide perovskite Nanoplatelets (NPls) and Nanosheets (NSs) in colloidal media.

| Study | Chemical phase | 2D morphology | Synthesis method/ temperature | Precursors and Ligands Tunable parameters | Thickness in nm or MLs | Lateral size (nm) | PLQY |
|---|---|---|---|---|---|---|---|
| | | | | Lead based all inorganic metal halides | | | |
| **Ligand assisted hot injection and precipitation** | | | | | | | |
| Bekenstein et al. 2015 [11] | CsPbBr$_3$ | NPls | Hot injection, 90-130 °C-open air | PbBr$_2$, Cs$_2$CO$_3$, CsBr, OA, Olam Tuning parameter: temperature | 3 nm | 20 -several hundred | 84.4± 1.8 % (NPls of 5 MLs) 44.7± 2.6 % (NPls of 4 MLs) 10± 0.5 % (NPls of 4 MLs) |
| Akkerman et al.,2016[12] | CsPbBr$_3$ | NPls | LARP, RT | PbBr$_2$, Cs$_2$CO$_3$, CsBr, OA, Olam, HBr Tuning parameter: amount of HBr | 3-5 MLs (tuning the acidity) | 50 | 31% (NPLs of 4 MLs) |
| Sun et al 2016,[97] | CsPbBr$_3$ | NPls | LARP, RT | PbBr$_2$, Cs$_2$CO$_3$, OA, octylamine Tuning parameter: type of the ligands | 5.2 nm (4 MLs) | 100 | - |
| Wang et al 2016[98] | CsPb$_2$Br$_5$ | NPls | LARP, RT | PbBr$_2$, CsBr, OA, Olam, hexylamine | 3 nm | 10–100 | 20–90 % |
| Samsi et al. 2016[21] | CsPbBr$_3$ | NPls, NSs | LARP, RT | PbX$_2$, Cs$_2$CO$_3$, octanoic acid and octylamine Tunable parameter: reaction time (NPls), ratio of shorter to longer ligands (NSs lateral size tuning) | 5 nm | NPLs: 50-200 NSs: 300 -1000 | ~33 % (NSs of the narrowest thickness distribution) |
| Lv et al 2016[116] | CsPbX$_3$ (X: Cl, Br, I) | NSs | Hot injection, 60-150 °C | PbBr$_2$, Cs$_2$CO$_3$, CsBr, OA, Olam Tunable parameter: reaction time | 3 nm (3 MLs) | 100 - 1000 | 6-10 % |
| Pan et al. 2016 [99] | CsPbX$_3$ | NPls | Hot injection, 140-170 °C | PbBr$_2$, CsOAc, combination of different ligands Tuning parameter: hydrocarbon chain composition of carboxylic acids and amines | 1.8- 4.5 nm | 20- 40 | 38 % (thickness 2.6 nm) 51% (thickness 4.5 nm) 61 % (thickness 3.5 nm) 46 % (thickness 2.6 nm) 36 % (thickness 1.8 nm) |

| Almeida et al 2018,[19] | CsPbBr$_3$ | NPIs | Hot injection, 190 °C | PbBr$_2$, Cs$_2$CO$_3$, OA, Olam Tuning parameter: Concentration of the ligands, acidity | 1.7 nm | 20 -1000 | - |
|---|---|---|---|---|---|---|---|
| Sheng et al. 2018[100] | CsPbX$_3$ (X: Cl, Br, I or their mixtures) | NPIs | Hot injection, 60-150 °C | PbBr$_2$, Cs$_2$CO$_3$, CsBr, OA, Olam, MX'$_2$ or MX'$_3$ (M = Cu, Zn, Al, Pb etc.; X' = Cl, Br or I) Tuning parameter: additional metal halide, reaction temperature, reaction time, ligand ratio | 1-5 nm (2-9 MLs) | 10-120 | - |
| Bohn et al. 2018[107] | CsPbBr$_3$ | NPIs | LARP, RT | PbBr$_2$, Cs$_2$CO$_3$, OA, Olam Tuning parameters: Cs/Pb ratio, amount of acetone | 2-6 MLs | 14 | 49 % (2 MLs) 60 % (6 MLs) 73% (6 MLs) |
| Wu et al. 2018[22] | CsPbBr$_3$ | NPIs | LARP, RT | PbBr$_2$, CsBr, OA, Olam, | 3 nm | 12 | 96 % |
| Zhao et al. 2018 [105] | CsPbBr$_3$ | NPIs | LARP, RT | PbBr$_2$, Cs$_2$CO$_3$ Amino acids (Trp), OA, Olam | 4-12 MLs | ~20-50 | 38-75 % |
| Bertolotti et al. 2019[101] | CsPbBr$_3$ | NPIs | Hot injection, 115°C | PbBr$_2$, Cs$_2$CO$_3$, OA, Olam | 3.5 nm (6 MLs) | 12-14 | 75 % |
| Chakrabarty et al., 2020[20] | CsPbBr$_3$ | NPIs | Hot injection, 90° C | PbBr$_2$, cesium cholate (CsCh), OA, Olam | 8-9 MLs | L = 12.4 ± 3.2 | 94 % |
| Samsi et al, 2020[102] | CsPbBr$_3$ | NPIs | Hot injection, 110 °C | PbBr$_2$, Cs$_2$CO$_3$, OA, short hexylphosphonate ligand (C$_6$H$_{15}$O$_3$P) | 2.4 ± 0.3 nm | 6 ± 2 × 26 ± 5 | 40 % |
| Zeng et al. 2021[103] | CsPbBr$_3$ | NPIs | LARP, RT | PbBr$_2$, Cs$_2$CO$_3$, OA, octylamine | 2.1 nm (3 MLs) | | 97.4 % |
| Zhang et al. 2019[117] | CsPbBr$_3$ | NSs | LARP, RT | PbBr$_2$, CsA, Octylamine, octanoic acid | 2.8 ~ 4.2 nm (5-7 MLs) | 100 | ~63 % |
| | | | | | | | |
| Sun et al. 2022[118] | Yb$^{3+}$ doped CsPbCl$_3$ | NSs | Hot injection method, 150°C | Pb(OAc)$_2$, CsOAc, OA, Olam, octylamine | 6.4 nm | 100-300 | ~128 % |
| **Spontaneous crystallization in nonpolar organic media** | | | | | | | |
| Huang et al 2019[29] | CsPbBr$_3$, CsPbI$_3$ | NPIs | Spontaneous crystallization, RT | PbBr$_2$, Cs$_2$CO$_3$, CH$_3$COOCs, OA, Olam | 1.1 nm (CsPbBr$_3$) and 1.2 nm (CsPbI$_3$) | ~20 | - |
| **Cation mediated colloidal method** | | | | | | | |
| Ding et al 2019.[30] | CsPbBr$_3$ and CsPbCl$_x$Br$_{3-x}$ | NPIs | Directly heating, 110 °C, open air | Tin salt (e.g., SnBr$_2$, SnBr$_4$, SnCl$_2$, SnCl$_4$, Sn(OAc)$_2$), PbBr$_2$, Cs$_2$CO$_3$, OA, Olam, | 2.7 nm (3 MLs) | ~20 | 29.3-80.2 % |
| Bonato et al. 2020[104] | CsPbX$_3$ | NPIs | Hot injection, 170 °C | Pb(acetate)$_2$.3H$_2$O, Cs$_2$CO$_3$, SnX$_4$, OA, Olam | 1-6 MLs | ~20 | - |

| **Solvothermal method** | | | | | | | |
|---|---|---|---|---|---|---|---|
| Chen et al 2018.,[108] | CsPbBr$_3$ | NPls | Solvothermal, 140°C Tuning parameter: time of the reaction | PbX$_2$, Cs$_2$CO$_3$, OA, Olam Tuning parameter reaction time | 3 nm | 100-200 | 28% |
| Zhai et al 2018[31] | CsPbBr$_3$ | NPls | Solvothermal, 100 °C | Cs$_2$CO$_3$, PbX$_2$, OA, Olam Tuning parameter reaction time | 4.2 nm | 44- 101 | 50 % |
| **Ligand- and laser-assisted exfoliation** | | | | | | | |
| Uddin et al 2020[16] | CsPbX$_3$ (X: Cl, Br, I or their mixtures) | NPls | Hot injection method, ligand mediated anion exchange, RT | Cs$_2$CO$_3$, PbCl$_2$, TOP, DDT-AlX3 | 2 nm (4 MNLs) for CsPbCl$_3$ 5-9 nm (15 MNLs) for CsPbBr$_3$, 3 nm (5 MNLs) for CsPbl$_3$ | ~10 | 50-65 % (CsPbCl$_3$) 100 % (CsPbBr$_3$) 81 % (CsPbl$_3$) |
| Kostopoulou et al[17] | CsPbBr$_3$ | NPls and NSs | Laser irradiation of as synthesized nanocrystals with LARP, RT | PbBr$_2$, CsBr, OA, Olam, Bad solvent: toluene | - | ~100-1500 | - |
| **External stimuli-triggered process** | | | | | | | |
| Tong et al., 2016[18] | CsPbX$_3$ | NPls | Ultrasonication, RT, ambient conditions | PbX$_2$, Cs$_2$CO$_3$, OA, Olam Tunable parameter: Cs$_2$CO$_3$/ Pbl$_2$ ratio | 3-6 MLs | ~10-20 | 20–90% |
| Pan et al., 2017[110] | CsPbX$_3$ | NPls | Microwave, 80 °C, ambient conditions | (CsOAc), (PbX$_2$, X = Cl, Br, I, or their mixture), trioctylphosphine oxide (TOPO), OA, Olam | 3.3 nm. | 23.4 | - |
| | | | | **Lead-free all inorganic metal halides** | | | |
| **Ligand assisted hot injection** | | | | | | | |
| Wang et al., 2016 [111] | Cs$_2$Snl$_6$ | NPls | Hot injection method, 220 °C | Snl$_4$, Cs$_2$CO$_3$, OA, Olam | 8 nm | 200-300 | - |
| Wong et al., 2018[112] | CsSnl$_3$ | NPls | Hot injection method, 135 °C | Cs$_2$CO$_3$, octanoic acid, oleylamine, octylamine | <4 nm | ~ 500 nm | - |
| Lian et al, 2018[113] | Cs$_3$Bi$_2$Br$_9$ | NPls | Hot injection method, 180 °C | Cs2CO3, OA, Olam | ~9 nm | 60–250 nm | - |
| Huang et al., 2021[114] | Cs$_2$AgBiBr$_6$ | NPls | Hot injection method, 150 °C | AgBr, BiBr$_3$, OA, Olam, octanoic acid, octylamine | 3–5 nm | ~ 200 nm | - |
| **Room temperature injection and heating** | | | | | | | |

| Liu et al., 2021[115] | Cs$_2$AgBiBr$_6$ | NPls | Room temperature precursor injection followed by a solution heating up process. | BiBr$_3$, AgNO$_3$, OA, Olam<br>Tuning parameter: ratio octylamine/oleylamine | 3-4 MLs | 180 ± 130 nm<br>630 ± 380 nm (8:2 octylamine/oleylamine) | |

RT: room temperature, LARP: Ligand-assisted reprecipitation, OA: oleic acid, Olam: oleylamine, NPls: Nanoplatelets, NS: nanosheets, Ml: monolayers

# 3. Metal Halide Perovskite nanocrystals/2D material Heterostructures

The inherently soft crystals lattice allowing greater tolerance to lattice mismatch making the metal halide perovskites ideal candidates for heterostructure formation for electronics and optoelectronics applications.[119] 2D metal halide perovskite epitaxial heterostructures [119–121] and heterostructures with metal halide materials on 2D materials such as graphene, graphene oxide, reduced graphene oxide or transition metal dichalcogenides[45,54,55,58–60,64] have shown efficient electron transfer and synergetic functionalities. In this section, we will focus on the latter heterostructures and a summary of the reported to date metal halide perovskite nanocrystals/2D material heterostructures will be presented. These heterostructures consist of two different parts, one is the metal halide perovskite nanocrystals of various morphologies and chemical phases and the second is a 2D material. New physics and synergetic effects are emerging from the coupling between the two different materials and new or improved functionalities have been arisen due to the interfacial phenomena. Solution-processed fabrication methods will be described for the different combinations of perovskite nanocrystals and the 2D materials. Heterostructures included all-inorganic or organic inorganic metal halide nanocrystals with 2D materials such as graphene, graphene oxide, reduced graphene oxide or transition metal dichalcogenides will be summarized (Table 3). The fabrication methods of the metal halide perovskite nanocrystals/2D material heterostructures can be categorized in six types: a) spin coating or deposition of metal halide nanocrystals on 2D material layers, b) in-situ crystallization through colloidal chemistry methods (LARP and hot injection methods), c) in situ coupling through mixing of the two materials, d) in situ coupling through mixing of the two materials and photoreduction, e) laser-induced conjugation in solution and e) solid state reaction .

## 3.1 Spin coating or deposition of metal halide nanocrystals on 2D material layers

This technique is widely used for the fabrication of the optoelectronic devices as it produces quite uniform nanocrystal films, and it is cost-effective and easily accessible to many laboratories. The metal halide nanocrystals which used for the deposition have been synthesized through LARP or hot injection methods and then after the purification steps were dispersed in an organic solvent. Then, the as-synthesized nanocrystals were spin coated or deposited on the layer of the 2D materials. This room temperature method provides unique opportunities for the design and development of metal halide perovskite nanocrystals-2D heterostructures, exhibiting synergetic functionalities by combining nanocrystals of different morphologies and chemical phases with various 2D materials.

All-inorganic[42, 45, 58, 55, 57] or organic- inorganic[57] metal halide nanocrystals were used to be spin coated on the 2D materials substrates such as bilayers of mechanically exfoliated graphene, [42] CVD grown single layer graphene, [45] and spin-coated $CdS_xSe_{1-x}$ nanosheets. [55]

Chen et al. unveiled through a PL blinking analysis the efficient electron transfer in the $CsPbI_3$/single-layer graphene heterostructure due to the interfacial interaction. [45] This strong interfacial electron transfer interactions found that was unaffected by the morphology of the perovskite nanocrystals. Two different nanocrystal morphologies, nanocubes and nanowires, have been included in this study. Furthermore, in the $CsPbI_{3-x}Br_x$/$MoS_2$ heterostructures reported by Wu et al. [58], a favorable energy band alignment facilitating interfacial photocarrier separation and efficient carrier injection into the $MoS_2$ which resulted to improved photocarrier generation efficacy and photogating effect. Very recently, efficient charge carrier transfer ability between the all-inorganic $CsPbBr_3$ perovskites and 2D cadmium sulfide selenide has been found by Peng et al. due to energy band alignment engineering. [55] The chemical phase of the perovskite nanocrystals seemed to be an important parameter to the final performance of the devices based on the heterostructures. [57]

## 3.2 In-situ crystallization through colloidal chemistry methods (LARP and hot injection methods)

The metal halide nanocrystals can be directly grown on the 2D materials during the nanocrystal synthesis (Figure 13). In the LARP approach, the 2D materials have been dispersed in the bad solvent[43, 64] or in the precursor stock solution[61]. Quite spherical $CsPbBr_3$[43] and $MAPbBr_3$[64] nanocrystals around 10 nm in size have been grown on GO flakes (Figure 13a, b). In a similar manner, $MAPbBr_3$ nanocrystals were grown on protonated graphitic carbon nitride (p-g-$C_3N_4$) (Figure 13c) [61] and CsPbBr₃ on Hexagonal Boron Nitride Nanosheet (h-BN) [52] (Figure 13d) or CsPbBr₃ on black phosphorus. [47]

In addition, the 2D materials can be dispersed in the reaction mixture with the Pb-precursor and ligands in the hot injection syntheses. Then, the Cs-precursor has been injected at high temperature and the metal halide perovskite nanocrystals started to grow on them (Figure 13e). [44] Following this approach CsPbBr₃ nanocrystals have been grown on GO and rGO sheets,[44,53] a-CsPbI₃ on rGO[4] and CsPbBr₃ on hexagonal boron nitride nanosheets (Figure 13f) [51].

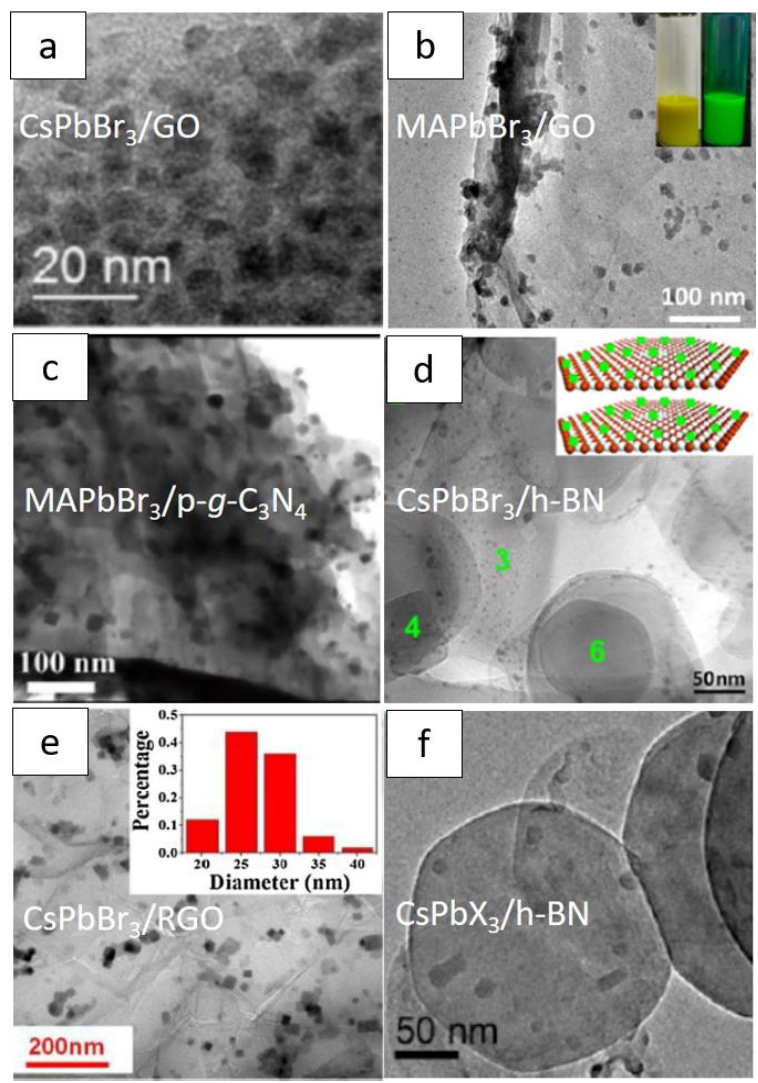

**Figure 13:** Metal halide perovskite nanocrystals/2D materials heterostructures fabricated with LARP and hot injection colloidal methods. a) Reprinted with permission from Ref.[43], copyright 2017, American Chemical Society. b) Reprinted with permission from Ref.[64], copyright 2019, Elsevier. c) Reprinted with permission from Ref. [61], copyright 2017, Royal Society of Chemistry. d) Reprinted with permission from Ref.[52], copyright 2019, American Chemical Society. e) Reprinted with permission from Ref.[44], copyright 2017, ScienceDirect. f) Reprinted with permission from Ref.[51], copyright 2019, Royal Society of Chemistry.

## 3.3 In situ coupling through mixing of the two materials

In this approach, the metal halide nanocrystals have been synthesized initially through colloidal methods and then dispersed in a proper solvent. The nanocrystals were mixed with the 2D materials that were dispersed also in the same solvent. The conjugation of the two materials were originated through chemical bonding or non-covalent interactions. [41] The challenge in this method is to select the suitable common solvent for both perovskite nanocrystals and 2D materials in order to mix them and prepare the heterostructures.

Muduli et al. succeed to bind CsPbbr₃ nanocrystals on a few layered black phosphorous sheets through this room temperature method (Figure 14a). [48] The CsPbBr₃ nanocrystals were synthesized through a hot injection method and the 2D through a probe sonication method independently. Then the two solutions were mixed followed by re-sonication at room temperature and the heterostructures were obtained. The common solvent in this case was the toluene. The selection of the common solvent for this process was not so easy because the few-layered black phosphorous sheets have to be prepared in the N-Methyl-2-pyrrolidone (NMP) solvent, but the perovskite nanocrystals were degraded in the presence of this solvent. So, the 2D materials were prepared in NMP and then the NMP was exchanged with the toluene by severe centrifugation steps to make sure that no remaining NMP was left. A reduction in the CsPbBr₃ nanocrystals band gap which is reflected as a blue shift of the PL peak thereby confirming the process of charge transfer in the heterostructure. Moreover, the facile coupling of few-layer MoS₂ NSs with MAPbI₃ nanocrystals in MAPbI₃-saturated aqueous HI solution towards a MAPbI₃/MoS₂ heterostructures has been reported by Wang et al. [59] In addition, the binding can be facilitating through proper ligands. As an example, perovskite nanocrystals have been conjugated on the porous g-C₃N₄ nanosheets through the abundant amino sites (NHₓ, x = 1, 2) existed on the edges of heptazine units (Figure 14b). The amino group can interact strongly with CsPbBr₃ nanocrystals. [49] Similarly the CsPbBr₃ nanocrystals found closely anchored on 2D graphitic carbon nitride nanosheets, containing titanium-oxide species (TiO-CN), via the N–Br and O–Br bonding (Figure 14c).[46] In the same direction, there are ligands which have selective affinity toward both the materials, can be quintessential for making heterostructures with close contact.[54] For example, 4-aminothiophenol (4-ATP) is among the bifunctional ligands having both thiol (−SH) and amine (−NH₂) functionalities. Therefore, the thiol group can easily get adsorbed on the surface of the MoSe₂ NSs, whereas the protonated amine group can form an electrostatic bond with the surface bromide of CsPbBr₃ nanocrystals (Figure 14d). [54]

Furthermore, all-inorganic metal halide perovskite nanocrystals were presented as non-destructive dispersants capable of dispersing TMD nanosheets in the liquid phase.[66] Specifically, diverse Transition Metal Dichalcogenide NSs such as MoSe₂, MoS₂, and WS₂ were successfully dispersed in a non-polar solvent with the aid of either CsPbCl₃ or CsPbBr₃ nanocrystals (Figure 14 e). Thus, a few-layered TMD nanosheets decorated with the perovskite nanocrystals were formed with long-term stability.

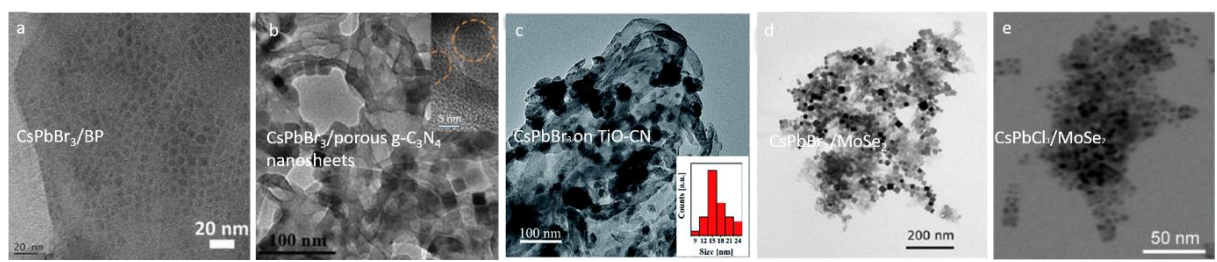

**Figure 14:** Metal halide perovskite nanocrystals/2D materials heterostructures synthesized by mixing the nanocrystal solution with the 2D material solutions. a) Reprinted with permission from Ref.[48], copyright 2018, Wiley. b) Reprinted with permission from Ref.[49], copyright 2018, Wiley. c) Reprinted with permission from Ref.[46], copyright 2019, Royal Society of Chemistry. d) Reprinted with permission from Ref.[54], copyright 2020, American Chemical Society. e) Reprinted with permission from Ref.[66], copyright 2022, Wiley.

### 3.4 In situ coupling through mixing of the two materials and photoreduction

Metal halide nanocrystals/rGO heterostructures have been fabricated through a photoreduction method. With this method, MAPbI₃/rGO and Cs₂AgBiBr₆/rGO have been obtained by mixing the perovskite nanocrystals with GO and irradiated with visible light produced by Xe lamp with 420 nm cut-off filter.[62,65] The whole process was kept by a cooling water system at 15 °C for the first system and 5 °C for the second and the irradiation duration was 10 h and 20 h respectively.

### 3.5 Laser-induced conjugation in solution

A rapid photo-induced process has been reported from our group to conjugate metal-halide perovskite nanocrystals on graphene-based materials.[60] It is revealed studying the heterostructures under different number of laser pulses that a small number of pulses is sufficient to decorate the 2D flakes with metal-halide nanocrystals without affecting their primary morphology and also heterostructures with higher density of nanocrystals decorated on GO flakes have be obtained. At the same time, the density of anchored nanocrystals can be finely tuned by the number of irradiation pulses (Figure 14a-g).

Specifically, a high repetition rate fs laser system using a directly diode-pumped Yb:KGW (ytterbium doped potassium gadolinium tungstate) was employed for the irradiation of the mixed perovskite nanocrystals-GO solution. The common solvent found for the dispersion of the two materials was the dichlorobenzene. The laser wavelength, the repetition rate and the pulse duration used for the irradiation experiment were 513 nm, 60 kHz and 170 fs, respectively. The laser fluence in all the experiments kept constant at the value of 0.5 mJ/cm$^2$ and only the number of the pulses was varied. The colloids obtained by this process, were stable and no-precipitation was observed. The nanocrystal density was increased with the number of the pulses retaining their primary size and the perovskite nanocrystals were assembled first on the periphery of the GO flakes and then on the basal plane when the number of the pulses is increased to over 10$^3$ pulses (Figure 14h).

The unique advantage of this technique is that it is rapid, performed at room temperature and provide unique opportunities for the design and development of perovskite-2D heterostructures; exhibiting synergetic functionality by combining nanocrystals of different morphologies and chemical phases with various 2D materials without the initial nanocrystals morphologies to be affected.

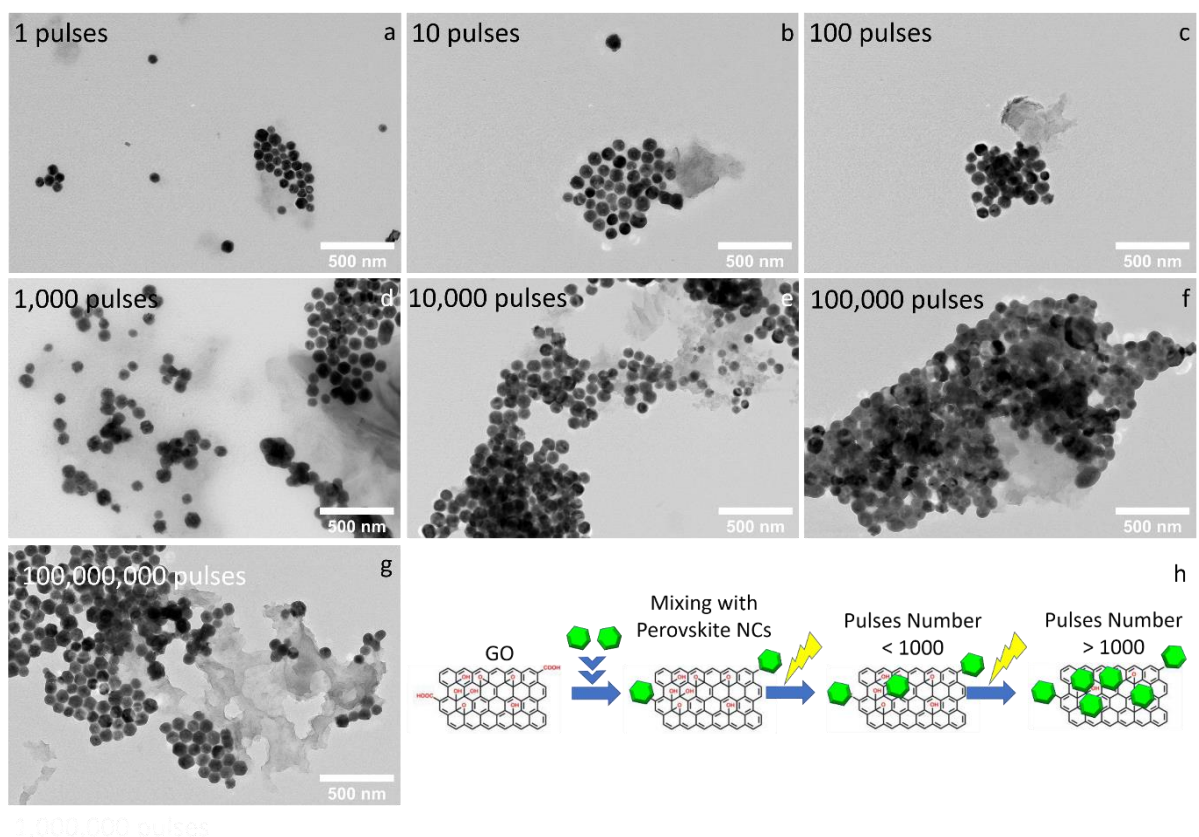

**Figure 15:** Laser-induced Cs$_4$PbBr$_6$/GO heterostructures for different number of pulses (a-g). Proposed conjugation mechanism. Reprinted with permission from Ref.[60], copyright 2020, MDPI.

### 3.6 Solid state reaction

Encapsulation of CsPbBr$_3$ nanocrystals in 2D C$_3$N$_4$ layer have been performed via a solid-state reaction by Bian et al (Figure 16).[56] The as-synthesized CsPbBr$_3$ nanocrystals were mixed thoroughly with urea in a ceramic crucible at 450 °C (optimum conditions) under N$_2$ atmosphere. The N$_2$ atmosphere is essential for the successful formation of the heterostructure. The same synthesis has showed very little polymerization of urea to carbon nitride.

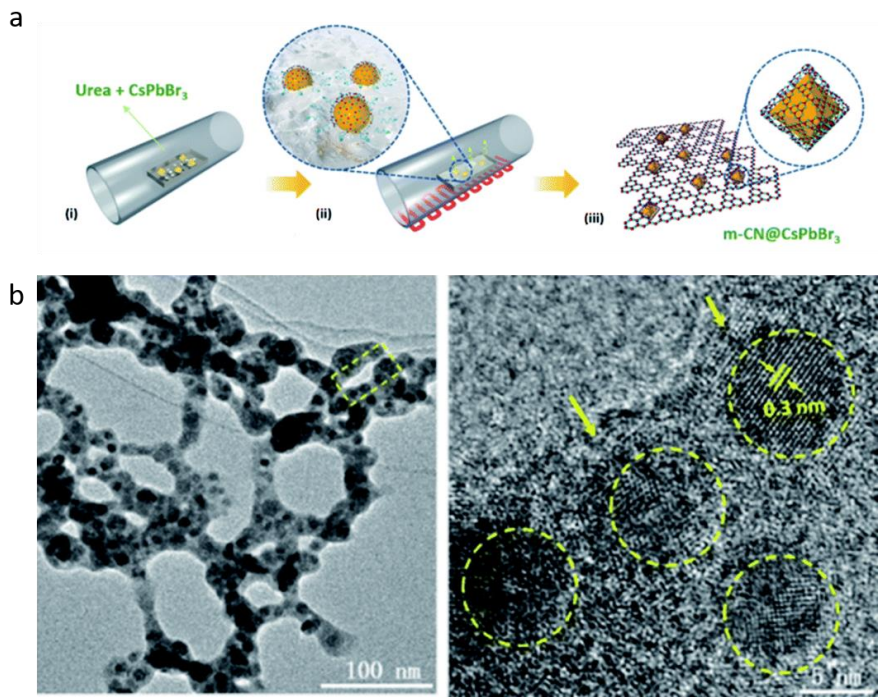

**Figure 16:** CsPbBr$_3$ encapsulated in 2D C$_3$N$_4$ layer fabricated through a solid-state reaction. Fabrication process (a) and TEM images. Reprinted with permission from Ref. [56], copyright 2022, Royal Society of Chemistry.

**Table 3.** Summary of the metal halide perovskite nanocrystals/2D material heterostructures.

| Study | Metal halide perovskite phase/ morphology | 2D material | Fabrication method of the heterostructure | Improved property/target application |
|---|---|---|---|---|
| Kwak et al. 2016[42] | $CsPbBr_{3-x}I_x$<br><br>Cubic-shaped nanocrystals | Graphene | Deposition of nanocrystals on the bilayer graphene flakes film. | Improved photosensitivity due to the rapid carrier transport of the graphene.<br><br>Photodetectors. |
| Xu et al. 2017[43] | $CsPbBr_3$<br><br>Quite spherical nanocrystals | Graphene oxide (GO) | Room temperature colloidal chemistry route (LARP). The GO is dispersed in the bad solvent for the precipitation of the perovskite nanocrystals. | Improved rate of electron consumption compared to that of net perovskite nanocrystals.<br><br>Photocatalytic $CO_2$ reduction to ethyl acetate. |
| Pu et al. 2017[61] | $CH_3NH_3PbBr_3$<br><br>Quite spherical nanocrystals | Protonated graphitic carbon nitride (p-g-$C_3N_4$) | Room temperature colloidal chemistry route (LARP). The p-g-$C_3N_4$ is dispersed in the precursor solution. | Significant charge separation.<br><br>Photocatalysis. |
| Tang et al., 2017[44] | $CsPbBr_3$<br><br>Cubic nanocrystals | Reduced graphene oxide | Hot-injection method at 150° C. | Enhanced strong photoresponse.<br><br>Photodetectors. |
| Wu et al. 2018[62] | $CH_3NH_3PbI_3$<br><br>Irregular-shaped nanocrystals | Reduced graphene oxide | Photoreduction method including visible light irradiation at room temperature for several hours. | Superb $H_2$ evolution activity and very stable in $MAPbI_3$-saturated aqueous HI solution.<br><br>Visible-light photocatalysts active in aqueous solution. |
| Chen et al. 2018[45] | $CsPbI_3$<br><br>Cubic nanocrystals and nanowires | Single-layer graphene | Deposition of nanocrystals on the graphene layer fabricated by chemical vapor deposition. | Increased light-harvesting properties.<br><br>Next generation photodetectors and photovoltaic devices, including polarization sensitive photodetectors. |
| Guo et al. 2018[46] | $CsPbBr_3$<br><br>Irregular-shaped nanocrystals | Graphitic carbon nitride nanosheets, containing titanium-oxide species (TiO-CN) | Sonication and stirring of the perovskite nanocrystals and 2D material in hexane. | Increased number of catalytic sites and improved charge separation efficiency.<br><br>Photocatalytic $CO_2$ reduction. |
| Huang et al. 2018[47] | $CsPbBr_3$<br><br>Irregular-shaped nanocrystals | Black phosphorus (BP) nanosheets | In situ growth of nanocrystals on BP exfoliated sheets. Mixing at room temperature of the BP sheets with the precursor of the nanocrystal. | Robust month-long air-stability.<br><br>Next-generation optoelectronic devices. |
| Muduli et al., 2018[48] | $CsPbBr_3$<br><br>Cubic nanocrystals | Black phosphorus (BP) nanosheets | Mixing of the as synthesized perovskite nanocrystals with black phosphorus (BP) nanosheets followed by the bath sonication at room temperature. | Improved charge transfer between the two materials.<br><br>Optoelectronic devices. |

| Wu et al., 2018[57] | $CH_3NH_3PbBr_3$ and $CsPbI_{3-x}Br_x$ Cubic and spherical nanocrystals | $MoS_2$ | Spin coated nanocrystals on the 2D-$MoS_2$ monolayer. | Improved photoresponsivity and specific detectivity. Phototransistors. |
|---|---|---|---|---|
| Wu et al., 2018[58] | $CsPbI_{3-x}Br_x$ Cubic nanocrystals | $MoS_2$ | Spin coated nanocrystals on the 2D-$MoS_2$ monolayer. | Photocurrent was enhanced by 15.3-fold. Photodetectors. |
| Ou et al., 2018[49] | $CsPbBr_3$ Cubic nanocrystals | g-$C_3N_4$ | Mixing of the as synthesized perovskite nanocrystals with porous g-$C_3N_4$ | Improved photocatalytic behavior. Photocatalytic reduction of $CO_2$ to CO under visible light irradiation |
| Casanova-Cháfer et al 2019[63] | $CH_3NH_3PbBr_3$ Quite spherical nanocrystals | Graphene | Spin coated perovskite nanocrystals on drop casted graphene nanoplatelet layer. | Enhanced gas sensitivity. Gas sensors. |
| Wang et al 2019[64] | $CH_3NH_3PbBr_3$ Quite cubic nanocrystals | Graphene oxide | Room temperature colloidal chemistry route (LARP). The GO is dispersed in the bad solvent for the precipitation of the perovskite nanocrystals. | Considerable electron-hole separation efficiency and high selectivity of solar fuels. Photoelectrochemical conversion $CO_2$ into solar fuels in nonaqueous media. |
| Zhang et al 2019[50] | a-$CsPbI_3$ Cubic nanocrystals | RGO | Hot-injection method at 140-170° C. | Improved stability and carrier transport quality. Optically active material for optoelectronic devices. |
| Qiu et al. 2019[51] | $CsPbBr_3$ Quite cubic nanocrystals | h-BN NSs | Hot-injection method at 180° C. | Superior thermal stability. LEDs. |
| Li et al 2019[52] | $CsPbBr_3$ Cubic nanocrystals | Hexagonal Boron Nitride Nanosheet (h-BN) | Room temperature colloidal chemistry route (LARP). The h-BN nanosheets are dispersed in the bad solvent for the precipitation of the perovskite nanocrystals. | Enhanced humidity stability and thermal stability. LEDs. |
| Wang et al, 2020[65] | $Cs_2AgBiBr_6$ Irregular-shaped nanocrystals | Reduced graphene oxide | Photoreduction method including visible light irradiation at 5 °C for severe hours. | Enhanced photocatalytic activity. Photocatalytic hydrogen generation. |
| Pu et al. 2020[53] | $CsPbBr_3$ Irregular-shaped nanocrystals | Reduced graphene oxide | Hot injection method at 160 °C | The local electric field around $CsPbBr_3$ QDs is increased by surface plasmon resonance on the surface of RGO nanosheet. LEDs. |
| Hassan et al. 2020[54] | $CsPbBr_3$ | $MoSe_2$ | The as-synthesized nanocrystals have been mixed with the 4-ATP-functionalized $MoSe_2$ NSs at 80 °C. | Faster charge diffusion across $CsPbBr_3$/ $MoSe_2$ interfaces. |

| | Cubic nanocrystals | | | Potential applications in photovoltaic devices. |
|---|---|---|---|---|
| Wang et al 2020[59] | MAPbI$_3$<br><br>Irregular-shaped nanocrystals | MoS$_2$ | In situ coupling method through mixing. | More than two orders of magnitude higher activity.<br><br>Visible-light-driven photocatalytic H$_2$ evolution. |
| Kostopoulou et al., 2020[60] | Cs$_4$PbBr$_6$<br><br>Hexagonal nanocrystals | Graphene | Laser induced conjugation in solution. | The nanocrystals morphology does not be affected and the density of the nanocrystals can be tuned by the number of the irradiation pulses. |
| Peng et al., 2021[55] | CsPbBr$_3$<br><br>Quite cubic nanocrystals | 2D non-layered Cadmium Sulfide Selenide | Spin coated perovskite nanocrystals on the top of the nanosheets layer. | Enhanced charge transport capability.<br><br>Photodetectors. |
| Lee et al., 2022[66] | CsPbCl$_3$ and lanthanide ion doped CsPbCl$_3$<br><br>Cubic nanocrystals | MoSe$_2$ | Mixing of the nanocrystals and 2D materials | The nanocrystals efficiently withdraw electrons from the nanosheets, and suppress the dark current of the MoSe$_2$ nanosheets.<br><br>Flexible near-infrared photodetectors with a high ON/OFF photocurrent ratio and detectivity. |
| Bian at al., 2022[56] | CsPbBr$_3$<br><br>Spherical nanocrystals | C$_3$N$_4$ | Solid state reaction. Encapsulation of CsPbBr$_3$ nanocrystals in 2D C$_3$N$_4$ (core shell) | Significant improvement in CO$_2$ capture and charge separation.<br><br>Thermocatalysis. |

# 4. Applications

## 4.1 Low-dimensional perovskite light emitting diodes (LEDs)

### 4.1.1. Metal halide perovskite nanosheets

A couple of studies on the development of LED devices based on perovskite NSs will be consider first. Parveen et al. demonstrated a solvothermal process for the precise thickness tuning of 2D $CH_3NH_3PbBr_3$ NSs [122]. The authors discovered that upon heating at 100 °C, the self-assembled and randomly oriented nanorods allows the growth of multilayered perovskite NSs. Moreover, when further heating was applied (150 °C), large area 2D few layer $CH_3NH_3PbBr_3$ NSs were obtained. The thermal process is depicted schematically in Figure 17a. Indeed, NSs with 14 down to 2 layers were formed. Figure 17b presents the change in optical bandgap and photoluminescence (PL) peak position against the number of layers. [122] The obtained change in the band gap with number of layers and thickness implied a strong quantum confinement effect in the constructed 2D multilayers. The authors exploited the so-formed $CH_3NH_3PbBr_3$ NSs for the fabrication of a photodetector with ultrafast response times of ms (Figure 17c). In addition, a white light converter based on compositionally tuned 2D $CH_3NH_3PbBrI_2$ NSs and a blue LED chip was developed (Table 4). Figure 17d depicts the optical characteristics of the device, while the inset shows a photo during operation [122]. On a rather different manner, Gao et al. adopted an alternative doping method for the fabrication of deep-blue LEDs based on all-inorganic perovskite NSs [123]. Namely, the authors employed a heteroatomic $Cu^{2+}$ doping route along with $Br^-$ ion exchange for the synthesis of deep-blue emitting $CsPb(Br/Cl)_3$ perovskites. The doping of the perovskite NSs is shown schematically in Figure 18a. The embedded $Cu^{2+}$ cations were found to decrease the intrinsic chlorine defects, resulting to the formation of 2D NSs of high crystalline quality. As a result, both the current density and the luminance of the developed 462 nm blue-LED enhanced significantly, when compared to the undoped device (Figure 18b) [123].

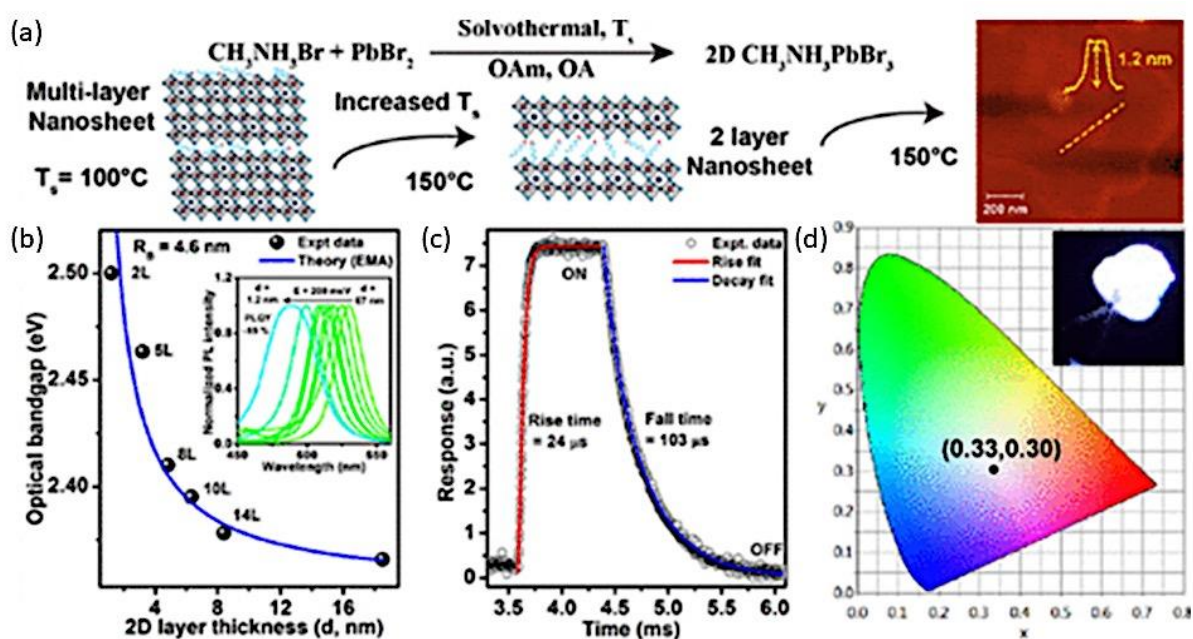

**Figure 17:** (a) Schematic representation of the solvothermal process for the synthesis of 2D $CH_3NH_3PbBr_3$ NSs. (b) $CH_3NH_3PbBr_3$ NSs optical band gap dependence on number of layers and thickness. (c) $CH_3NH_3PbBr_3$ NSs photodetector characteristics. (d) $CH_3NH_3PbBr_3$ NSs white light emitter in operation and chromaticity coordinates. a-d have been reproduced from ref. [122] with permission from ACS American Chemical Society, copyright 2020.

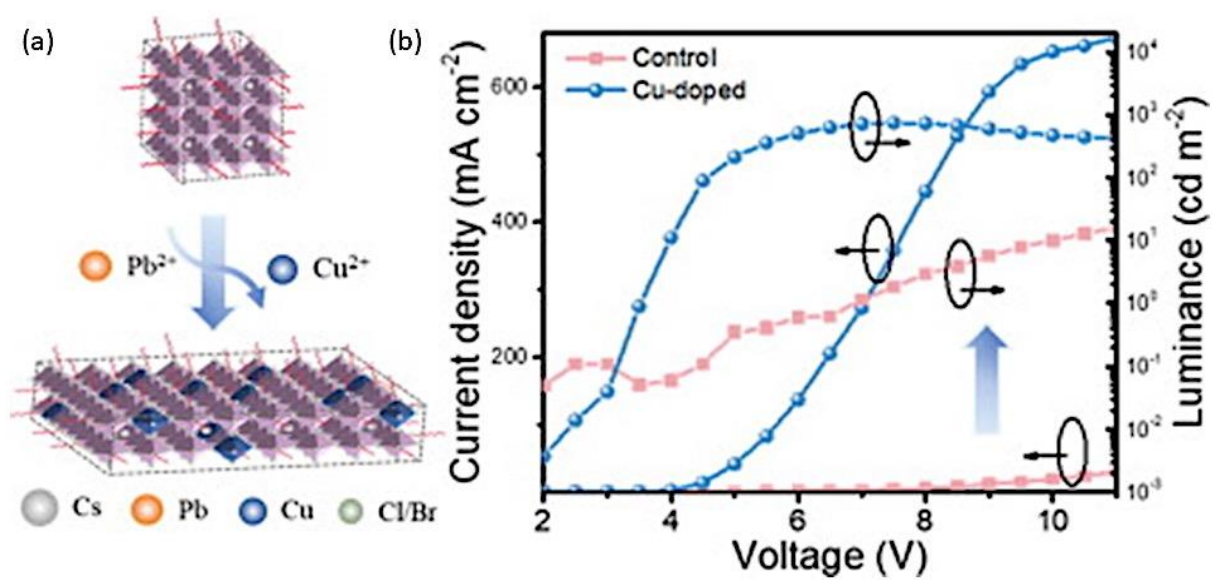

**Figure 18:** (a) Schematic animation of the CsPb(Br/Cl)$_3$ perovskite NSs copper doping process. (b) Current density and luminance versus voltage characteristics of the developed blue LED device based on copper-dopped and pristine CsPb(Br/Cl)$_3$ NSs. a and b have been reproduced from ref. [123] with permission from Elsevier, copyright 2022.

### 4.1.2. Metal halide perovskite nanoplatelets

An overview of the more popular LEDs based on perovskite NPls. Notably, the main light emitting component of the fabricated devices consists of several perovskite NPls monolayers (with n>7). Overall, the external quantum efficiency (EQE) of the most sufficient green NPls-based LEDs has been enhanced drastically from 0.48 % back in 2016, up to 23.6 % in 2020 (Table 4) [34]. Ling et al. reported on the first bright LEDs based on solution-processable MAPbBr$_3$ perovskite NPls [124]. The devices exhibited narrow band electroluminescence (EL) emission at 529 nm (Table 4), while the employed ligand-capped NPls showed enough stability to allow out-of-glove device fabrication. Figure 19a shows schematically the developed Glass/ITO/PEDOT:PSS/MAPbBr$_3$/BCP/LiF/Al device architecture, along with an indicative TEM image photo of the MAPbBr$_3$ NPls and the LED in operation. [124] In the same year, Kumar et al. exploited 2D CH$_3$NH$_3$PbBr$_3$ perovskites for the development of LEDs emitting in various wavelengths of the visible. [82] In particular, the EL wavelength can be tune upon altering the number of monolayers (n) of the synthesized 2D perovskites. Figure 19b presents indicative TEM photos of the 2D PNCs with n=7-10 (left), and with n equal to as low as 3 (right). The induced differences in the optical properties of the 2D perovskites upon changing the layer stacking were striking. Figure 19c depicts the optical absorbance and PL spectra of the bulk single crystal and the colloidal solutions with various values of n, i.e. ranging from 1 to 10. As the number of layers is reduced the PL blue shifts drastically. Moreover, Figure 19d shows the characteristics of the fabricated LED devices as expressed in terms of normalized EL intensity, current density versus voltage, and luminescence versus voltage plots [82]. The reported EQE was 0.23% and 2.31% for the pure blue (n=3) and pure green (n=7-10) LEDs, respectively (Table 4). Nevertheless, the great highlight of this study was to achieve for the first-time room temperature EL below 490 nm, i.e. in the deep blue region.

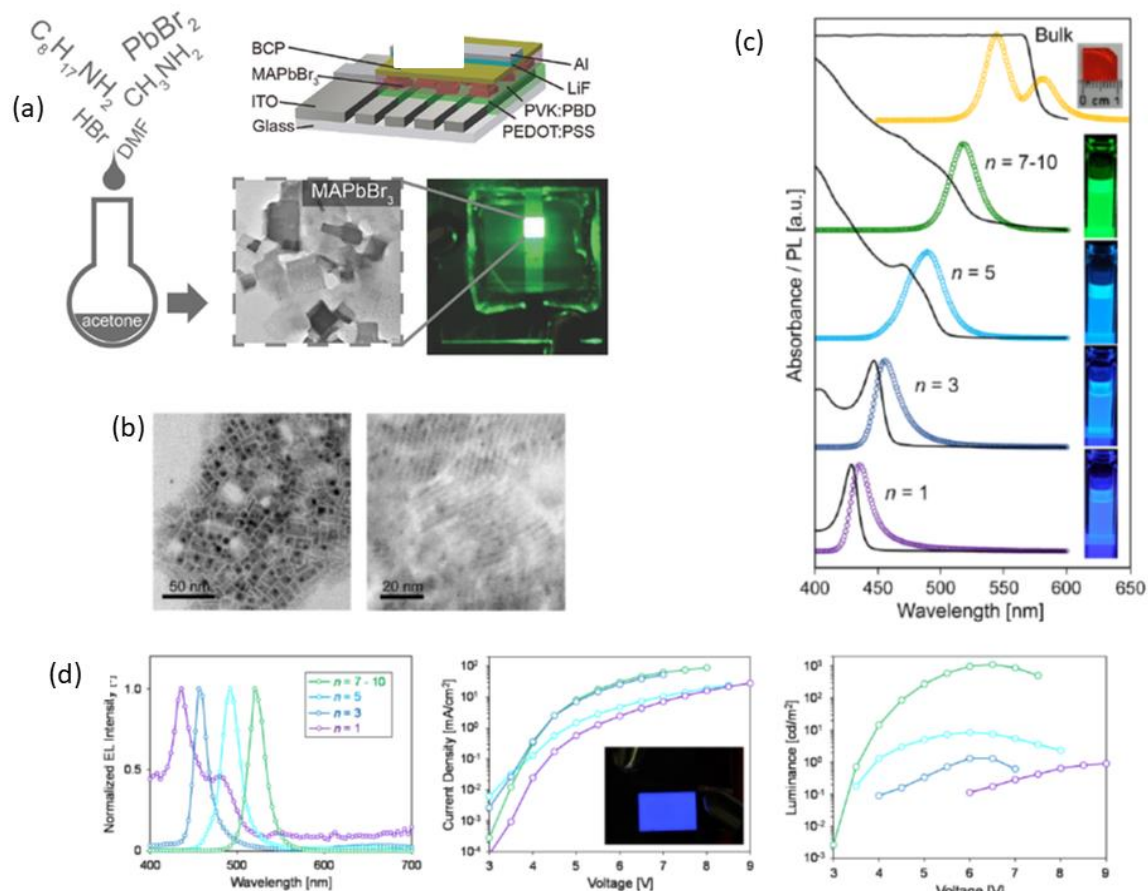

**Figure 19:** (a) Schematic representation of the solution process for the synthesis of MAPbBr₃ NPls used in the first NPls-based LED device which is also shown schematically and in operation. (a has been reproduced from ref. [124] with permission from Wiley Online Library, copyright 2016). (b) Indicative transmission electron microscopy (TEM) profiles of $CH_3NH_3PbBr_3$ crystals with layers n=7-10 (left), and n=3 (right). (c) Absorbance and photoluminescence (PL) spectra of $CH_3NH_3PbBr_3$ bulk single crystal and colloidal solution with various n. (d) 2D $CH_3NH_3PbBr_3$ LEDs device characteristics as expressed by normalized electroluminescence (EL), current density versus voltage, and luminescence versus voltage plots. The inset in (d) slows a pure blue LED. b-d have been reproduced from ref. [82] with permission from American Chemical Society, copyright 2016.

The following year, a great enhancement on the EQE of pure green LEDs based on monolayers of $CsPbBr_3$ NPls was reported by Si et al. [125] In particular, $CsPbBr_3$ NPls that are passivated by bulky phenylbutylammonium (PBA) cations have been fabricated, for the development of 2D layered $PBA_2(CsPbBr_3)_{n-1}PbBr_4$ perovskite films. Figure 20a depicts typical TEM images of the so-formed perovskite nanocrystals, whereas Figure 20b presents the structure schematic animation of the synthesized NPls. The thickness of the perovskite structure is of n=12-16 monolayers. Notably, the formation of these thickness-controlled quantum-well (TCQW) structures resulted to crystalline films with smooth surface features and narrow emission line widths. Figure 20c presents the EL spectrum of the LED upon applying a bias voltage of 4 V, while the inset shows a photo of a device in operation. [125] An EQE up to 10.4 % was achieved for the developed green LED. Moreover, on a similar basis, the authors synthesized $CsPbI_3$ NPls for the development of red-light emitting devices, that exhibited EQEs up to 7.3% (Table 4). [125]. Kumar et al. exploited further the synthesis of 2D formamidinium lead bromide ($FAPbBr_3$) PNCs for the development of green LEDs, while the highlight of the study was the demonstration of an ultra-flexible large-area (3 cm²) LED device (Table 4). [126]

In 2018, Yang et al. developed a blue LED device based on ultrathin $CsPbBr_3$ NPls. [127] The NPls were synthesized through a simple large-scale one-pot approach that allowed plausible thickness control upon varying the duration of heating. Figure 20d depicts indicative photographs of the synthesized NPls dispersions and the architecture of the developed device, while the LED features are shown in Figure 20e, i.e. luminescence emission, current density-voltage-luminescence (J-V-L plot), and current efficiency-current density-power efficiency (E-J-P plot). The reported EQE of the blue LED device was small (0.1%), however the study by Yang et al. paves a nice way towards large-scale production of the widely used ultrathin perovskite NPls in LED applications (Table 4). Similarly small EQEs were reported for other two cases of blue LED devices within the same year. In particular, Bohn et al. developed a blue LED based on CsPbBr₃ NPls passivated by PbBr₂-ligands [128], and Wu et al. showed the benefits of in-situ HBr passivation on CsPbBr₃ NPls-based LEDs [129]. The reported EQEs for the two blue LED devices were 0.05 % and 0.12 %, respectively.

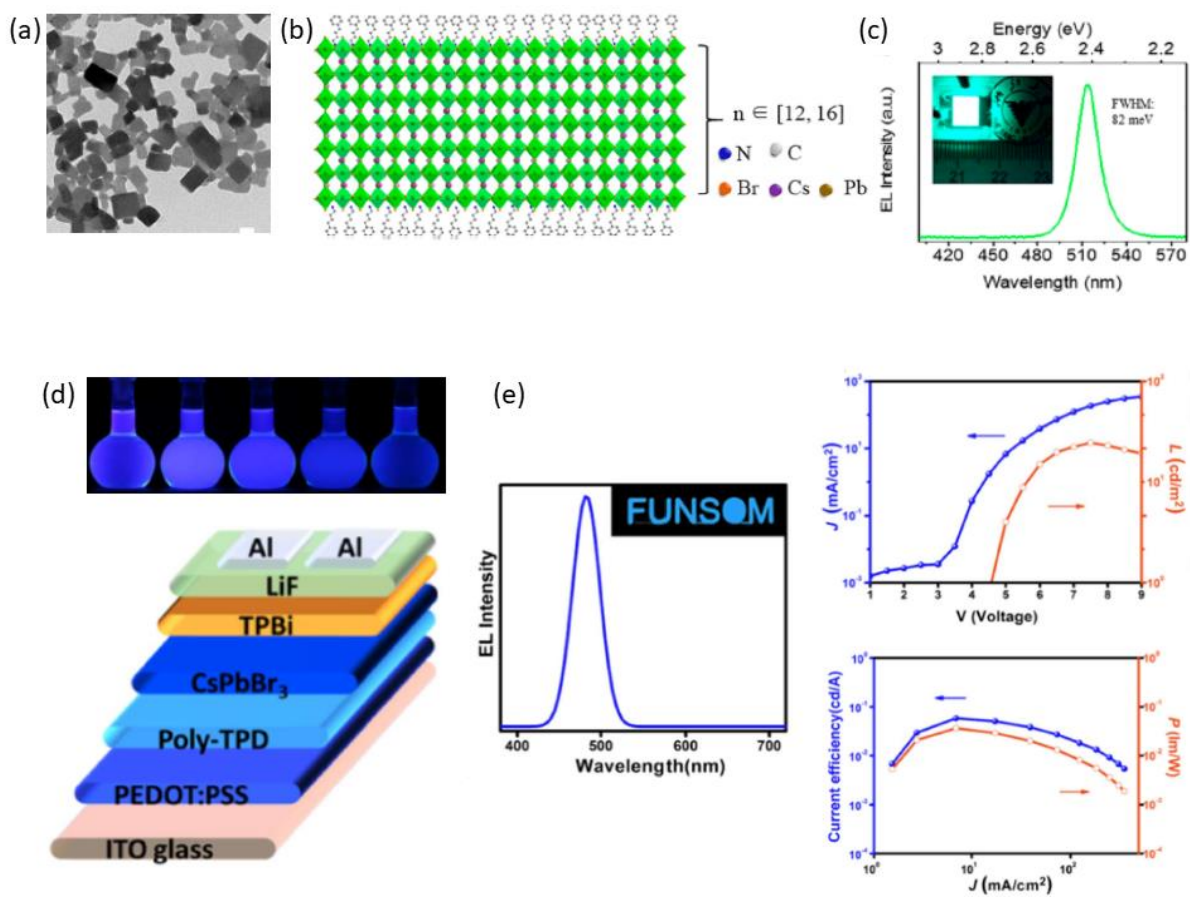

**Figure 20:** (a) Transmission electron microscopy (TEM) photo of the synthesized CsPbBr₃ NPls. (b) Schematic animation of the layered CsPbBr₃ NPls structure. (c) Electroluminescence (EL) spectrum at an applied bias voltage of 4 V. The inset shows the LED device in operation (d) Photos of the CsPbBr₃ NPls dispersions and schematic animation of the LED device architecture. (e) LED features as expressed by luminescence emission spectrum, current density-voltage-luminescence (J-V-L plot), and current efficiency-current density-power efficiency (E-J-P plot). a-c have been reproduced from ref. [125] with permission from American Chemical Society, copyright 2017. d and e have been reproduced from ref. [127] with permission from Elsevier, copyright 2018.

A year later, another two studies focused on the development of blue LEDs. Hoye et al. demonstrated the fabrication of blue emitting devices based on CsPbBr₃ NPls, while employing poly(triarylamine) interlayers. [130] The incorporation of the polymer was found to reduce effectively the nonradiative losses within the PNCs, suggesting a promising strategy towards improving the EQE of blue light emitting devices. LEDs emitting at 464 nm (blue) and at 489 nm (sky-blue) were fabricated, upon following the device configuration depicted in Figure 21a. Figure 21b shows the corresponding EL spectra along with photographs of the devices in operation. [130] The achieved EQEs of blue and sky-blue LEDs were 0.3 % and 0.55 %, respectively (Table 4). Following this, Fang et al. reported on the development of green LEDs based on FAPbBr₃ nanocrystals [90]. As in a previous occasion [126], the formamidinium perovskite was selected also by Fang et al. due to its better thermal stability and greener PL when compared to the more popular methylammonium component. Figure 21c presents the green LED architecture, along with the energy band diagram of the components and a cross-section SEM image of the device. The fabricated LED emitting at 532 nm exhibited an EQE of 3.53% (Table 4) [90]. Figure 21d shows an actual device in operation, whereas Figure 21e depicts the EL stability of a resin encapsulated FAPbBr₃-based LED over operational time. The corresponding stability test of the MAPbBr₃ device is included for comparison. Inspection of Figure 21e reveals the superiority of the former LED in terms of stability.

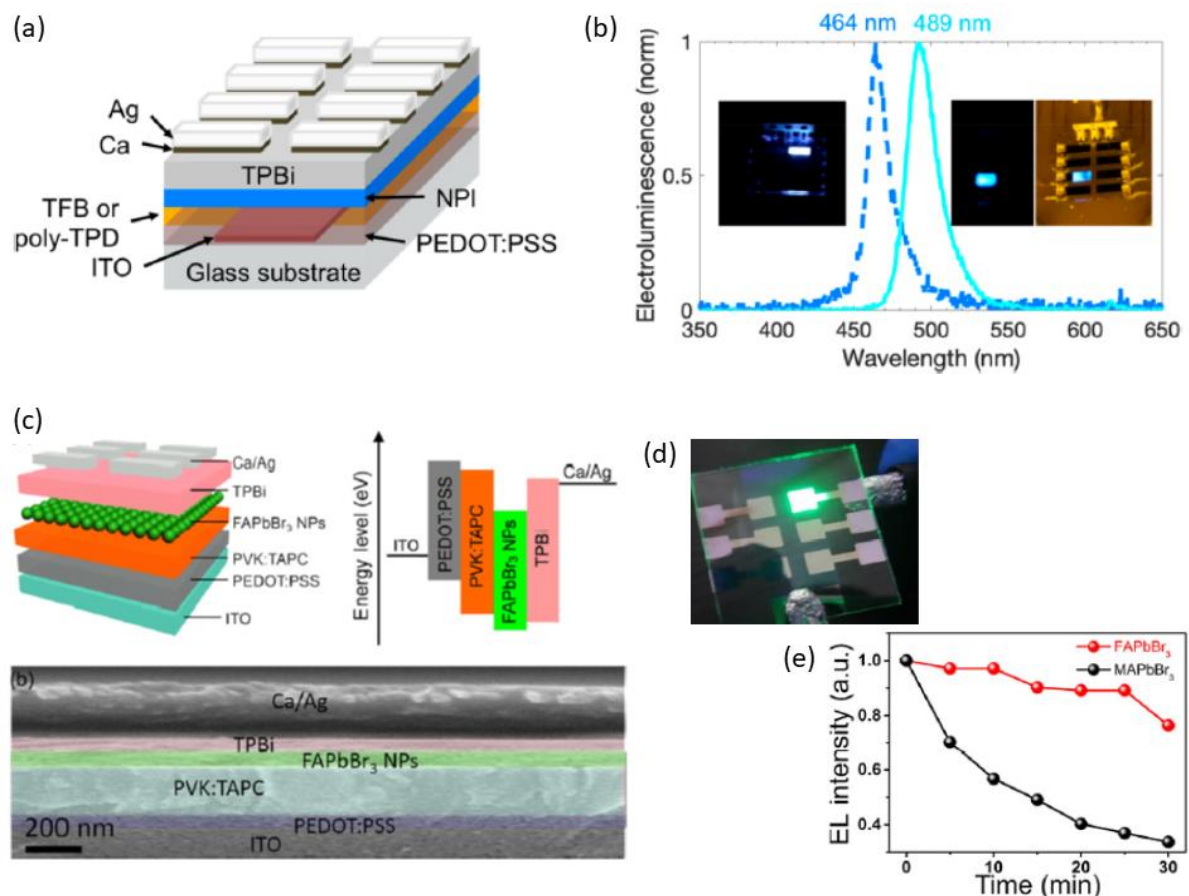

**Figure 21:** (a) CsPbBr₃ NPIs blue LED device configuration. (b) Electroluminescence (EL) spectra along with photos of blue and sky-blue LEDs in operation. (c) FAPbBr₃ NPIs green LED device configuration, energy band diagram, and scanning electron microscopy (SEM) cross-sectional image. (d) FAPbBr₃ NPIs green LED in operation. (e) Electroluminescence (EL) stability of resin encapsulated FAPbBr₃ and MAPbBr₃ devices. a and b have been reproduced from ref. [130] with permission from American Chemical Society, copyright 2019. c-e have been reproduced from ref. [90] with permission from Springer, copyright 2019.

The following year, Peng et al. presented another deep-blue LED emitting at 439 nm [91]. The device was based on FAPbBr₃ NPIs treated by trioctylphosphine oxide (TOPO) for enhancing stability and charge transport. Figure 22a presents the EL spectra of the LED, and the corresponding PL of the perovskite NPIs. The inset shows the deep-blue LED in operation. Furthermore, Figure 22b depicts the energy gap schematic of the constructed device, whereas Figure 22c presents the chromaticity coordinates [91]. An EQE of 0.14% was obtained with a turn-on voltage of 3.6 V (Table 4). Yin et al. in 2021 reported on another blue LED emitting at 465 nm [131]. In this device CsPbBr₃ NPIs were used, treated by polyethylenimine (PEI) in order to synthesize larger NPIs without changing the thickness. The authors explain that this lateral size enhancement reduced the coalescence features of the NPIs and decreased the density of their trap states. Consequently, higher PL quantum yields were obtained along with better color saturated blue emission. Figure 22d depicts the device configuration, while the corresponding energy diagram of the components is shown in Figure 22e [131]. Figure 35f shows the normalized intensity of the EL of the LED along with the PL of the perovskite NPIs. The fabricated LED device exhibited an EQE of 0.8% with a turn-on voltage of 2.6 V (Table 4). Moreover, an inset in Figure 22f depicts a photo of the blue LED device in operation. In the same year, Lin et al. developed all-perovskite white light LEDs based on K-Br passivated CsPbBr₃ NPIs, with CsPbBr₃ and CsPbBr₁.₅I₁.₅ components as green and red phosphor, respectively [132]. It is worth noting at this point, a different approach on enhancing the PL properties of 2D perovskites towards LEDs and other optoelectronic applications, by means of employing hyperbolic metamaterials. Indicatively, Tonkaev et al. predicted theoretically and demonstrated experimentally the enhancement of external quantum yield upon depositing quasi-2D perovskite films on hyperbolic metamaterial substrates consisting of gold and aluminum oxide layers.[133] Along similar lines, Guo et al. achieved hyperbolic dispersion arising from anisotropic dispersion in 2D perovskites.[134]

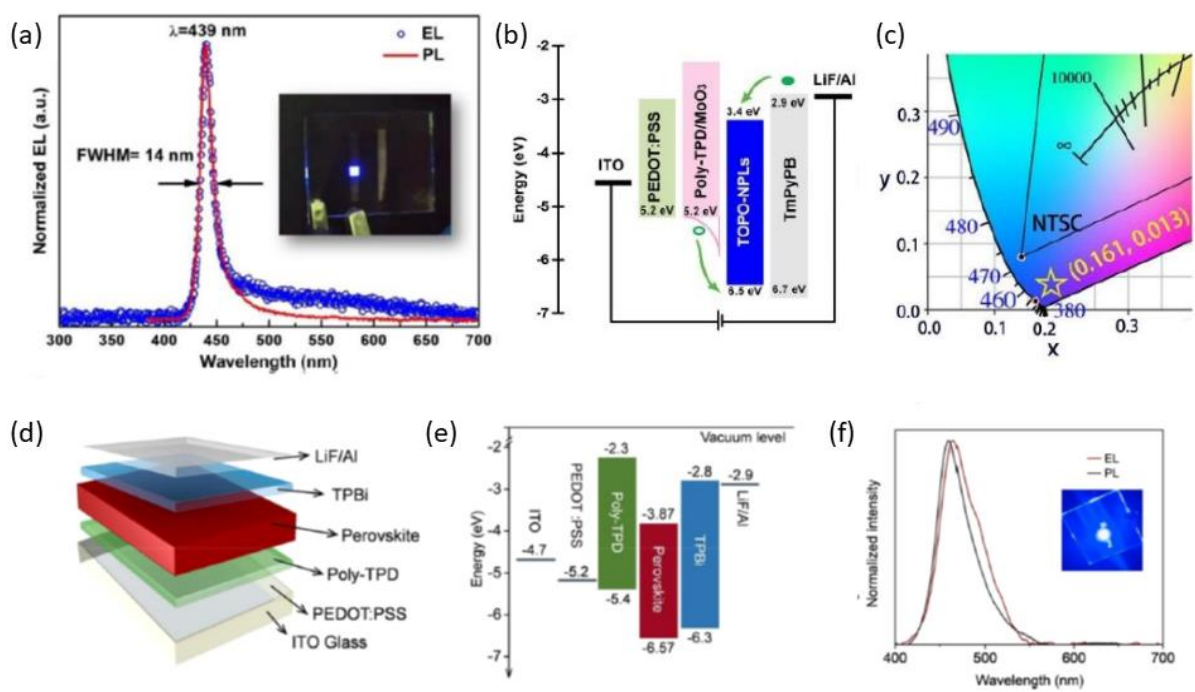

**Figure 22:** (a) Electroluminescence (EL) and photoluminescence (PL) of the LED devices and the perovskite NPls. (b) Energy diagram of the device structure. (c) LED device chromaticity coordinates. (a-c have been reproduced from ref. [91] with permission from American Chemical Society, copyright 2020). CsPbBr₃ NPls LED device configuration (d), and energy band diagram (e). (f) Normalized EL and PL intensity of the LED device and the perovskite NPls. The insets show the developed blue LED devices in operation. (d-f have been reproduced from ref. [131] with permission from American Chemical Society, copyright 2021).

### 4.1.3. Metal halide perovskite nanocrystals/2D materials

Finally, it is worth pointing out that 2D/3D perovskite heterostructures offer a nice platform for the development of LEDs with superior characteristics. As for instance, Heo et al. reported that the employment of a 2D/3D multidimensional interface induces carrier transmission, enhances the density of electrons and holes, and increases recombination.[135] Consequently, the EQE of the LED device is advanced. Similarly, Jiang et al. developed 2D/3D heterojunction perovskite LEDs with tunable ultrapure blue emissions.[136] Namely, by means of organic halide salt post treatment, a 2D capping layer is formed on the original 3D perovskite surface. The formation of the 2D layer assisted the passivation of defects and improved the exciton radiative recombination rate. These studies indicate that the employment of perovskite 2D/3D heterojunctions appears a promising tool towards efficient and stable LEDs.

Moreover, the design of advanced heterostructures consisting of PNCs and 2D non-perovskite materials has also been demonstrated towards the fabrication of efficient LEDs. Indicatively, Qiu et al. reported on the formation of CsPbX3 PNCs (with X=Cl, Br, I) on top of previously exfoliated 2D hexagonal boron nitride (h-BN) NSs, by means of one-pot in-situ growth synthesis.[51] The developed protocol allowed homogeneous distribution of the so-formed PNCs on the surface of the h-BN NSs, while resulting to significant improvement of the PNCs/h-BN heterostructures thermal stability. Namely, at 120 °C nearly 80% of the initial PL intensity was maintained. In addition, the PNCs/h-BN NSs exhibited improved environmental stability, upon storage in ambient atmosphere for several days without noticing severe degradation of the PL intensity. The authors demonstrated the construction of a LED device emitting warm white light, based on the employment of the synthesized blue, green, and red PNCs/h-BN powders.[51] Along similar lines, Li et al. employed successfully the 2D h-BN for improving the stability of CsPbBr₃ quantum dots (QDs).[52] Such approaches open another window towards practically expanding the application of PNCs/2D heterostructures towards advanced, stable, and efficient LEDs.

**Table 4:** Summary of low-dimensional perovskite nanosheets (NSs) and nanoplatelets (NPls) based configurations in light-emitting devices (LEDs). *NR stands for not reported.

| Study-year | Synthesis protocol | LED device configuration | LED device characteristics (EL peak, EQE, turn on voltage) | Key features / Remarks |
|---|---|---|---|---|
| **Nanosheets (NSs) based LEDs** | | | | |
| Parveen et al., 2020[122] | Solvothermal process | $CH_3NH_3PbBrI_2$ NSs/blue LED chip | 470-570 nm, NR, NR | White light converter based on blue LED |
| Gao et al., 2022[123] | Post-treatment ion exchange | $CsPb(Br/Cl)_3$ NS/blue LED chip | 462 nm, NR, NR | Copper doping, deep-blue LED |
| **Nanoplatelets (NPls) based LEDs** | | | | |
| Ling et al., 2016[124] | Solution-processed | Glass/ITO/PEDOT:PSS/MAPbBr$_3$/BCP/LiF/Al | 529 nm, 0.48%, 3.8V | First organic-inorganic perovskite NPls LED |
| Kumar et al., 2016[82] | Solution-processed | Glass/ITO/PEDOT:PSS/HTL/EML/ETL/LiF/Al | 456 nm, 0.23%, 3.5 V | First LED emission below 490 nm (pure blue) |
| Kumar et al. 2016[82] | Solution-processed | Glass/ITO/PEDOT:PSS/HTL/EML ETL/LiF/Al | 530 nm, 2.31%, 3.5 V | Pure green LED, n=7-10 monolayers |
| Si et al., 2017[125] | Solution-processed | Glass/ITO/NiO/TFB/PVK/CsPbBr$_3$ TCQW/TPBi/Ca/Al | 514 nm, 10.4%, 2.8 V | Pure green LED, n=12-16 monolayers |
| Si et al., 2017[125] | Solution-processed | Glass/ITO/NiO/TFB/PVK/CsPbBr$_3$ TCQW/TPBi/Ca/Al | 683 nm, 7.3%, 3 V | Pure red LED, CsPbI$_3$ NPls |
| Kumar et al., 2017[126] | Solution-processed | Glass/ITO/PEDOT:PSS/Poly-TPD/2D FAPbBr$_3$ PMMA/3TPyMB/LiF/Al | 530 nm, 3.04%, 2.75V | Large area (3 cm$^2$) flexible LED |
| Yang et al., 2018[127] | Solution-processed | ITO glass/PEDOT:PSS/Poly-TPD/CsPbBr$_3$/TPBi/LiF/Al | 480 nm, 0.1%, 3 V | Blue LED, large scale production |
| Bohn et al., 2018[128] | Solution-processed | ITO/PEDOT:PSS/Poly-TPD/CsPbBr$_3$/TPBi/Ca/Ag | 464 nm, 0.05%, 3.8 V | Blue LED, PbBr$_2$-ligands passivation |
| Wu et al., 2018[129] | Solution-processed | ITO/PEDOT:PSS/Poly-TPD/CsPbBr$_3$ NPls/TPBi/LiF/Al | 463 nm, 0.12%, 3.5 V | Blue LED, HBr passivation |
| Hoye et al., 2019[130] | Solution-processed | Glass/ITO/PEDOT:PSS/poly-TPD/CsPbBr$_3$ NPls/TPBi/Ca/Ag | 463 nm, 0.3%, 4 V | Blue LED, Poly-TPD band alignment |
| Hoye et al., 2019[130] | Solution-processed | Glass/ITO/PEDOT:PSS/poly-TPD/CsPbBr$_3$ NPls/TPBi/Ca/Ag | 489 nm, 0.55%, 4 V | Sky-blue LED, Poly-TPD band alignment |
| Fang et al., 2019[90] | Solution-processed | ITO/PEDOT:PSS/PVK/TAPC/FAPbB$_3$ NPls/Ca/Ag | 532 nm, 3.53%, NR | Green LED, FA-based |
| Peng et al., 2020[91] | Solution-processed | Glass/ITO/PEDOT:PSS/poly-TPD/FAPbBr$_3$ NPls/TmPyPB/LiF/Al | 439 nm, 0.14%, 3.6 V | Deep-blue LED, TOPO treatment |
| Yin et al., 2021[131] | Solution-processed | Glass/ITO/PEDOT:PSS/poly-TPD/CsPbBr$_3$ NPls/TPBi/LiF/Al | 465 nm, 0.8%, 2.6 V | Blue LED, PEI treatment |

| Lin et al., 2021[132] | Solution-processed | NR | NR, NR, NA | White LED, K-Br passivated, $CsPbBr_3$ green and $CsPbBr_{1.5}I_{1.5}$ red phosphor |

## 4.2 Photodetectors

### 4.2.1. Metal halide perovskite nanoplatelets/nanosheets as photodetection element

Metal halide perovskite nanocrystals in the form of 2D architectures, NPls or NSs, combine the excellent optoelectronic properties of the perovskite nanomaterials with their unique 2D geometry and large lateral dimensions make them ideal building blocks for optoelectronic devices. Among these devices, photodetectors with high on/off ratios and fast response times have been design and fabricated using such geometries. The features of these devices are summarized in the Table 5.

**Rigid photodetectors.** The first photodetector using all-inorganic metal halide perovskite NSs was reported in 2016 by Lv et al. [116] The simple structured device was constructed as following: the photodetection element which was the $CsPbBr_3$ NSs synthesized by hot injection method was drop casted between two Au electrodes pre-patterned on $SiO_2$/Si wafers followed by drying (Figure 23a). The $CsPbBr_3$ NSs were selected due to their better stability against moisture compared to that of $CsPbI_3$. Time-dependent photocurrent was measured under a 1 Hz pulse laser (450 nm) at a fixed light intensity (13.0 mW/cm$^2$) with a bias voltage of 1 V (Figure 23 b). The optical switching and stability of the perovskite NS photodetector was confirmed by the prompt and reproducible photocurrent response to on/off cycles (Figure 23c). The rise time was determined to be 17.8 ms, while the decay times were determined to be 14.7 and 15.2 ms (Figure 23d). In the same year, a bit later, $CH_3NH_3PbI_3$ NSs have been used as photodetection element as well (Figure 23e).[137] The NSs have been prepared by a combined solution process and vapor-phase conversion. 2D $PbI_2$ NSs were firstly nucleated onto the substrate by drop casted a saturated $PbI_2$ aqueous solution and subsequently heated at an elevated temperature and then the 2D $CH_3NH_3PbI_3$ NSs were formed by intercalating the $CH_3NH_3I$ molecules into the interval sites of $PbI_6$ octahedron layers through a CVD process. The photodetector including these NSs found that was sensitive to a broad-band light from the ultraviolet to the entire visible spectral range and presented an effective optical switching (Figure 23f). Nevertheless, the 2D perovskite-based photodetector is sensitive to a broad-band light from the ultraviolet to the entire visible spectral range. Time dependent photocurrent was measured under the illumination of natural light (inset of Figure 26f). The rise and decay times for this device were found slightly higher (Figure 23h) than that of using $CsPbBr_3$ and reported by Lv et al. [116] while its photoresponsivities with a voltage bias of 1 V were calculated to be 22 AW$^{-1}$ under a 405 nm laser and 12 AW$^{-1}$ under a 532 nm laser (Figure 23g), higher than that of the devices included bulk perovskite film[138] and lower than that of perovskite microplates[139] of the same chemical phase. The dependence of the photoresponse as a function of NS thickness was studied by Niu et al and the photodetector including NSs of 30–40 nm thickness showed the best electronic and optoelectronic performance (Figure 23i). [32] Furthermore, the nanoplatelet-based devices demonstrated a relatively higher performance compared to that of the nanowires of the $CH_3NH_3PbI_3$ chemical phase. [140]

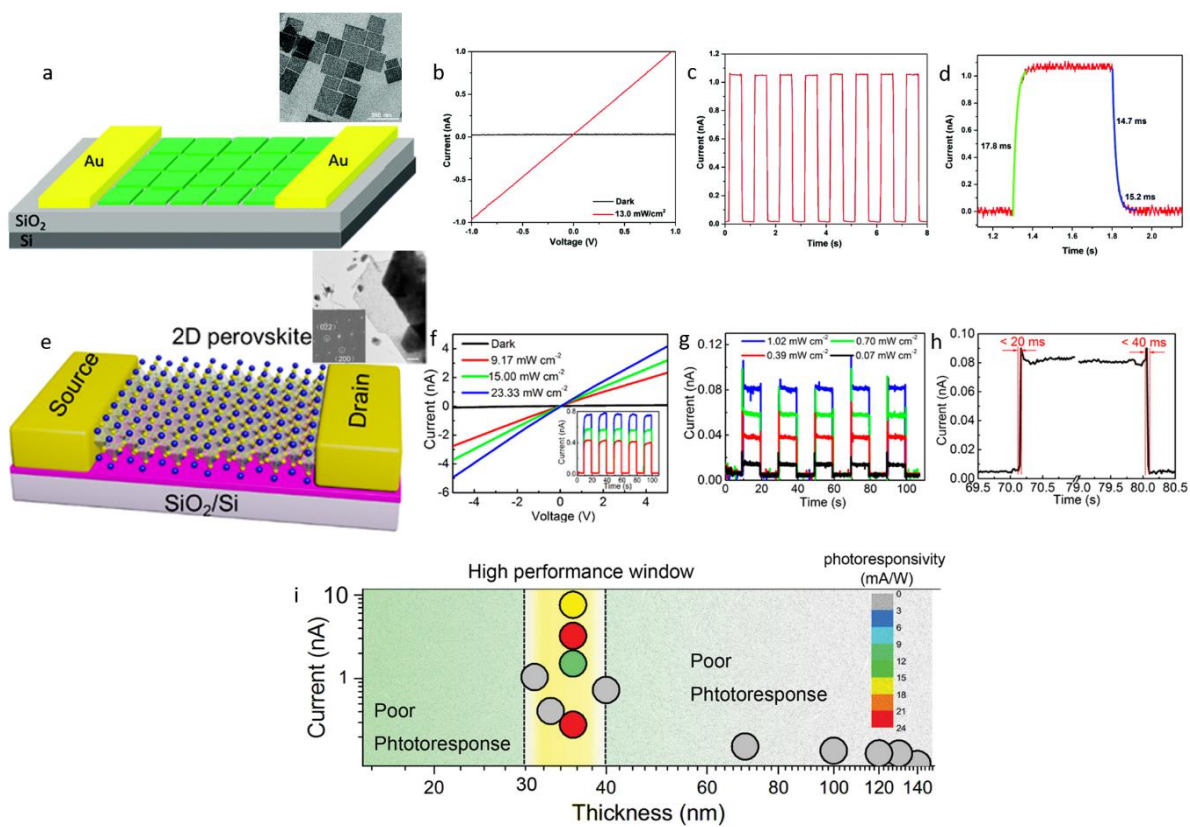

**Figure 23:** (a-d) Schematic of the first photodetector based on all inorganic metal halide (CsPbBr₃) nanosheets and its characteristic, *I-V* curves measured in the dark and under illumination using a 450 nm laser diode, photocurrent-time response measured in the dark and under pulsed laser (450 nm, 1 Hz) with a bias of 1 V ($P = 13.0$ mW/cm²) and rise and decay times of the photodetector. Reprinted with permission from Ref. [116], copyright 2016, The Royal Society of Chemistry. (e-h) Schematic of the first photodetector based on organic-inorganic metal halide nanosheets and its characteristic, I−V curves of the 2D perovskite-based device under the irradiation of natural light with different power, time-dependent photocurrent measurement on the 2D perovskite phototransistor under the different power of a 405 nm laser with a voltage bias of 1 V, temporal photocurrent response excited at 405 nm. Inset f: Time-dependent photocurrent measurement over five on−off periods of operation under different power of natural light with a voltage bias of 1 V. Reprinted with permission from Ref.[137], copyright 2016, American Chemical Society. (i) Different electrical performances with perovskite nanoplatelets depending on different thickness. Derived bias is 1 V for both dark current and photoresponse. Reprinted with permission from Ref.[32], copyright 2016, Wiley.

Later, two works showed that the design of the photodetector (planar or vertical-type) and also the method that the 2D perovskites were fabricated were crucial for the final efficiency of the photodetector. Firstly, Li et al. fabricated a vertical-type photodetector to compare its efficiency with those of planar-type. The CH₃NH₃PbI₃ perovskite NSs used as photodetection elements were fabricated with the same method used previously by Liu et al. [137] The comparison revealed that the vertical-type photodetector showed the advantages of low-voltage operation and large responsivity compared to the planar-type design. Furthermore, photodetectors based on solution-processed scattered CsPbBr₃ NPls with lateral size of 10 μm developed though an ion-exchange soldering mechanism showed higher efficiency compared to that used nanoplatelets/nanosheets synthesized by different methods. [141] The large lateral size facilitates a single CsPbBr₃ nanoplatelet to bridge across the channel between two metal electrodes. In order to study the effect of using these scattered single nanoplatelets as photodetector elements, a drop casted densely packed NPls were also fabricated and tested. The photoresponsivities of these densely packed CsPbBr₃ NPls were approximately two orders of magnitude lower than the responsivities calculated from scattered single CsPbBr₃ NPls due to the significantly enhanced light absorption and the facilitated subsequent electron–hole transport directly between metal electrodes without boundaries among different NPls. Furthermore, post annealed procedure has been performed in order to obtain strongly coupled ligand-capped nanosheets films, to improve the charge transfer rate while maintaining their stability in the NSs film. [142]

A comprehensive study of the effect of the thickness of the NSs on the photodetection capability using these materials were reported by Mandal et al recently (Figure 24a).[33] This investigation revealed that the CsPbBr$_{1.5}$I$_{1.5}$ NSs with thickness of ~6.1 nm showed the best responsivity in photodetectors with in plane configuration. The photodetector's properties including the CsPbBr$_{1.5}$I$_{1.5}$ NSs are improved compared to those of CsPbBr$_3$ or CsPbI$_3$ (Figure 24b) and nanocrystals having the same stoichiometry (Figure 24c). The current density, the responsivity and the detectivity between the different devices were illustrated in Figures 24 c-f and the bar plots of the maximum obtained responsivity and detectivity at different applied biases were in Figure 27h. The highest responsivity and detectivity at 2 V were demonstrated for device including the Br$_{1.5}$I$_{1.5}$ NSs (2121 A W$^{-1}$) and 1.5 V (3.8 × 10$^{10}$ Jones), respectively. In addition, the photodetector parameter found that can be enhanced by further altering the thickness of the NSs. While the NSs with 4.9 ± 0.5 nm were the thinner one prepared by using 18-carbon ligands, the NSs with 6.1 ± 0.3 by combining short-chain (8-carbon) ligands (octanoic acid and octylamine) showed the best photodetector performance (photocurrent, responsivity and detectivity) (Figure 24g-i). In addition, this device showed lower rise (116 ms) and decay time (147 ms) than the thinner Br$_{1.5}$I$_{1.5}$ NS devices (150 and 156 ms respectively).

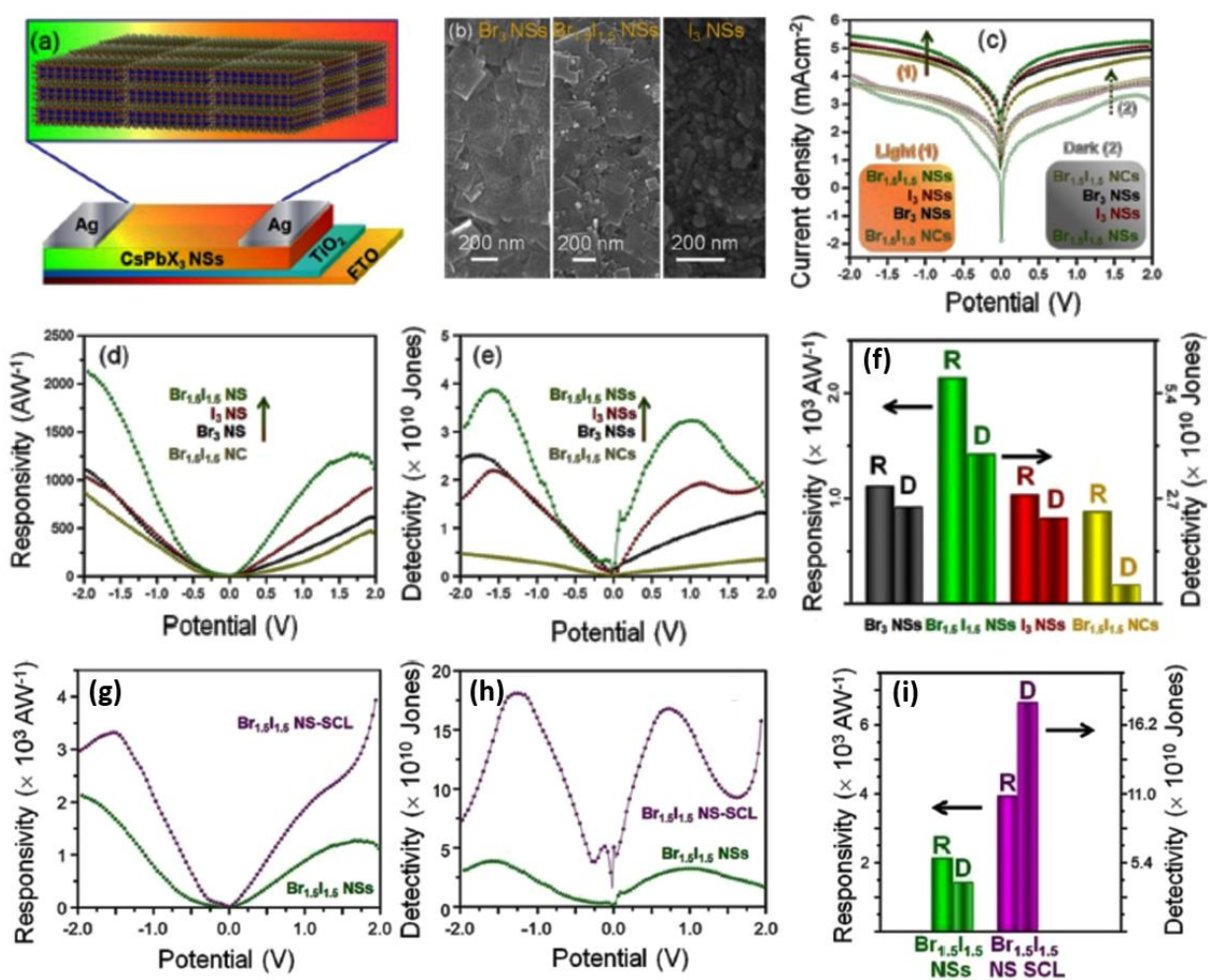

**Figure 24:** (a) Schematic diagram of the CsPbX$_3$ (X= Br, Br/I, I) NSs photodetector device. (b) Top view FESEM images of the nanosheets film. (c-f) Responsivity plot, detectivity plot and Bar plot of the optimum values. (g-i) The corresponding diagrams for the optimum device including the NSs capped with short chain ligand (nanosheets thickeness= 6.1 nm). Reprinted with permission from Ref.[33] , copyright 2021, American Chemical Society.

**Flexible photodetectors.** The first flexible photodetector using similar structured materials was reported by Song et al. in 2016. [143] All-inorganic CsPbBr$_3$ NS dispersion was used to assemble films by centrifugal casting on the patterned ITO/polyethylene terephthalate (PET) substrates to form the flexible device (Figure 25a). The responsivity of this flexible photodetector under 5 V at 517 nm reached the value of 0.25 A W$^{-1}$, which matched that of commercial Si photodetectors (<0.2 A W $^{-1}$). The reproducible character in response to the on-off cycles also revealed and extracted rise and decay times were 19 and 25 μm respectively which are much shorter than the previously reported rigid photodetectors including metal halide perovskite NSs. Also, the flexible photodetectors were stable at various bending curvatures even after bending for 10000 times but also after irradiating for 12 h with a 442 nm laser (10 mW cm$^{-2}$). The current fluctuations were <3%.

In order to improve the film conductivity of the CsPbBr$_3$ NSs and hence boost the performance of the flexible photodetector, the construction of a perovskite NSs/carbon nanotubes (NSs/CNTs) composite has been proposed by Li et al. (Figure 25b). [144] Such strategy would contribute to both high response speed and high responsivity. According to the energy level of CsPbBr$_3$ NSs and CNTs the photogenerated electrons would be extracted quickly to CNTs resulting in a fast response speed and efficient usage of excited electrons and also the electron transport could be improved as these could be drift along the CNTs smoothly with little scattering and recombination due to the high conductivity of them. Hence, the improved responsivity was 31.1 AW$^{-1}$ under a bias of 10V, and the rise and decay times were 16 μs and 0.38 ms. The composite showed good flexibility (> 10000 bending cycles) and photostability (< 3.8% after 10000 on/off switching cycles). Smaller fluctuations in photocurrent (~2%) after 10000 bending cycles have been succeed using a "double solvent evaporation inducing self-patterning" strategy to generate high-quality patterned thin NSs films in selected areas automatically after drop-casting in flexible substrates. [145] This "self-patterning" approach led to large continuous area, micro-cracks-free, dense and high-quality CsPbBr$_3$ NSs films and to photodetectors with responsivity of 9.04 A/W.

In a different approach, all-inorganic metal halide NSs were blended also with PCBM to form a photodetector element for flexible photodetectors. [146] The photodetectivity in the device using the hybrid element found much higher compared to that using the bare perovskites NSs. Efficient conductive layers such as CuSCN and NiO were employed also in order to increase the charge transfer rates. [147]

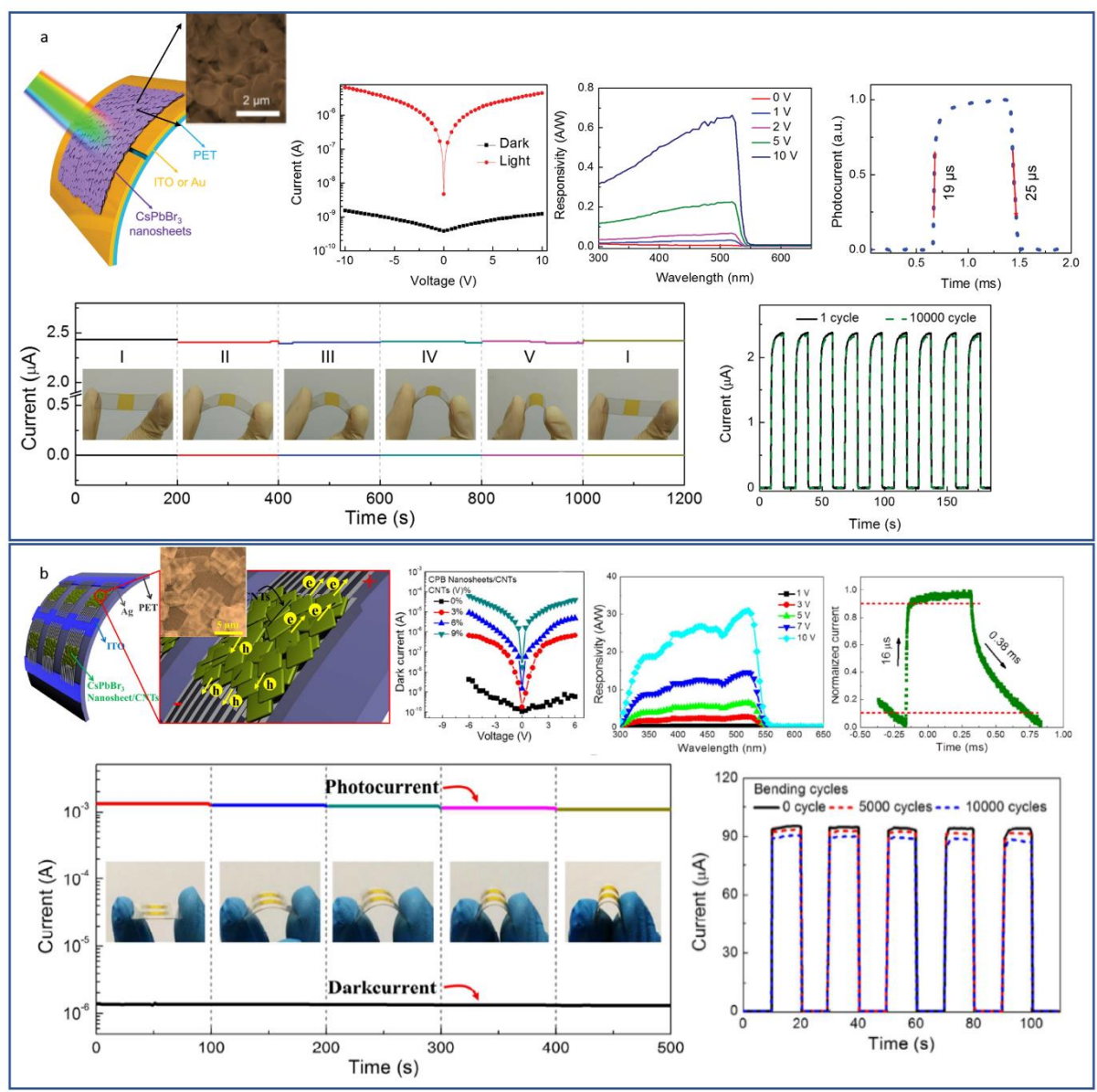

**Figure 25:** Flexible photodetectors based on 2D metal halide NSs (CsPbBr₃) (a) and perovskite NSs/carbon nanotubes composite (b). Device structure and basic features. (a) Reprinted with permission from Ref. [143], copyright 2016, Wiley . (b)Reprinted with permission from Ref. [144], copyright 2017, American Chemical Society.

### 4.2.2. Metal halide perovskite nanocrystals/2D materials heterostructures as photodetection element

**Rigid photodetectors.** The first photodetector including metal halide perovskite nanocrystals/2D materials heterostructures reported in 2015 by Wang at al. [72] In particular, a high performance phototransistor based on organic-inorganic metal halide nanocrystals/graphene heterostructure was constructed in which CH₃NH₃PbBr₂I perovskite islands instead of continuous film have been grown through a fast crystallization deposition method on monolayer graphene. Perovskite nanoislands with well-controlled size and distribution were grown directly on the graphene film that was deposited on Si/SiO₂ substrates. The electrodes were fabricated to facilitate the collection of photocarriers. The photodetector showed an extremely high responsivity $6.0 \times 10^5$ A W$^{-1}$ and a photoconductive gain of $\approx 10^9$ electrons per photon because of effective photogating effect applied on graphene along with increased lifetime of trapped photocarriers in separate perovskite islands. This photodetector was also capable for broadband detection from 250 to 700 nm. Furthermore, almost simultaneously photodetectors using heterostructured materials synthesized with wet chemistry approach was reported by He et al. 2015. [148] In this device, CH₃NH₃PbI₃/rGO heterostructures were synthesized by in situ growth of the perovskite on the 2D materials through a LARP method. Then this $CH_3NH_3PbI_3$/rGO solution was drop-casted on Si/SiO$_2$ substrate, above which the Au/Cr electrodes were thermally deposited. $CH_3NH_3PbI_3$/rGO hybrids formed p–n molecular junctions with rich interfaces, which lead to efficient charge separation while the insertion of rGO lowered the energy offset between $CH_3NH_3PbI_3$ and Au, which facilitates charge collection. Thus, despite the dark current of the photodetectors including the pure $CH_3NH_3PbI_3$ and those including the perovskite/2D material heterostructure were very close, the second exhibited a photocurrent almost 6.5 times higher. In addition. the ON/OFF ratio found 168, about 6 times higher than the $CH_3NH_3PbI_3$ photodetector (ON/OFF= 23.5) and the responsivity presented a 6-fold improvement (73.9 mA W$^{-1}$).

The photoresponsivity was improved by using all-inorganic nanocrystals/graphene materials heterostructures.[42] The hybrid photodetector CsPbBr$_{3-x}$I$_x$/graphene results in a responsivity as high as 8.2×10$^8$ AW$^{-1}$ and detectivity of 2.4×10$^{16}$ Jones at incident power of 0.07 μW/cm$^2$ under 405 nm illumination. The ligand-caped nanocrystals in this case were prepared by a hot injection method and then deposited on a bilayer graphene film mechanically exfoliated from the graphene flakes on the Si/SiO$_2$ substrate. The slow rise and decay time of the hybrid photodetector calculated in the order of seconds found that could be originated from the blockage of carrier transport due to the long-chain organic ligands on metal halide perovskite nanocrystals. Higher photoresponsivity found also by combining all inorganic metal halides (CsPbBr$_{3-x}$I$_x$) with MoS$_2$ compared to those using organic inorganic ones (CH$_3$NH$_3$PbBr$_3$) (Figure 26a). [57] Much shorter rise and decay times were measured in the photodetectors based on CsPbBr$_3$/MoS$_2$ heterostructures. [149]

Moreover, a solution-processed surface ligand density control strategy has been proposed by Wu et al. to increase the interfacial charge carrier extraction and injection efficiency by removing the residue ligands that exists due to the fabrication approach used and therefore to improve the efficiency of the phototransistors based on CsPbI$_{3-x}$Br$_x$/MoS$_2$ heterostructures (Figure 26b). [57] A treatment with 1-octane/ethyl of the perovskite nanocrystals was included in this approach and no effect on their stability was observed after the treatment. The improved photocurrent and photoresponsivity as well as the rise and decay time of the phototransistor were presented in the Figure 26b. These results can be attributed to the prevention of the carriers to be recombined at CsPbI$_{3-x}$Br$_x$ nanocrystals/MoS$_2$ interfacial traps as well as to probable contribution from the reduced recombination rate in MoS$_2$ region via the n-doping effect of perovskite nanocrystals.

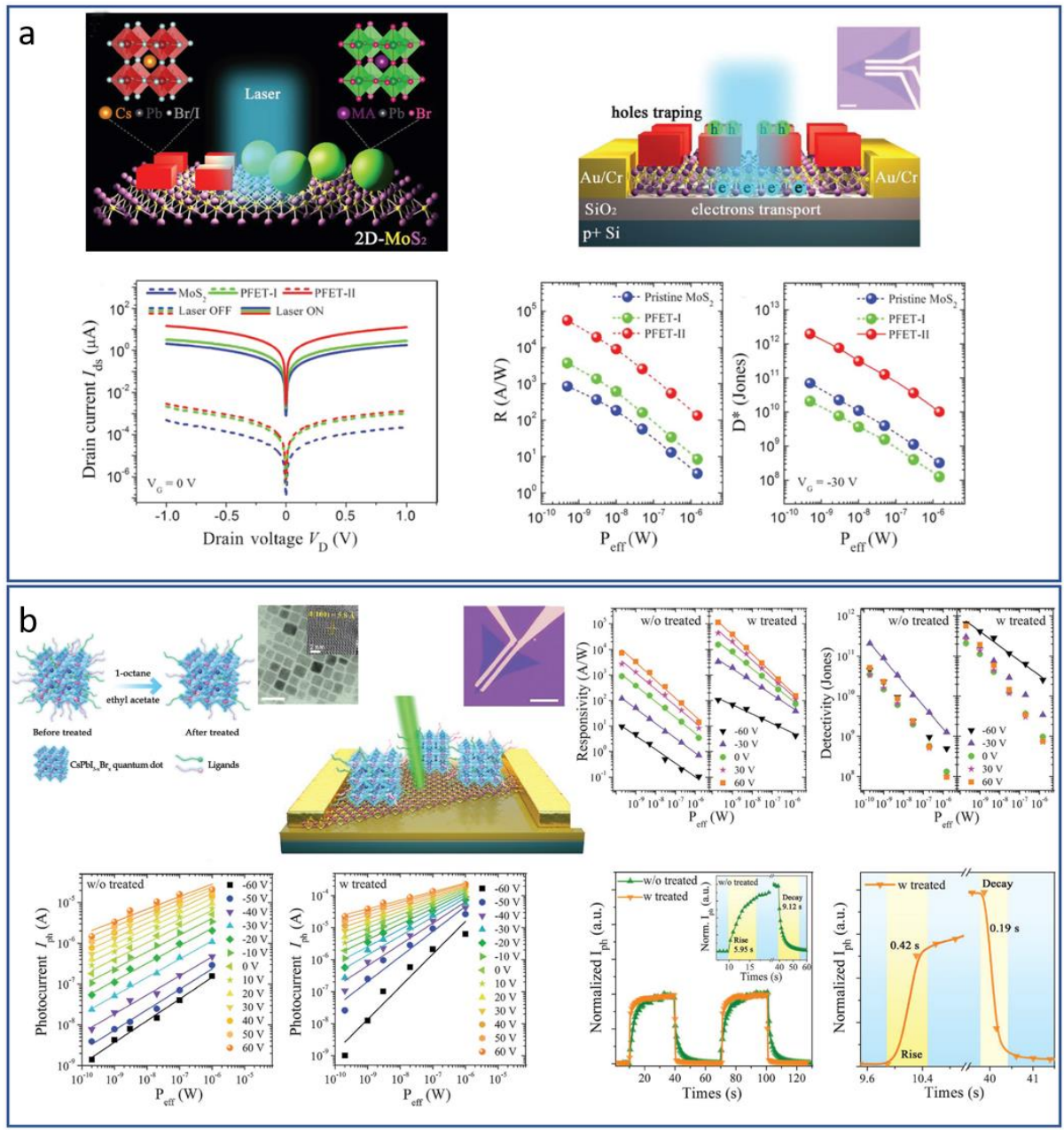

**Figure 26:** Phototransistor including metal halide nanocrystals/2D materials heterostructures. Effect of the chemical phase (a) and surface treatment (b) of the nanocrystals on the photodetector performance. (a) Reprinted with permission from Ref. [57], copyright 2018, Wiley and (b) Reprinted with permission from Ref.[150], copyright 2019, Wiley.

**Table 5.** Summary of the photodetectors including metal halide perovskite nanoplatelets/ nanosheets and perovskite/2D heterostructures as photodetection element and their optoelectronic features.

| Study | Photodetection material | Photodetector fabrication description | Type of the photodetector | Preparation method of the perovskites | Responsivity | On/Off ratio | Rise/decay time | Test conditions |
|---|---|---|---|---|---|---|---|---|
| **Nanoplatelets/Nanosheets** | | | | | | | | |
| Lv et al., 2016[116] | $CsPbBr_3$ NSs | $CsPbBr_3$ NSs have been drop-casted between the electrodes on the Si/$SiO_2$ substrate. | Rigid | Hot injection method | - | $1\times10^2$ | 17.8 ms/ 14.7 and 15.2 ms | Pulsed laser (450 nm, 1 Hz) at a fixed light intensity (13.0 mW/$cm^2$) with a bias of 1 V |
| Liu et al. 2016[137] | $CH_3NH_3PbX_3$ NSs | 2D $PbI_2$ was used as the template, resulting in the growth of a non-van der Waals-type 2D perovskite $CH_3NH_3PbX_3$. A lithography-free technique with electrode materials with a patterned $Si_3N_4$ shadow mask was used. | Rigid | Combined solution process and vapor-phase conversion | 22 $AW^{-1}$ (405 nm, 1V) 12 $AW^{-1}$ (532 nm, 1V) | - | Rise time:< 20 ms Decay time: < 40 ms | 405 and 532 nm lasers, voltage bias of 1V |
| Song et al. 2016[143] | $CsPbBr_3$ NSs | $CsPbBr_3$NSs dispersion was used to assemble films by centrifugal casting on the patterned ITO/polyethylene terephthalate (PET) substrates. | Flexible | Hot injection method | 0.64 mA $W^{-1}$ | >$10^3$ | 0.019 ms/0.024 ms | 442 nm laser light, voltage bias of 1V |
| Niu et al. 2016[32] | $CH_3NH_3PbI_3$ NPls | Vapor phase deposition of 2D triangular or hexagonal $CH_3NH_3PbI_3$ perovskite nanoplatelets on $SiO_2$/Si substrates. For the fabrication of the electrodes, a copper grid was placed on top of as-grown flake as a shadow mask for resist-free metal evaporation. | Rigid | Vapor phase deposition | 23.3 mA $W^{-1}$ | - | 150 ms | 633 nm laser light, voltage bias of 1 V |
| Qin et al., 2016[140] | $CH_3NH_3PbI_3$ NPls | The $CH_3NH_3PbI_3$ NPls synthesized by a solution recrystallization strategy were transferred from the growth surface to the Si/$SiO_2$ through the probe mechanical transferring method. Then, Au electrodes were deposited by a shadow mask method. | Rigid | Solution recrystallization strategy | - | 1210 | - | Various light intensities from 13.0 to 73.7 mWcm$^{-2}$ at 2 V. |
| Li et al. 2017[144] | $CsPbBr_3$ NSs/CNT | Drop-casting the $CsPbBr_3$ NSs/CNT dispersion on the interdigitated electrodes. | Flexible | LARP (NSs). Mixing and sonication (CNTs+ NSs) | 31.1 $AW^{-1}$ | - | 16 μs/0.38 ms | 442 nm laser light, 10V voltage bias |
| Li et al., 2017[151] | $CH_3NH_3PbI_3$ NSs | Combined solution process and vapor-phase conversion method to grow the $CH_3NH_3PbI_3$ NSs on the $SiO_2$ substrate. | Rigid | Solution crystallization process to make the $PbI_2$ nanosheets and then CVD to | 36 mA $W^{-1}$ | | 320 ms/ 330 ms | 635 nm laser light, 0.5 V voltage bias |

| | | | | | | | | convert them to $CH_3NH_3PbI_3$ NSs |
|---|---|---|---|---|---|---|---|---|
| Liu et al. 2017[141] | CsPbBr₃ NPls | The CsPbBr₃ NPls were drop casted onto SiO₂/Si substrates followed by annealing at 60 °C. | Rigid | LARP | 34 A W⁻¹ | - | 0.6 ms/ 0.9 ms | 442 nm laser irradiation, 1.5 V |
| Shen et al. 2018[146] | CsPbBr₃ NSs | The CsPbBr₃ NS solution were blended with PCBM and then dropped onto Si/SiO₂ and ITO/PET electrodes. | Flexible | LARP | 10.85 A AW⁻¹ | - | 44 μs/390μs | 442 nm laser irradiation, 1.5 V |
| Yang et al. 2018[142] | CsPbBr₃ NSs | The CsPbBr₃ NSs were drop casted onto interdigitated electrodes and then annealed. | Rigid | Hot injection | 0.53 AW⁻¹ | 10⁴ | 0.333 ms/0.418 ms | 525 nm LED, 5V |
| Yang et al. 2018[147] | CsPbBr₃ nanosheets | The CsPbBr3 NSs were spin casted onto interdigitated electrodes and then annealed. | Flexible | Hot injection | 11 AW⁻¹ | 11 | 2650 ms/4000 ms | 365 nm LED, 1V |
| Xin et al., 2019[145] | CsPbBr₃ NSs | The CsPbBr₃ nanosheets solution were drop casted on the patterned Au/Ti/polyimide films substrates and then annealed | Flexible | Hot injection | 9.04 AW⁻¹ | - | - | 242 |
| Mandal et al 2022[33] | CsPbBr₁.₅I₁.₅ NSs | CsPbBr₁.₅I₁.₅ NS solution was spin coated on the top of the HTL followed by dipping in a saturated solution of $Pb(NO_3)_2$ in methyl acetate for the removal of the excess ligands. Then the substrates transferred to thermal evaporation for the deposition of the electrodes. | Rigid | Hot injection | 3946 A W⁻¹ | - | 116 ms/147 ms | AM 1.5 G solar illumination of power density 100 mW cm−2, |
| *Metal halide perovskite nanocrystals/2D materials heterostructures* | | | | | | | | |
| Wang at al 2015[72] | CH₃NH₃PbBr₂I/GO | The perovskite nanoislands were grown directly on the graphene film that was deposited on Si/SiO₂ substrates and the electrodes were fabricated. | Rigid | Direct crystallization | 6.0 × 10 5 A W ⁻¹ | | 120 ms/750ms | 405 nm laser, 1V |
| He et al. 2015[148] | CH₃NH₃PbI₃/rGO | CH₃NH₃PbI₃/rGO solution was drop-casted on Si/SiO₂ substrate, above which the Au/Cr electrodes were thermally deposited. | Rigid | In situ growth of nanocrystals on the 2D materials in LARP process | 73.9 mA W⁻¹ | 23.5 | 40.9 ms/28.8 ms | Light irradiation at 520 nm, 5V |
| Kwak et al 2016[42] | CsPbBr₃₋ₓIₓ/graphene | CsPbBr₃₋ₓIₓ nanocrystals were deposited on bilayer graphene mechanically exfoliated from the graphene flakes on the Si/SiO₂ substrate. | Rigid | Hot injection method | 8.2×10⁸ AW⁻¹ | | 0.81 sec /3.65 sec | 405 nm at bias 1 V |
| Song et al. 2017[149] | CsPbBr₃/ MoS₂ | Ligand-capped CsPbBr₃ NSs were drop-casted on MoS₂ monolayer film growth by CVD on SiO₂ substrate, which was followed by annealing the device at 60 °C. | Rigid | LARP | 4.4 AW⁻¹ | 10⁴ | 0.72 ms/1.01 ms | 442 nm laser, 10V |

| Wu et al. 2019[150] (treated nanocrystals) | CsPbI$_{3-x}$Br$_x$/MoS$_2$ | CsPbI$_{3-x}$Br$_x$ nanocrystals were spin-coated onto MoS$_2$ layer and then annealed at 60 °C. | Rigid | Hot injection method and treatment with 1-octane/ethyl acetate | 1.13 × 10$^5$ A W$^{-1}$ | | 0.42/0.19 s | 532 nm illumination |

## 4.3 Photocatalysis

### 4.3.1 Metal halide perovskite nanoplatelets/nanosheets for photocatalysis

**Photocatalysts for organic chemistry.** Oxidation of toluene which is an organic pollutant into useful chemical products like benzaldehyde and benzoic acid as essential materials for organic synthesis, antibacterial, biological and pharmaceutical applications is of great interest nowadays.[152] The ligand on the surface of the nanoparticulate photocatalysts could limit their performance in photocatalytic conversion applications.

Lead free bismuth-based metal halide platelets free of caping ligands used as active photocatalysts and found that can continuously and stably convert toluene to benzaldehyde with a high selectivity (≥ 88%) after 36 h light irradiation in air (Figure 27a).[36] In the presence of dilute $H_2SO_4$ solution and ethyl acetoacetate as the directing agent, $Cs_3Bi_2Br_9$ platelet microcrystals with different thicknesses ranging from 100 to 500 nm could be formed after rapid cooling in liquid nitrogen within one minute. The high activity and long-term stability of the $Cs_3Bi_2Br_9$ platelet photocatalysts are revealed by time-dependent photocatalytic reaction experiments, in which 232 µmoles of toluene in total has been converted after 36 h light irradiation in air. In addition, a high selectivity (≥ 88%) toward benzaldehyde was achieved and benzyl alcohol was produced as the main byproduct, while less than 3% of benzoic acid was detected. The conversion rate was increased with the irradiation time during the first 8 h and stayed relatively stable for another 28 h. This high stability with the time was observed for the platelet's morphology and not for the nanocrystals of the same chemical phase which have been deactivated after short reaction time (8h) due to the complete hydrolysis of the bismuth halide perovskite to the BiOBr phase.

In the same direction, 2D metal halide perovskite morphologies have been used as visible light photocatalysts to convert styrene into benzaldehyde.[153] $CsPbBr_3/Cs_4PbBr_6$ NSs synthesized by LARP method have been chemically modified by $ZrCl_4$ to simultaneously achieve the Cl doping and the surface modification with Zr species (Figure 27b). The production rate of the modified nanosheets found to be 1098 µmol $g^{-1} h^{-1}$, in comparison to the value of 372 µmol $g^{-1} h^{-1}$ of the untreated platelets. Both the Cl doping and the surface treatment with the Zr species found that was crucial for the improvement of the catalytic performance. According to the proposed mechanism, the excitons were activated by the visible light, and the electrons generated by the separation of electron-holes pairs were moved to the surface of the platelets and react with the $O_2$. The activated oxygen species then selectively oxidized the cation radicals produced by the oxidation of the styrene by the holes leading to the formation of the benzaldehyde. The photocatalytic production rate of benzaldehyde was accelerated in the case of the treated platelets as the generation and transfer of excitons was promoted.

**Photocatalytic $CO_2$ reduction.** The room temperature synthesized nanosheet-like morphology of the metal halide nanocrystals due to their large proportion of low-coordinated metal atoms together with the short carrier diffusion distance proved that facilitates and improves the photocatalysis of the $CO_2$ reduction without any organic sacrificial agent compared to the nanocrystals of the same chemical phase.[37] The activity of the photocatalytic $CO_2$ reduction to CO seemed to be thickness dependent (Figure 27c). The generation rate of CO was decreased gradually along with the thickness reducing of the NSs and the highest CO generation rate found to be for the NSs with 4 nm thickness (21.6 µmol $g^{-1} h^{-1}$). Furthermore, anion exchanged $CsPbBr_{3-x}I_x$ NSs showed enhanced photocatalytic $CO_2$ reduction compared to that of $CsPbBr_3$ NSs, which can be attributed to the enhanced visible light-harvesting capacity of the mixed halide perovskites. The highest CO generation rate found for the $CsPbBr_{2.4}I_{0.6}$ NSs (electron consumption rate of 87.8 µmol $g^{-1} h^{-1}$, which is over seven and two times higher than that of the $CsPbBr_3$ nanocrystals and $CsPbBr_3$ NSs. Also, the partially replacement of the $Br^-$ with $I^-$ only increased the photocatalytic activity without affecting their stability.

The first report on using lead-free 2D perovskite materials in photocatalytic $CO_2$ reduction was recently by Liu et al. (Figure 27d).[115] The photocatalytic performance found enhanced in the case of $Cs_2AgBiBr_6$ nanoplatelets compared with their nanocube counterpart. Nearly no $H_2$ can be detected throughout the entire reaction, suggesting that the selectivity for the $CO_2$ reduction was >99%. Furthermore, both the CO and $CH_4$ production rates were significantly higher when using the NPls as catalysts compared to the NCs.

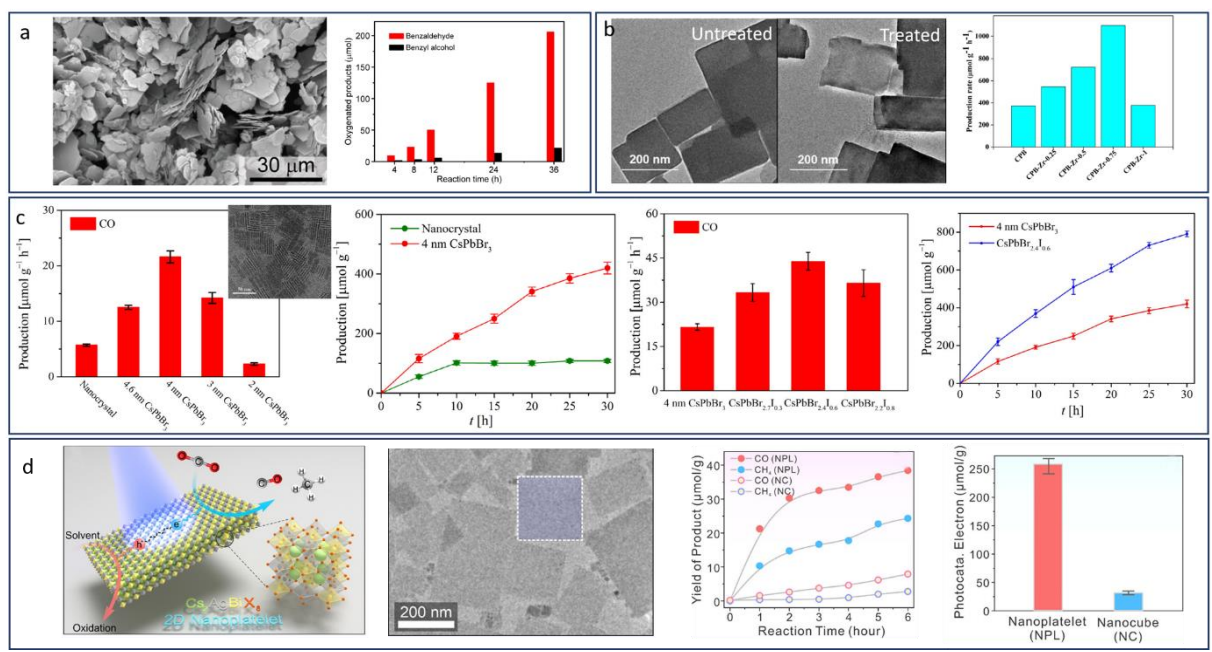

**Figure 27:** Metal halide perovskite nanoplateles/nanosheets as photocatalysts for (a) oxidation of toluene, (b) conversion of styrene into benzaldehyde and (c-d) CO₂ reduction. (a) Reprinted with permission from Ref. [36], copyright 2021, Wiley. b) Reprinted with permission from Ref. [153], copyright 2020, Frontiers. c) Reprinted with permission from Ref. [37], copyright 2021, Wiley, d) Reprinted with permission from Ref.[115], copyright 2021, American Chemical Society.

### 4.3.2 Metal halide perovskite nanocrystals/2D materials heterostructures for photocatalysis

Metal halide perovskite nanocrystals/2D materials heterostructures are unique photocatalysts because they offer increased catalytic sites and sufficient charge separation compared to the pure perovskite nanocrystals and the 2D materials that are characterized from strong radiative recombination and insufficient stability limiting their catalytic performance and application.

**Photocatalysts for CO₂ reduction.** The first report of using metal halide perovskite/2D materials heterostructures for CO₂ reduction was in 2017 from Xu et al. [43] CsPbBr₃ quantum dots was conjugated on graphene oxide through a room temperature LARP method (Figure 28a). The CO₂ reduction experiments were conducted with ethyl acetate as solvent. Ethyl acetate was selected among different solvents because its mild polarity can stabilize the CsPbBr₃ quantum dots and the CO₂ is highly soluble in this. Light was provided by a 100-W Xe lamp with an AM 1.5G filter to simulate solar light illumination. Figure 28a illustrates the amounts of the CH₄ and H₂ produced after 12h of the photocatalytic reaction. The selectivity for CO₂ reduction was greater than 99%. Neither phase transformation or degradation were occurred after 12 h of the photocatalytic reaction. The rate of electron consumption for the heterostructure found to be 29.8 μmol/g h compared to the 23.7 μmol/g h for the individual CsPbBr₃ quantum dots. Much improved yield was observed for the conjugation of CsPbBr₃ nanocrystals with the NHₓ-rich porous g-C₃N₄ nanosheets (Figure 28c).[49] The photocatalytic CO₂ reduction reaction in the later heterostructure was carried out in acetonitrile/water or ethyl acetate/water mixture. Acetonitrile and ethyl acetate were chosen because CsPbBr₃ nanocrystals was more stable in them than in water and their CO₂ solubility is high under mild reaction conditions. The existence of the NHₓ on the surface on the g-C₃N₄ found to be crucial for the formation of the N-Br bond leading to higher photocatalytic CO₂ reduction activity. The activity of heterostructured photocatalyst with 20% anchored nanocrystals for reduction of CO₂ to CO was calculated 15 and 3 times higher compared to the pure perovskite nanocrystals and 2D materials. The stability under consecutive runs has been also evaluated, and only 10.3 and 2.4% was incurred after three runs in acetonitrile/water and ethyl acetate/water, respectively. In addition, the yield of GO for the electrochemical CO₂ reduction was improved when organic-inorganic metal halide nanocrystals were conjugated on graphene oxide. In this case the yield of the products was 1.05 μmol cm⁻² h⁻¹ compared to that of the 0.268 μmol cm⁻² h⁻¹ of the net perovskite nanocrystals (Figure 28b).[64]

Various 2D materials have been used in the heterostructures for catalysis except of GO[43,64] and $C_3N_4$ [49] such as graphitic carbon nitride (containing titanium-oxide species),[46] $SnS_2$ nanosheets[67], $MoS_2$ [68] while the most of them are conjugated with $CsPbBr_3$ nanocrystals due to their higher stability at ambient conditions. The $CsPbBr_3/MoS_2$ heterostructure reached the high yield of products for CO of 25.0 μmol·g$^{-1}$ h$^{-1}$ (Figure 28d). [49] The only report found using lead free perovskite nanocrystals reported by Wang et al (Figure 28e).[67] The $Cs_2SnI_6/SnS_2$ heterostructures showed 5.4-fold and 10.6-fold enhancements of the $CO_2$ reduction and photoelectrochemical performance compared to the unadorned $SnS_2$.

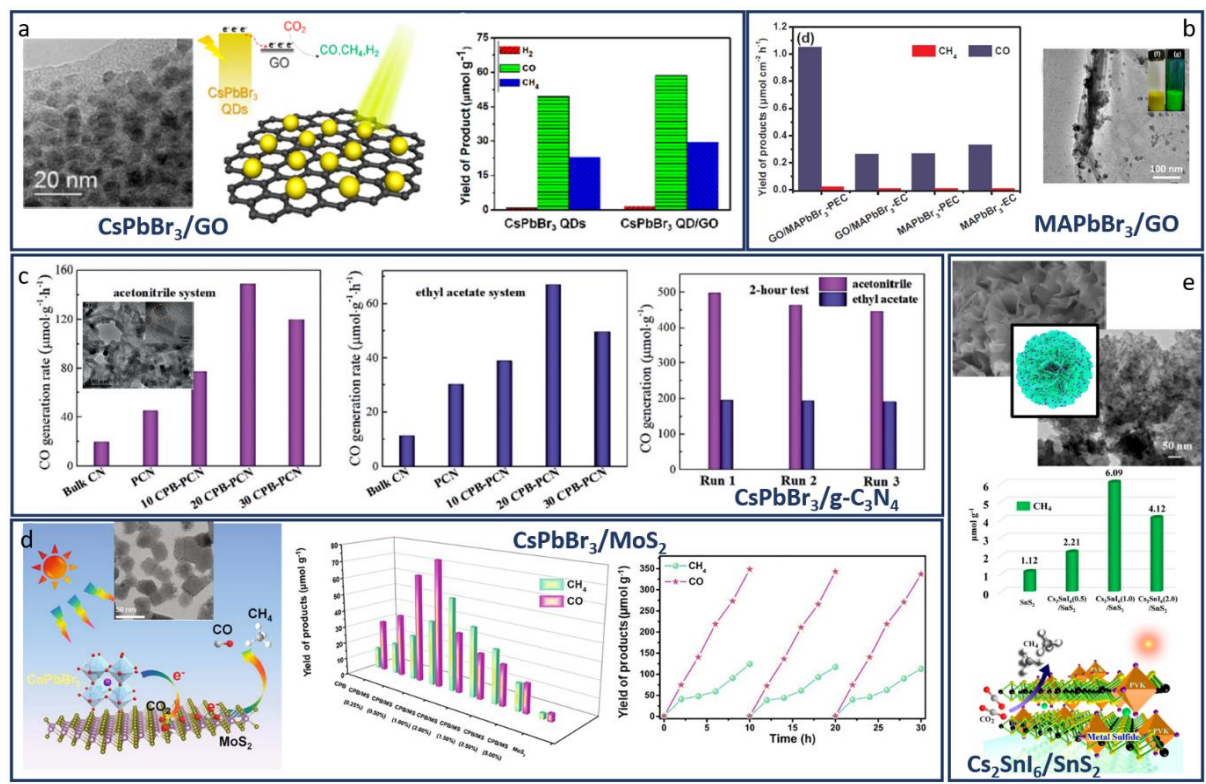

**Figure 28:** Photocatalysts for $CO_2$ reduction based on metal halide perovskite nanocrystals/2D materials heterostructures. Reprinted with permission from Ref. [43], copyright 2017, American Chemical Society (a), Reprinted with permission from Ref. [64], copyright 2018, Elsevier (b), Reprinted with permission from Ref.[49], copyright 2018, Wiley (c), Reprinted with permission from Ref.[68], copyright 2018, Elsevier (d), Reprinted with permission from Ref.[67], copyright 2019, American Chemical Society (e).

**Photocatalysts for hydrogen splitting.** Photoinduced splitting of hydrohalic acids (HX) to generate $H_2$ is a growing research field and effective photocatalysts of low cost and easily scalable fabrication process are demanded. In 2016 was the first report on a strategy for photocatalytic hydrogen iodide (HI) splitting using methylammonium lead iodide (MAPbI$_3$) in aqueous HI solution [69] while two years later the photocatalytic $H_2$ evolution of their heterostructures with rGO was evaluated.[62] These heterostructures showed 67 times faster $H_2$ evolution rate (93.9 μmol h$^{-1}$) than that of pristine MAPbI$_3$ nanocrystals under 120 mW cm$^{-2}$ visible-light (λ ≥ 420 nm) illumination. In addition, these heterostructured photocatalysts found that were highly stable showing no significant decrease in the catalytic activity after 200 h (i.e., 20 cycles) of repeated $H_2$ evolution experiments. The conjugation of the same perovskite materials with 2D few-layer black phosphorus yielded the superb Hydrogen evolution reaction rate of 3742 μmol h$^{-1}$ g$^{-1}$ [154] while with $MoS_2$ the evolution rate of 206.1 9 μmol h$^{-1}$. [59]

Despite the high photocatalytic $H_2$ evolution of the pure metal halide nanocrystals or their heterostructures with 2D materials, the lead toxicity issue could limit their use in photocatalysis. Wang et al. reported the heterostructure of lead-free metal halide nanocrystals with rGO for application in $H_2$ evolution in saturated HBr aqueous solution.[65] Using the $Cs_2AgBiBr_6/RGO$ heterostructure as photocatalyst, 489 μmol h$^{-1}$ $H_2$ can be produced within 10 h under visible light irradiation (λ≥420 nm, 300W Xe lamp). The heterostructure with 2.5 % rGO found to be the optimum concentration and also showed the high stability of the 120 h. In addition, hydrogen evolution rate of 380 μmol h⁻¹ found for $Cs_2AgBiBr_6$/Nitrogen-doped carbon heterostructures in aqueous HBr solution.[70]

# 4.4 Low-dimensional perovskite solar cells

### 4.4.1. Metal halide perovskite nanosheets/nanoplatelets

The introduction of low-dimensional perovskite nanosheets in the field of solar cells was demonstrated by Bai et al. in 2018. [35] The authors presented a recrystallization method for the formation of ultrathin high-quality lead-free $Cs_3Bi_2I_9$ perovskite NSs film. Figures 29 a-d depict top-view scanning electron microscopy (SEM) images of different $Cs_3Bi_2I_9$ films. The morphology of the so-formed NSs was varied according to the concentration of DMF solvent during the recrystallization process. Namely, the NSs edge could reach a length of around 2 μm, whereas their thickness was about 400 nm. [35] The perovskite solar cell (PSC) devices with ultrathin $Cs_3Bi_2I_9$ NSs exhibited a remarkable improvement in the power conversion efficiency (PCE) and stability, when compared to the corresponding devices with typical films by means of conventional spin-coating methods. Figure 29e shows the developed PSC configuration along with the SEM cross-section of the fabricated device. A planar FTO/c-TiO₂/$Cs_3Bi_2I_9$/HTL/Au architecture was employed, while three types of hole transport layer (HTL) materials were tested. Figure 29f presents the indicative current density versus voltage plots for the PSCs with CuI, Spiro-OMeTAD, and PTAA polymers as HTL. The determined values of PCE for the three devices were 3.2 %, 1.77 %, and 2.30 % (Table 6), respectively [35]. Notably, all values are improved when compared to the highest previously reported PCE of $A_3Bi_2I_9$-based devices without nanosheets (1.64 %). More importantly, the nanosheet-based PSCs exhibit good PCE stability when exposed to ambient air of 45 % relative humidity (RH) without encapsulation (Fig. 29g). Indeed, the CuI device retained 57 % of the initial PCE after 38 days of exposure.

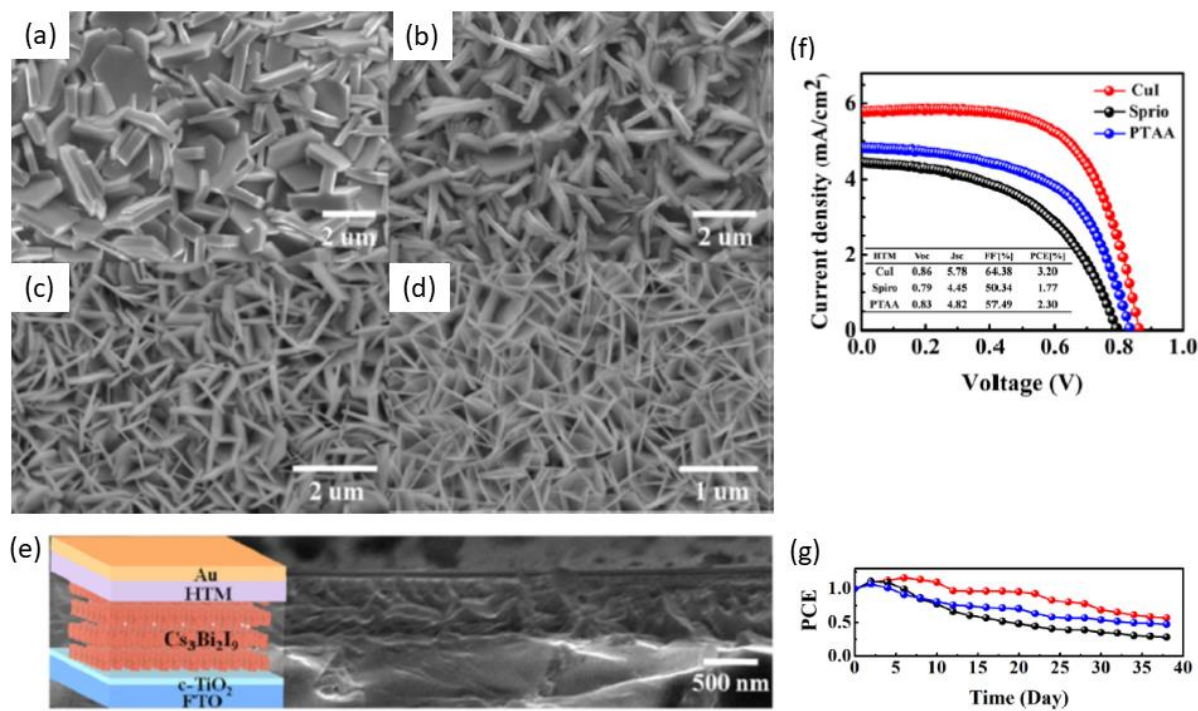

**Figure 29:** (a) Top-view scanning electron microscopy (SEM) image of $Cs_3Bi_2I_9$ film obtained by means of conventional spin-coating process. Recrystallization films with different DMF concentrations 50 μL (b), 100 μL (c), and 200 μL (d). (e) Schematic of the developed $Cs_3Bi_2I_9$ perovskite solar cells (PSCs) and SEM cross-section. (f) Current density versus voltage (J-V) curves of the PSCs with CuI, spiro-OMeTAD, and PTAA as hole transport layer (HTL). (g) Normalized power conversion efficiency (PCE) of unencapsulated devices after subjected to ambient air (45% RH). (a-g) have been reproduced from ref [35] with permission from Elsevier, copyright 2018.

One year later, Liu et al. reported on the layer-by-layer self-assembly of two-dimensional (2D) perovskite NS building blocks into ordered Ruddlesden-Popper perovskite phases [155]. Figure 30a shows a schematic representation of the process where $C_8H_{17}NH_3$-capped $CsPb_2Br_7$ NSs self-assemble into layered ($C_8H_{17}NH_3)_2CsPb_2Br_7$ NSs superlattice nanocrystals. The disassembly of the blocks is achieved upon controllable sonication in toluene. Figures 30b-e presents the corresponding TEM images of the $C_8H_{17}NH_3$-capped $CsPb_2Br_7$ NSs in their initial state, and after several minutes where the superlattice intermediates (after 20 and 70 min), and the final nanocrystals have been formed (after 120 min) [155]. Following the self-assembling process, uniform superlattice nanocrystals with an average lateral size of around 500 nm were formed. EDX analysis revealed an atomic ratio of Cs:Pb:Br of approximately 1:2:7, which matches well with the two-dimensional (2D) layered Ruddlesden-Popper phase crystal stoichiometry. The developed method paves the way towards increasing order in a system of nanoparticles for the synthesis of multi-layered low-dimensional perovskite configurations for next-generation photovoltaic applications.

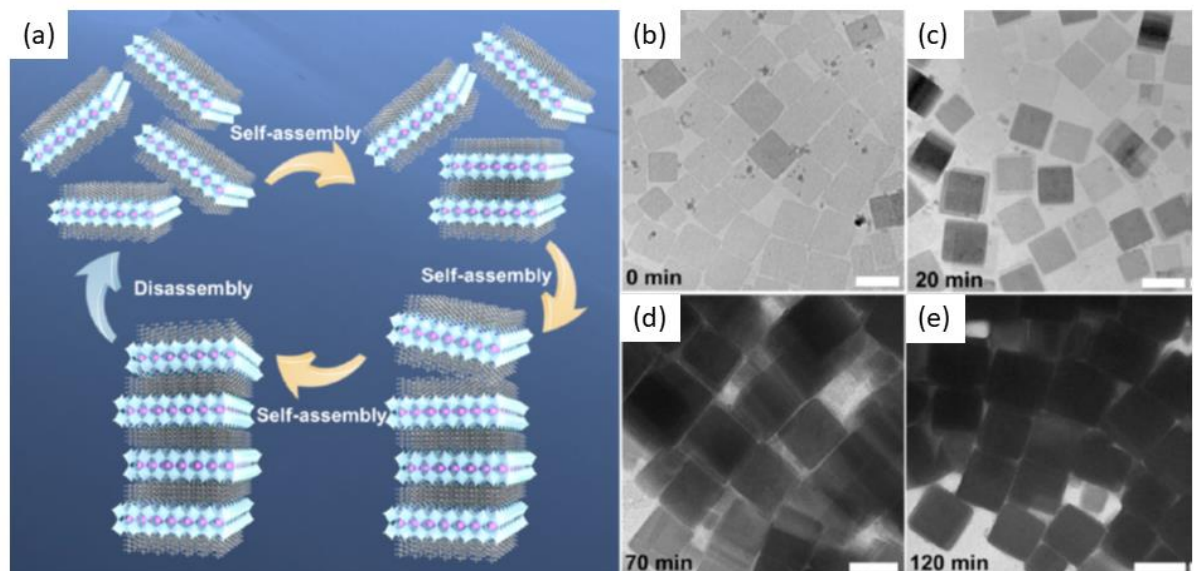

**Figure 30:** (a) Schematic illustration of the layer-by-layer self-assembly $C_8H_{17}NH_3$-capped $CsPb_2Br_7$ NSs into layered ($C_8H_{17}NH_3)_2CsPb_2Br_7$ superlattice nanocrystals. (b) Transmission electron microscopy (TEM) image of the initial $C_8H_{17}NH_3$-capped $CsPb_2Br_7$ NSs. Superlattice nanocrystal intermediates after 20 min (c), and 70 min (d). (e) Final ($C_8H_{17}NH_3)_2CsPb_2Br_7$ superlattice nanocrystals after 120 min. The scale bar is 500 nm. (a-e) have been reproduced from ref. [155] with permission from American Chemical Society, copyright 2019.

Notably, the incorporation of other types of low-dimensional layered perovskites like Ruddlesden-Popper in solar cell architectures was pre-existing and motivated by their ability to form high crystalline films that appeared to exhibit better stability when compared to typical three-dimensional (3D) crystals, while an order of magnitude enhancement of the PCE was achieved [156–162]. Smith et al. back in 2014 demonstrated the first perovskite solar cell (PSC) device in which 2D perovskites were used as light absorber elements [163]. In particular, $(PEA)_2(MA)_2[Pb_3I_{10}]$ layered perovskites were derived from the 3D crystals by means of slicing along specific crystallographic planes, whereas the interlayer spacing is controlled by the selection of organic cations. The first-generation devices of this type exhibited an open-circuit voltage ($V_{oc}$) of 1.18 V and a power conversion efficiency (PCE) of 4.73% (Table 6) [163].

The importance of employing low-dimensionality organic-inorganic perovskites in solar cells was thoroughly considered by Quan et al. [164]. Namely, upon modifying the stoichiometry multiple PSCs with low-dimensional intermediates between 3D and 2D layered $(PEA)_2(MA)_{n-1}Pb_nI_{3n+1}$ perovskites were fabricated. Figure 31a shows the corresponding unit cell structures of $(PEA)_2(MA)_{n-1}Pb_nI_{3n+1}$ perovskites with different n values, while the energetics of perovskite formation and stability are also presented [164]. A clear picture emerges from Figure 31a. Moving from the strict 2D structure with n=1, to the quasi-2D with n>1, and towards the cubic 3D perovskite with n=∞, the perovskite stability deteriorates. On the other hand, the PCE increases drastically with n as depicted in Figure 31b, with the n=60 device exhibiting a PCE of around 17 % (Table 6). Moreover, the same configuration with n=60 was the first certified hysteresis-free planar PSC obtaining a certified 15.3 % PCE. In terms of stability tests, the performance of the n=60 device dropped to 11.3 % after 60 days in low humidity atmosphere, whereas under humid air (55% RH) the PCE decreased to 13 % after two weeks. Notably, the corresponding 3D perovskite device exhibited an initial PCE of 16.6 %, but degraded to less than 3 % after the same period under ambient air. Overall, a tradeoff between device stability and performance of 2D-based PSCs is generated (Figure 31c) [164]. Based on this, a great scientific challenge emerged, namely to discover methods and advanced device architectures in order to enhance more the PCE of the low-dimensional PSCs, while taking advantage of their remarkable stability when compared to conventional 3D perovskite-based devices.

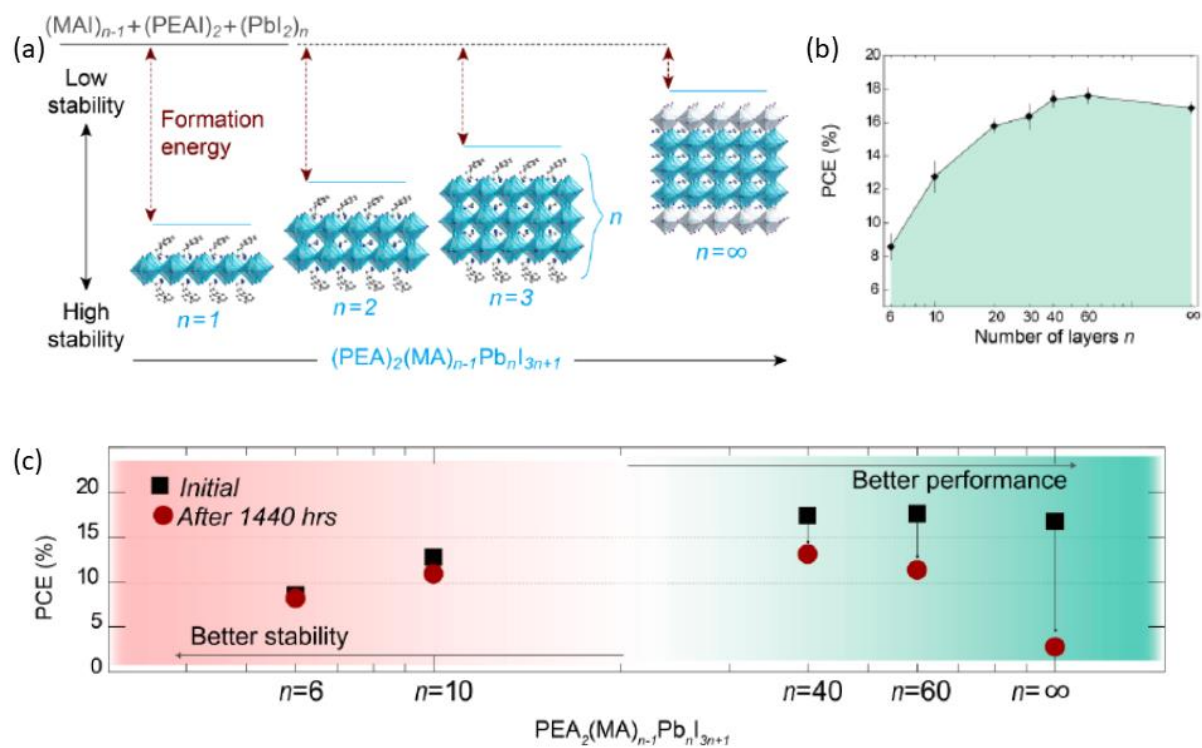

**Figure 31:** (a) Unit cell structure of $(C_8H_9NH_3)_2(CH_3NH_3)_{n-1}Pb_nI_{3n+1}$ perovskites with different n values. (b) Power conversion efficiency (PCE) of the perovskite solar cells (PSCs) versus number of layers. (c) PSC device performance as a function of n value. (a-c) have been reproduced from ref. [164] with permission from American Chemical Society, copyright 2016.

One year later, Zhang et al. reported on the development of stable and efficient PSCs upon introducing cesium cation (Cs$^+$) doping of the main 2D $(BA)_2(MA)_3Pb_4I_{13}$ perovskite [165]. The Cs$^+$ doped PSCs exhibited a PCE of 13.7% (Table 6), while showing excellent humidity resistance. The authors found that upon partially replacing MA$^+$ cation with Cs$^+$ the grain size and surface quality of the perovskite films were improved, and an almost ideal crystalline orientation was retained. As a result, the trap-state density is reduced, and the charge-carrier mobility is enhanced. Moreover, the fabricated devices maintain 89 % of their initial PCE after 1400 operating hours under ambient conditions [165]. Impressive tolerance to high humidity conditions was also reported with the developed PSCs maintaining more than 80 % of the initial PCE after several hours under 65% and 85% RH. More than 85% of the PCE was maintained also during a thermal stability test with the Cs-doped 2D perovskite device operating at 80 °C for several hours. On a rather difference approach, in the same year, Zhang et al. emphasized the importance of orientation regulation of the PEA cation within 2D PSCs. [166] For the formation of the 2D perovskite films, the authors employed one-step spin-coating method while using ammonium thiocyanate (NH$_4$SCN) additive. The addition of NH$_4$SCN intensified the crystallinity of the so-formed films, while resulted to the preferable vertical orientation growth with improved electron and hole mobility. The optimized 2D PSC of this study exhibited a PCE of 11.01% (Table 6), with an unsealed device maintaining 78.5 % of the initial PCE after 160 h of storage in air atmosphere (55% RH) [166].

Rather differently, Yang et al. managed to improve significantly the PCE up to 18.2% (Table 6) [167], upon exploiting a combination of different quasi-2D crystalline configurations. The high performance was attributed to the formation of a self-assembled multilayered microstructured light absorber film, consisted of vertically oriented low-layers (small-n) covered by a layer of large-n perovskite components, as depicted schematically in Figure 32a [167]. Figure 32b shows a top view SEM image of the quasi-2D film. The synthesis of the quasi-2D multilayer perovskite structure relies on the use of 3-bromobenzylammonium iodide precursor. The reported unique configuration is energetically ordered so that improved charge transport and reduced nonradiative recombination occur within the devices. Good stability is also reported as 82 % of the PCE is retained following 2400 h of storage at a RH of 40%, while only negligible hysteresis effects are noticed in the current-voltage plot (Figure 32c) [167].

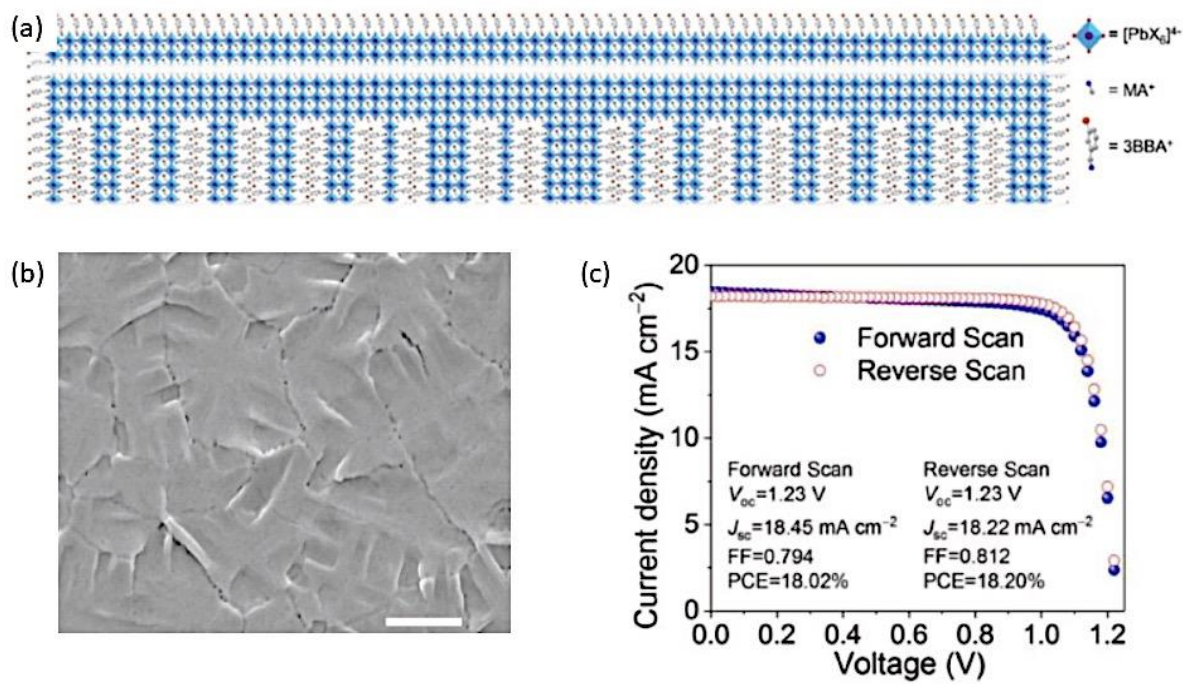

**Figure 32: (a)** Schematic illustration of the self-assembled quasi-2D perovskite structure. **(b)** Top view scanning electron microscopy (SEM) image of the quasi-2D self-assembled perovskite film. **(c)** Current density versus voltage (J-V) curves of the developed PSC showing both forward and reverse scans. (a-c have been reproduced from ref. [167] with permission from Wiley Online Library, copyright 2018).

Along similar lines, Cho et al. demonstrated the benefits of using low-dimensional perovskites with an enhanced water-resistant character as a protective layer on top of typical 3D light absorber films in PSC devices [168]. The success of the proposed 2D/3D perovskite composites relies on a saturated highly fluorinated organic cation which is incorporated in the perovskites either by direct blending with the precursors, or by controlled in-situ layer-by-layer approach as depicted schematically in Figure 33a. In both cases a thin layer of the fluorous low-dimensional perovskite resembles on the upper surface of the bulk 3D perovskite films (MFPI, $MA_{0.9}FA_{0.1}PbI_3$ and CFMPIB, $Cs_{0.1}FA_{0.74}MA_{0.13}PbI_{2.48}Br_{0.39}$). The crystalline structures of the fluorinated organic cation and the so-formed low-dimensional perovskites are also shown in Figure 33a. The water-repelling characteristics upon the introduction of the low dimensional component was probed by contact angle (c.a.) measurements presented in Figure 33b. The 3D MFPI film exhibits a c.a. of 55.3°, whereas the mixed and the layer-by-layer 2D/3D architectures show c.a. values of 83.8° and 98.4°. Moreover, the formation of the 2D layer enhances the PCE of the MFPI device from 17.98% to 20.13%, whereas for the CMFPIB device from 18.78% to 20.0%. [168] Impressively enough, the 2D/3D devices showed outstanding stability. Figure 33c presents the normalized PCE against exposure time under inert gas. Remarkably, the mixed 6% A43 device maintains around 82% of the initial PCE after 450 h, whereas the layer-by-layer device retains almost 96% of the original PCE (Table 6). Overall, the study of Cho et al., demonstrated an alternative role of the low-dimensional perovskites, namely, to act as water protective shield, apart from the ordinary role of being the main light absorber component of the PSC. [168]

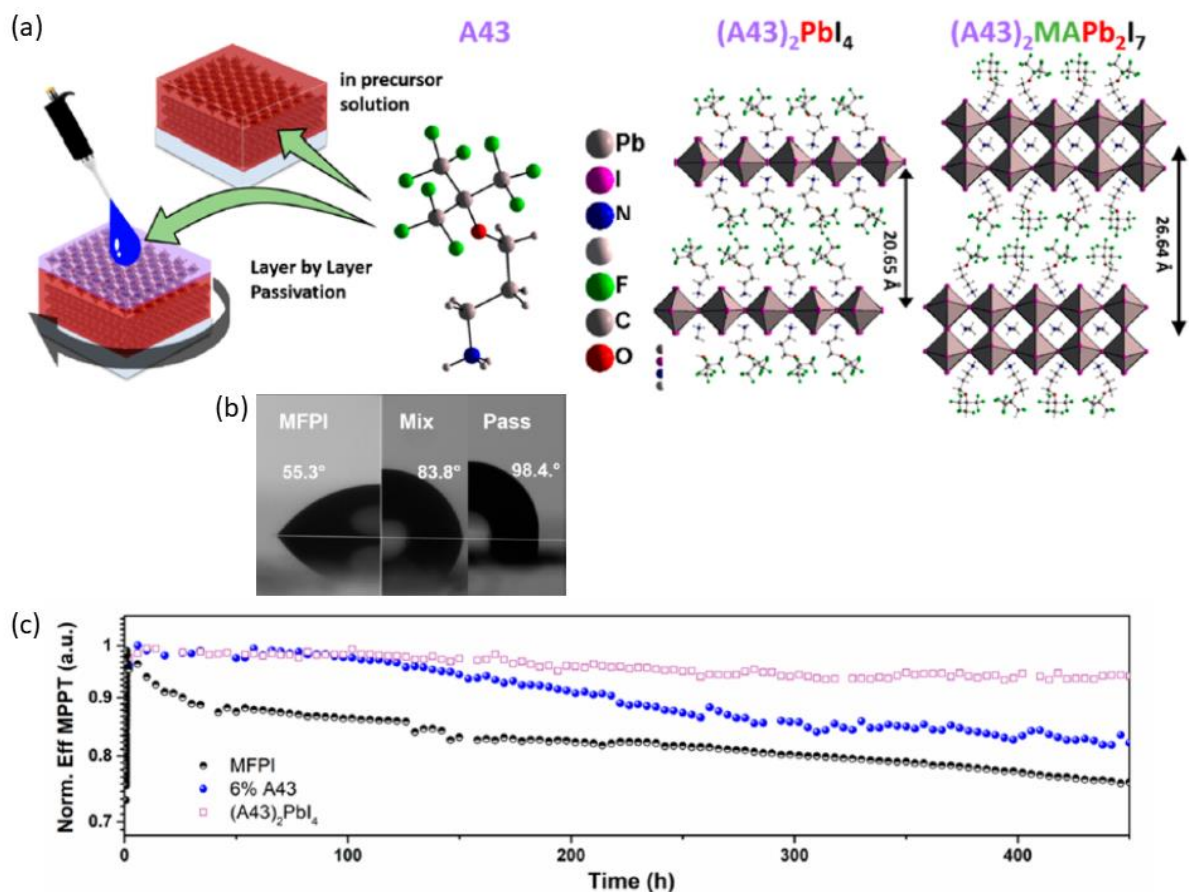

**Figure 33:** (a) Schematic diagram showing the incorporation method of low-dimensional fluorous perovskite on top of 3D perovskites. The crystal structures of the fluorous organic cation and the low-dimensional perovskites are also shown. (b) Water contact angle measurements on top of pristine MFPI perovskite and the developed 2D/3D configurations. (c) Normalized power conversion efficiency (PCE) over time upon exposure under inert gas. (a-c) have been reproduced from ref. [168] with permission from American Chemical Society, copyright 2018.

On a similar manner, Abuhelaica et al. developed a 2D/3D configuration in which the 2D perovskite is formed upon mixing two alkyl-based cations. [169] The so-formed mixed 2D perovskite was placed between the 3D perovskite film and the hole transport layer polymer of the PSC, and found to improve the carrier dynamics and the photovoltaic efficiency of the devices. Jang et al. reported on another outstanding 2D/3D-based PSC design by means of a solvent-free solid-phase in-plane growth (SIG) fabrication route. [170] SIG allowed the successful formation of a stable and highly crystalline 2D ($C_4H_9NH_3$)$_2$PbI$_4$ film on top of a 3D film. The cross-sectional scanning electron microscopy (SEM) of the SIG-processed 2D/3D architecture is shown in Figure 34a, while the device performance and stability tests are presented in Figures 34b and 34c, respectively [170]. The intact 2D/3D perovskite heterojunction enhances the PCE of the device to a certified value of 24.35% (Table 6). Moreover, the unencapsulated device maintained around 95% of the initial PCE after subjected to 85% RH for 1100 h, whereas the encapsulated cell retained 94% of the initial PCE after 1056 h under the dump heat test (85 ℃, 85% RH). [170] Huang et al. presented another multilayer design towards the improvement and stability of 2D perovskite films [171]. Namely, the authors fabricated PSC devices in which guanidinium bromide (GABr) was placed on top of the 2D GA$_2$MA$_4$Pb$_5$I$_{16}$ perovskite light absorber. The proposed multifunctional interfacial engineering led to devices with PCE of 19.3%, while the GABr addition optimized stability since 94% of the initial PCE was retained following 3000 h exposure to ambient conditions.

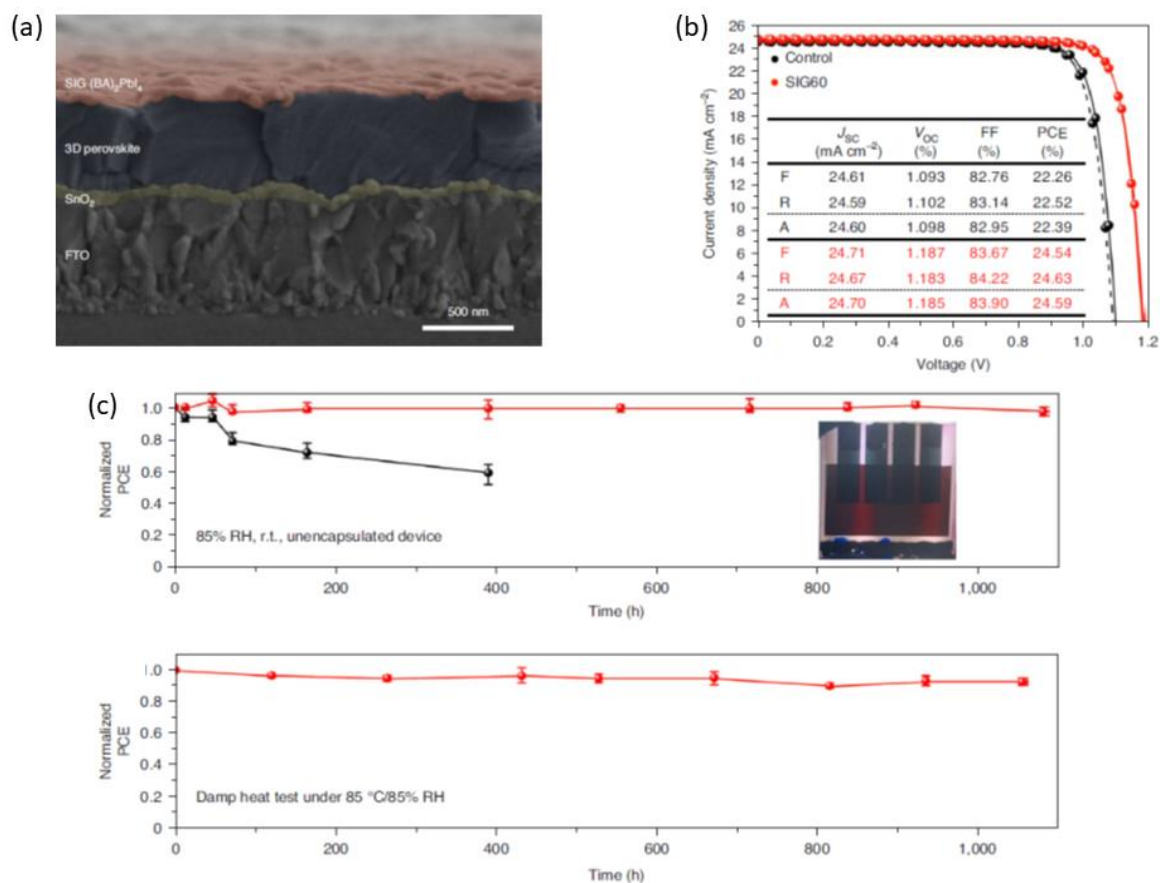

**Figure 34:** (a) Cross-sectional scanning electron microscopy (SEM) image of the SIG-processed 2D/3D device. (b) Photovoltaic performance of the devices as expressed by the current density versus voltage (J-V) curves. (c) Normalized power conversion efficiency (PCE) over time of the unencapsulated device subjected to 85% relative humidity (RH), and damp heat test at 85 °C under 85% RH of the encapsulated device. (a-c) have been reproduced from ref. [169] with permission from Nature, copyright 2021.

In the meantime, apart from the interlayer interactions, other authors explored the interaction effects on an atomic level towards improving performance and stability. As for instance, Ren et al. exploited sulfur-sulfur interactions within a newly employed alkylammonium [172], i.e. 2-(methylthio)ethylamine hydrochloride (MTEACL). It was shown that the interactions between sulfur atoms in the two MTEA molecules of the $(MTEA)_2(MA)_4Pb_5I_{16}$ perovskite, results to weaker van der Waals interactions and enhanced charged transport within the crystalline framework. Consequently, PSCs with PCE of 18.06% were developed, while having sufficient moisture tolerance up to 1512 h under 70% RH (Table 6) [172]. Similarly, Liang et al. focused on the effect of the ionic coordination of the employed n-butylamine salt within low-dimensional perovskites (layers n≤5) [173]. The authors discovered that upon using n-butylamine acetate instead of the common n-butylamine iodide, a gel of a uniformly distributed intermediate phase is formed. Eventually this leads to the formation of high crystalline quality vertically aligned grains. By means of this approach the authors reported PCEs of 16.25%, while stability tests showed less than 10% deterioration after storing the devices for 4680 h under humid air (65% RH). The same holds when the devices were subjected to 85 °C for 558 h, and under realistic operational conditions with continuous light illumination for 1100 h [173].

Recently, Li et al. developed an all-2D perovskite heterojunction consisting of a 2D Ruddlesden-Popper and a 2D Dion-Jacobson perovskite [174]. The combination of these 2D perovskites was found to enhance the PCE of the device up to 18.34% due to enhanced interfacial charged extraction and suppressed surface charge recombination. The fabricated PSCs showed nearly zero degradation of the PCE after 800 h of continuous thermal aging at 60 °C. Other authors achieved similar PCE (18.3%) upon studying the role of light-activated interlayer contraction in 2D PSCs (Table 6). [175] In particular, X-ray photoelectron spectroscopy revealed the accumulation of positive charges in the terminal iodine atoms, leading to I-I interactions across the organic barrier that activated out-of-plane contractions. Following this, an increased in charge carrier mobility is noticed, leading to the enhanced photovoltaic performance of the 2D perovskite devices. Liang et al. fabricated a high-quality FA-MA mixed 2D perovskite multilayer film, in which lower n-value phases are concentrated at the upper part of the film, while 3D-like phases are located in the film interior. An outstanding PCE value of 20.12% is achieved (Table 6) [176]. Finally, Shao et al. developed the 2D Ruddlesden-Popper perovskite-based solar cell that holds the PCE record of 21.07% (Table 6). [177] Remarkably, such PCE is close to that of the current state-of-the-art 3D-based devices that exhibit 25%, i.e. approaching the Shockley-Queisser limit. In summary, the considered studies highlight the importance of employing 2D perovskite NCs and related multilayer heterostructures in PSCs, towards achieving PCEs that bring the technology closer to the market, while securing operational stability of the devices.

### 4.4.2. Metal halide perovskite nanocrystals/2D materials

Due to the strong coupling interactions between mixed halide perovskites and 2D materials, another promising method towards improving the performance and stability of next-generation PSCs relies on the introduction of advanced perovskite/2D materials heterostructures.[41] The 2D material is either incorporated within the perovskite light absorbing film, or as a separate interface layer within the device architecture. We will now consider some typical examples. Hadadian et al. back in 2016 introduced nitrogen-doped reduced graphene oxide (N-RGO) into an organic-inorganic $FA_{0.85}MA_{0.15}Pb(I_{0.85}Br_{0.15})_3$ perovskite in order to enhance the PCE of the devices.[178] The 2D material additive improved the crystalline quality of the perovskite film, larger grains were formed, and the carrier dynamics of the device were improved. The crystalline quality of the light absorber film is a known factor towards advancing the charge carrier dynamics and the performance of PSCs. [179,180] As result, the PCE efficiency of the N-RGO containing device was increased from 17.3% to 18.7%, while the hysteresis effects were diminished. [178] Along similar lines, enhancement of the PCE was achieved by Jiang et al. [181], and Guo et al. [182], upon incorporating g-$C_3N_4$ NSs and $Ti_3C_2T_x$ NSs in the perovskite active layer, respectively. On a rather different manner, Wu et al. employed graphene oxide (GO) as hole transport layer (HTL) material in a PSC [183], whereas the same method was reported for RGO in other devices by Yeo et al. [184]. In both cases enhanced PCEs were observed when compared to the reference devices, i.e. without the 2D additives, pointing out the beneficial role of the perovskite/2D materials interlayer heterojunctions in PSCs.

Nevertheless, the employment of PNCs (NSs and NPIs)/2D materials heterostructures in the field of solar cells, appears to be limited until today. Interestingly enough though, the design of novel quantum dots (QDs) based heterostructures, appears to be promising for the development of efficient and stable devices when compared to the typical thin-film ones. Based on this approach, Zhao et al. reported on the fabrication of PSCs with enhanced photocarrier harvesting and performance [185]. Namely, the authors demonstrated the formation of inorganic and inorganic-organic perovskite QDs light absorption interlayers by means of layer-by-layer deposition. This resulted to improved charge carrier lifetime and better carrier mobility within the perovskite light absorption region of the device, leading to better charge extraction and enhanced photovoltaic performance when compared to the reference devices without the QDs-based interlayers. Moreover, the authors claim that this heterostructure approach can also be expanded to other 2D perovskite structures, thus maybe promising in the case of perovskite NSs and NPIs among other cases. [185,186]

**Table 6:** Summary of low-dimensional perovskite configurations in advanced perovskite solar cells (PSCs). *NR stands for not reported.

| Study-year | Synthesis protocol | PSC device configuration | PCE | Demonstrated PCE stability | Key features / Remarks |
|---|---|---|---|---|---|
| Bai et al., 2018[35] | Dissolution-recrystallization | FTO/c-TiO$_2$/Cs$_3$Bi$_2$I$_9$/HTL/Au | 3.2% | 57%, 38 days, 45% RH | Nanosheet films |
| Smith et al., 2014[163] | Crystallographic slicing | FTO/TiO$_2$/(PEA)$_2$(CH$_3$NH$_3$)$_2$[Pb$_3$I$_{10}$]/Spiro-OMeTAD/Gold | 4.73% | No | First-generation 2D PSC |
| Quan et al., 2016[164] | Spin-coating | FTO/TiO$_2$/Perovskite/SpiroOMeTAD/Au | 17.0% | From 17% to 11.3%, 1440 h, ambient | Low-dimensionality vs. PCE, certified hysteresis-free |
| Zhang et al., 2017[165] | NR | FTO/TiO2/Cs$_x$-2D Perovskites/Spiro-OMeTAD/Au | 13.7% | 89%, 1400 h, ambient >80%, 24 h, 65% and 85% RH >85%, thermal at 80 °C | Cation doping, negligible hysteresis |
| Zhang et al., 2018[166] | One step spin-coating | ITO/(PEA)$_2$(MA)$_{n-1}$Pb$_n$I$_{3n+1}$/butyric acid methyl ester/bahocuproine/Ag | 11.01% | 78.5%, 160 h, 55% RH | PEA cation crystal orientation |
| Yang et al., 2018[167] | Spin-coating, 3BBAI | ITO/PTAA/Perovskite/PCBM/Cr/Au | 18.2% | 82%, 2400 h, 40% RH | Multilayered quasi-2D perovskite film |
| Cho et al., 2018[168] | Spin-coating, layer-by-layer deposition | FTO/TiO$_2$/(2D/3D) perovskites/Spiro-OMeTAD/Au | 20,1% | 96%, 450 h, inert gas 82%, 450 h, inert gas | Water repelling 2D/3D architecture |
| Abuhelaiqa et al., 2022[169] | Spin-coating | FTO/TiO$_2$/3D perovskite/mixed-cation 2D perovskite/SpiroOMeTAD | 21.17% | 95%, 1000 h, inert gas | Mixed-cation 2D/3D configuration |
| Jang et al., 2021[170] | Solid-phase in-plane growth (SIG) | FTO/SnO$_2$/3D perovskite/SIG (BA)$_2$PbI$_4$/Spiro-OMeTAD/Au | 24.35% | 95%, 1100 h, 85% RH 94%, 1056 h, 85°C, 85% RH | 2D/3D intact heterojunction |
| Huang et al., 2021[171] | Spin-coating | Glass/ITO/SnO$_2$/GA$_2$MA$_4$Pb$_5$I$_{16}$/GABr/Spiro-OMeTAD/Ag | 19.3% | 94%, 3000 h, ambient | GABr interlayer |
| Ren et al., 2020[172] | Spin-coating | ITO/PEDOT:PSS/(MTEA)$_2$(MA)$_4$Pb$_5$I$_{16}$/PC$_{61}$BM/Cr/Au | 18.06% | 87.1%, 1000 h, operational with 0.9V, nitrogen | Sulfur-sulfur interactions in MTEA |
| Liang et al., 2021[173] | Spin-coating | ITO/SnO$_2$/pure phase perovskite/Spiro-OMeTAD/Au/MoO$_3$ | 16.25% | >90%, 4680 h, 65% RH >90%, 558 h, 85 °C >90%, 1100 h, operational | Pure phase, vertically aligned grains |
| Li et al., 2022[174] | Spin-coating | ITO/PTAA/RP-2D perovskite/BDAI$_2$/C$_{60}$/BCP/Ag | 18.34% | 98%, 800 h, 60 °C, inert gas 87%, 800 h, 60% RH | RP 2D/DJ 2D heterojunction |

| Li et al., 2022[175] | Spin-coating | ITO/PEDOT:PSS/2D perovskite/ PCBM/Au | 18.3% | NR | I-I interactions, out-of-plane contractions |
|---|---|---|---|---|---|
| Liang et al., 2022[176] | Alcoholic antisolvent deposition method | ITO/SnO$_2$/2D-3D perovskite/ Spiro-OMeTAD/Ag | 20.0% | 98%, 2000 h, 30% RH 96%, 360 h, operational | Multilayered quasi-2D perovskite film |
| Shao et al., 2022[177] | One-step hot-casting | ITO/PTAA/2D perovskite/ PC$_{61}$BM/Ag | 21.07% | 97%, 1500 h, 85 °C, inert gas | FA/MA mixed 2D, PCE record |

## 4.5 Sensing

### 4.5.1. Metal halide perovskite nanoplatelets/nanosheets as humidity and gas sensing elements

Ren et al turned the disadvantage of the metal halide perovskite nanocrystals to be moisture sensitive to advantage and constructed a humidity detector based on high-quality $CH_3NH_3PbI_{3-x}Cl_x$ nanosheet arrays (Figure 35a). [38] The vertically oriented NSs showed a terrace-like smooth surface, indicating that the nanosheet may have grown epitaxially from a set of ultra-thin nanosheets (Figure 35 g-h). The impressive sensitivity, repeatability and specificity were due the specific sensor device. Two advantageous considerations are involved in this design: i) the perovskite NSs arrays showed greater morphological and orientational uniformity than did the previously reported perovskite nanostructures and ii) the chlorine-based sensing element was decomposed less upon being exposed to moisture. In particular, the sensor device was very simple with two stripes of Ag electrodes deposited on its top surface (Figure 35 a). This sensor was placed in a chamber whose humidity was monitored by using a psychrometer in order to evaluate the humidity sensing capability (Figure 35b). The I–V curves in static air of 30–90 % RH at 27 °C showed a good linearity, indicating an ohmic contact between the sensing element and Ag. Interestingly, as the RH was increased, the current increased dramatically (Figure 35c). The resistance was calculated to sharply dropped from $1.28×10^8$ Ω to $7.39×10^4$ Ω, as the RH was increased from 30 % to 90 % and the sensitivity, defined from the ratio $R_{30\%RH}/R$, was found to be increased dramatically, from 1 to 1422, as the RH was increased from 30 % to 90 % (Figure 35d). Similar results also observed for humidity below 30 %. These results indicated a strong conductivity of the perovskite nanostructure arrays to the water molecules in the air and the device can also work in low-humidity conditions. The good reversibility of the sensor was revealed from the resistance plots by increasing and decreasing the humidity (Figure 35e) while the strictly modulation of the current by the changing the RH value indicating the high sensor's efficiency (Figure 35f).

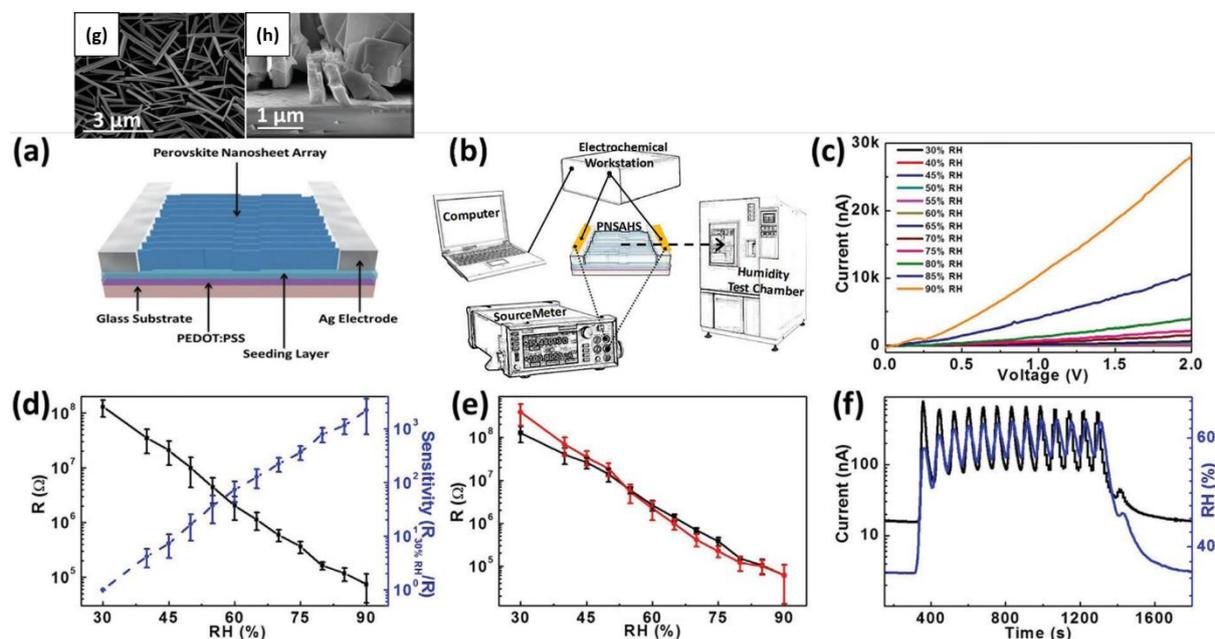

**Figure 35:** a) Perovskite nanosheet array sensor. (b) System for testing the sensor device. (c) I–V curves of the sensor in atmospheres with various relative humidity values (RHs) from low humidity (30% RH) to 90% RH at 27 1C. (d) The average resistance of the sensor (black line) and the resistance response sensitivity, which is defined by R30%RH/R (blue line), in the various RH conditions. A resistance hysteresis loop of the sensor. (f) Real-time current response properties of the sensor in different RH (35–65%) gases at 27 °C. Reprinted with permission from Ref. [38], copyright 2017, The Royal Society of Chemistry.

Metal halide NSs have been proposed also as sensing element for the detection of $O_2$. The $CsPbBr_3$ NSs showed PL enhancement in the presence of $O_2$.[39] The fast and reproducible PL response indicating that the NSs could be an effective sensing element for these types of applications. Comparing the sensitivity of the NSs structures with single crystal materials of the same chemical phase revealed that the responsiveness is much higher in the case of the NSs as the single crystal materials is characterized by less surface defects.

### 4.5.2 Metal halide perovskite nanocrystals/2D materials heterostructures as gas sensing elements

The 2D materials loading with metal halide perovskite nanocrystals have been shown to be promising materials for sensing application due to synergetic effects and enhanced sensing capability. The presence of the perovskite nanocrystals found that enhanced the reactivity of the 2D materials that usually performed a poor sensing performance. For example graphene has been used as sensing element for the detection of the atmospheric pollutants such as $NO_2$ and $NH_3$,[187] Volatile Organic Compounds (VOCs),[188] $H_2S$,[189] $SO_2$,[190] and CO.[191] The response of the metal halide perovskite nanocrystals/2D materials heterostructures has been attributed to the effect of the perovskite nanocrystals which play the role of the chemical receptor. The 2D materials play mainly the role of the efficient charge conductor, ensuring that the charges generated upon the interaction between gas molecules and perovskite nanocrystals reach the device electrodes. [71] Furthermore, the use of heterostructures in gas sensors can become a promising alternative to other gas sensitive materials, due to the protective character of graphene resulting from its high hydrophobicity.[192]

The metal halide perovskite nanocrystals/2D materials heterostructures have been used as sensing element to detect toxic gases such as ammonia ($NH_3$) and nitrogen dioxide ($NO_2$) without requiring high working temperature or UV irradiation to activate the sensing process (Figure 36a).[63] A film of graphene NPls was deposited onto quartz substrates by drop casting method and then the $MAPbBr_3$ nanocrystals were spin coated onto the film of the graphene. Graphene doped with perovskite nanocrystals presented a higher response (up to 3-fold) than bare graphene, even under a hundred ppb of $NO_2$ exposure. A slight decrease in the response towards 500 ppb of $NO_2$ was observed after 6 months of sensor operation indicating the stability of the perovskite nanocrystals which was originated from their protection by the hydrophobic graphene. Furthermore, the same sensor has been tested for detecting $NH_3$ at ppm level. The enhancement in the response when the heterostructures were exposed to the gases was associated with the creation of the electron-hole pairs by the perovskites. The $NO_2$ was getting absorbed on the graphene due to the interaction between the gas and the oxygen defects and functional groups of the graphene while the presence of the nanocrystals improved the sensitive because the electron-hole pairs were separated. The holes generated at the nanocrystals were transferred to the graphene sheets. Inversely, by the interaction with an $NH_3$ (an electron-donating gas), an excess of electron was generated and transferred to graphene. The proposed mechanism is illustrated in Figure 35a.

Recently, similar heterostructures including lead-free metal halide perovskites have been proposed by Casanova-Chafer et al. for the detection of $H_2$, $H_2S$, $NH_3$ and $NO_2$.[183] This is the first time that a lead-free perovskite-based heterostructure as an alternative, environmentally friendly and harmless option has been used for the detection of gas pollutants. The use of nanocrystals enables achieving excellent sensitivity toward gas compounds and presents better properties than those of bulky perovskite thin films, owing to their quantum confinement effect and exciton binding energy. Two lead-free perovskite nanocrystals have been presented in this work: (a) $Cs_3Cu_2Br_5$ characterized by its direct band gap, outstanding stability, and nontoxicity of Cu(I) and (b) $Cs_2AgBiBr_6$ with its indirect band gap, nontoxic character, and remarkable thermal and environmental stability. The heterostructures produced by anchoring these perovskite nanocrystals on the graphene were found that exhibits the higher sensing performance to date among the lead-free perovskite sensing elements. $H_2$ and $H_2S$ gases were detected for the first time by utilizing lead-free perovskites, and ultrasensitive detection of $NO_2$ was also achieved at room temperature (Figure 36c). In particular, the $Cs_3Cu_2Br_5$ nanocrystals/graphene heterostructures were present higher sensitivity towards gases than those included the $Cs_2AgBiBr_6$ nanocrystals (Figure 36b), showing lower Limit of Detection (LOD) and Limit of Quantification (LOQ) values for all the gases tested. Two main reasons can probably explain these experimental findings. On the one hand, the $Cs_3Cu_2Br_5$ nanocrystals presents a direct band gap, which is favorable from the gas sensing point of view compared to the $Cs_2AgBiBr_6$ which is an indirect band gap semiconductor. On the other hand, the strength of exciton–phonon coupling of self-trapped excitons in both materials seemed that plays an important role, being stronger in the case of $Cs_2AgBiBr_6$ nanocrystals, limiting the efficient generation of separate charges and reducing the interaction with gas molecules.

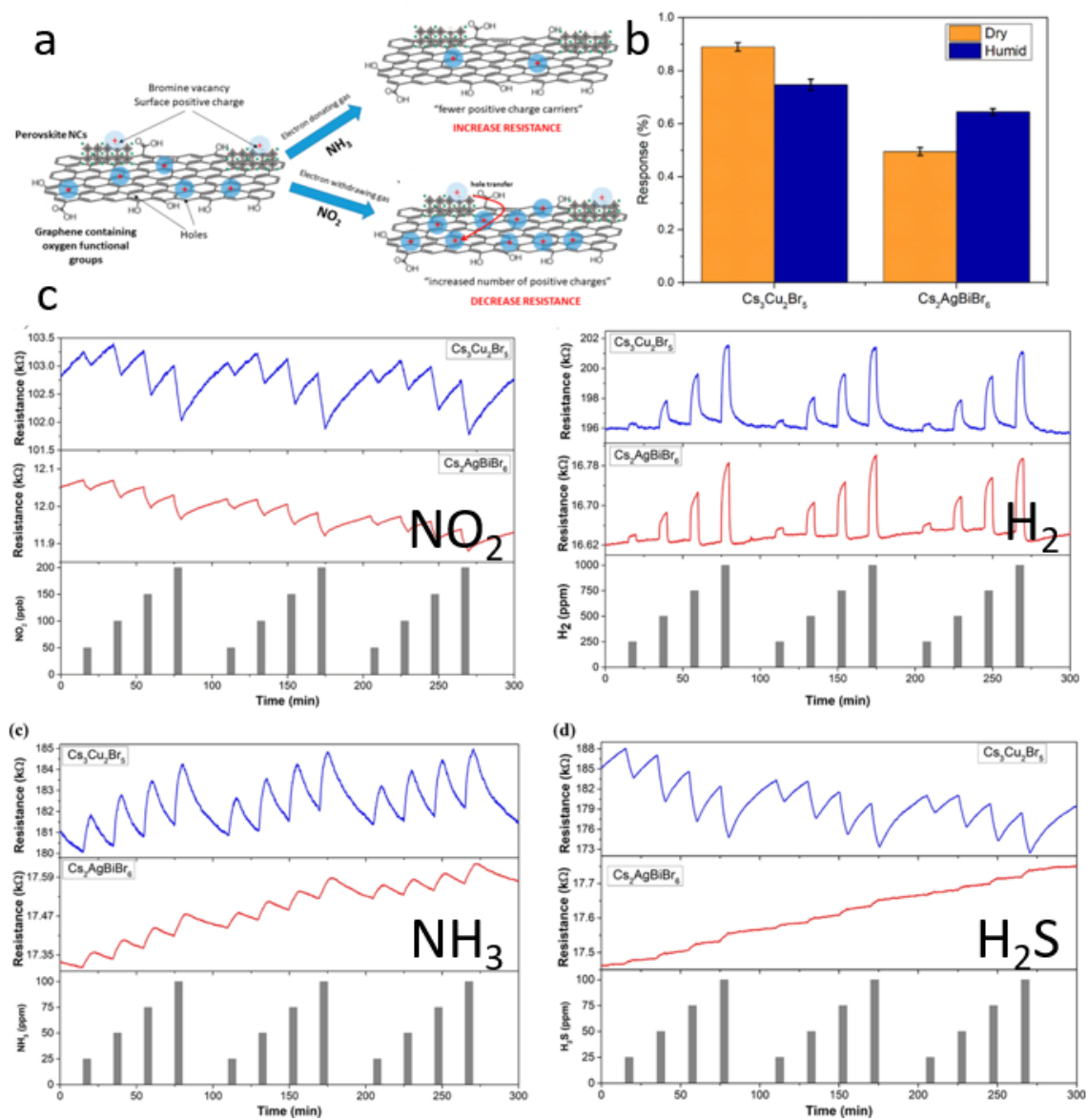

**Figure 36:** a) Proposed sensing mechanism of the metal halide nanocrystals/2D materials heterostructures in the presence of NH₃ (electron donating gas) and NO₂ (electron withdrawing gas). Two adsorption processes are proposed, one at the graphene surface and another at the perovskite nanocrystals. During the exposure to an electron-donating gas, an excess of positive charges is neutralized at the defective perovskite surface and the local hole concentration of the p-type graphene is decreased, which results in an increase in film resistance. While during the exposure to an electron-withdrawing gas, positive charges (holes) in the nanocrystals are formed, which are transferred to the graphene layers from the NCs, decreasing the overall resistance of the heterostructures film. Reprinted with permission from Ref.[63], copyright 2019, MDPI. b) Comparison of the sensing responses at room temperature toward NO₂ for Cs₃Cu₂Br₅ and Cs₂AgBiBr₆ nanocrystals supported on graphene under a dry and a humid (70% R.H.) environment and c) electrical responses when detecting NO₂, H₂, NH₃, and H₂S at room temperature. All gases were tested in the ppm range except NO₂, which was detected at ppb concentrations. Blue and red lines correspond to Cs₃Cu₂Br₅ and Cs₂AgBiBr₆ supported on graphene. b-c Reprinted with permission from Ref.[193], copyright 2022, American Chemical Society, https://pubs.acs.org/doi/10.1021/acssensors.2c01581, further permissions related to the material excerpted should be directed to the ACS.

Furthermore, such heterostructures have been evaluated to detect organic compounds (VOCs) such as benzene and toluene at ppb level. Highly reproducible, reversible, sensitive and ultrafast detection of VOCs at room temperature was achieved with such heterostructures. Additionally, in this study reported by Casanova-Chafer et al., the effect of the different cations (A) and halide anions (B) of the ABX₃ perovskite structure on the sensing capability of the heterostructures with the graphene has been elucidated. [71] Perovskite nanocrystals with three cations (methylammonium, MA (CH₃NH₃⁺); formamidinium, FA ((NH$_2$)$_2$CH$^+$); and cesium (Cs$^+$)), and three halide anions (Cl$^-$, Br$^-$ and I$^-$) have been synthesized for the fabrication of the heterostructures with the graphene flakes. These nanocrystals were bound on the exfoliated graphene by mixing the two solutions in an ultrasonic bath. Furthermore, the sensing capability of the heterostructures including the mixed halide perovskite, MAPbBr$_{2.5}$I$_{0.5}$ has been tested instead of MAPbI$_3$ due to its non-stability even at room temperature. In particular, regarding the cations effect, MA showed a clear enhancement in the responses (up to 3-fold) against FA and Cs and in sensitivity (the slope of the calibration curve) (Figure 37) while for the different halide anions, revealed that the Br$^-$ anions offer a higher response and sensitivity than the Cl$^-$ and I$^-$ anions. Equivalent behaviour was observed for toluene vapours. The different electrical response for the different heterostructures was originated from the energy level positions and the concentration of the trap states. The enhanced electrical response of the MAPbBr$_3$ perovskite/graphene heterostructure could be originated from the better energy-level alignment with the graphene for hole extraction, in comparison to the other two cation substituted perovskites (CsPbBr$_3$ and FAPbBr$_3$) and also the MAPbBr$_3$ nanocrystals have a higher trap density than CsPbBr$_3$ and FAPbBr$_3$ nanocrystals indicated from the slightly lower photoluminescence quantum yield. Increased active sites for interacting with vapour molecules was produced by increased number of defects thus increasing the electrical response. Nevertheless, MAPbBr$_3$ led to the best responses towards benzene and toluene vapours, in spite of the fact that the MAPbBr$_{2.5}$I$_{0.5}$ perovskite showed a better energy-level alignment with the graphene material for hole extraction. This led that there is also another factor responsible for the high response except of the energy-level alignment between the nanocrystals and the graphene and the higher number of superficial defects. The addition factor was the difference in the carrier mobility for electrons and holes as a function of the halide anion used. FAPbBr$_3$ nanocrystals/graphene heterostructures showed also high response in NO$_2$ and no response in NH$_3$. Moreover, MAPbCl$_3$ /graphene heterostructures showed 2-fold higher response to NH$_3$ compared to MAPbBr$_3$-based heterostructures.

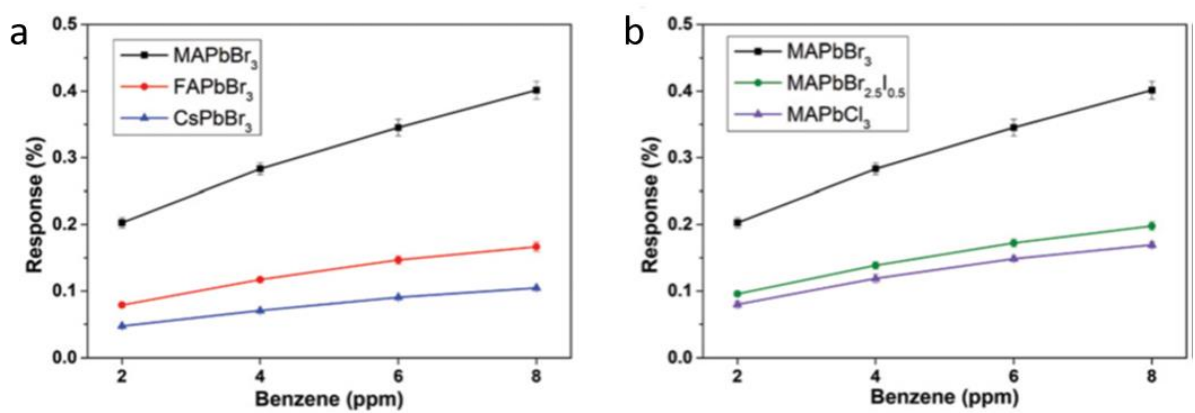

**Figure 37:** Response of metal halide nanocrystals/2D materials heterostructures using lead halide perovskites with different cations (a) and anions (b) in the detection of benzene. Reprinted with permission from Ref.[71], copyright 2020, The Royal Society of Chemistry.

# 4.6 Low-threshold lasers based on metal halide perovskite NPls and NSs

Finally, we would like to present a brief overview on the application of all-inorganic and hybrid quasi 2D perovskites for lasing features. Amplified spontaneous emission and a large-gain mechanism in lead halide perovskite materials were reported for the first time in 2004 in microcrystalline CsPbCl$_3$ films and found that is attributed to giant oscillator strength effect characteristic of excitonic superradiance.[194,194] It is worth noting that quasi 2D perovskites and quasi 2D/3D heterostructures appear promising components for amplified spontaneous emission (ASE) and lasing devices. Indicatively, Alvarado-Leanos et al. demonstrated lasing features in 2D tin-based organic-inorganic perovskites.[195] By changing the spacer cation the authors were able to improve the optical gain properties of 2D PEA$_2$SnI$_4$ films, and obtain ASE with a low threshold of 30 μJ/cm$^2$, and a high optical gain above 4000 cm$^{-1}$ at 77 K. On a similar manner, Raghavan et al. showed low-threshold lasing from homogeneous 2D organic-inorganic Ruddlesden-Popper perovskite single crystals.[196] The reported high-quality 2D crystals were synthesized by means of a slow evaporation solution-growth approach, while different emission wavelengths were achieved within the visible upon tuning the perovskite composition. Zhang et al. reported on the formation of high-density large-area of 2D Ruddlesden-Popper perovskite micro-ring arrays that exhibit high quality lasing factors. Indeed, the reported gain coefficient for the developed 2D components were found four times larger than the corresponding values obtain for the typical 3D perovskites.[197] In addition, a few years ago Qin et al. obtained stable room-temperature continuous-wave lasing in quasi 2D-perovskite films.[198] The authors fabricated surface-emitting lasers upon creating structures of spin-coated 2D perovskite films onto substrates with 2D gratings. Remarkably, the authors were capable of tuning the emission wavelength of the $FAPbBr_3$-based perovskites by changing the grating period on the surface of the employed substrate. Moreover, Li et al. demonstrated fascinating lasing features from laminated quasi-2D/3D $CsPbBr_3(BABr)_x/CsPbBr_3$ perovskite planar heterostructures.[199]

At the same time, whispering gallery mode lasing was demonstrated for the case of 2D metal halides perovskite NPls and NSs.[201] Zhang et al., demonstrated in 2014 planar room-temperature NIR nanolasers under femtosecond-pulsed laser excitation based on organic–inorganic perovskite $CH_3NH_3PbI_{3-a}X_a$ (X = I, Br, Cl) NPls of well-defined hexagonal and triangular morphologies.[202] Adequate gain and efficient optical feedback for low-threshold optically pumped in-plane lasing with threshold of 37 $\mu J/cm^2$ were ensured by the large exciton binding energies, long diffusion lengths, and naturally formed high-quality planar whispering-gallery mode cavities formed by the organic-inorganic metal halide perovskites NPls on Mica substrates (Figure 38). Two years later, Zhang et al. reported whispering-gallery-mode (WGM) microcavities using high-quality single-crystalline cesium lead halide $CsPbX_3$ (X = Cl, Br, I) NPls.[203] By tuning the halide composition, multi-color (400–700 nm) WGM excitonic lasing was achieved at room temperature with low threshold (~ 2.0 $\mu J/cm$) and high spectra coherence (~0.14–0.15 nm). Finally, square-like nanosheets of 2D all-inorganic metal halide ($CsPbI_3$) perovskite demonstrated one-photon and two-photon excitation-pumped lasing with low-threshold (0.3 $mJ/cm^2$) and high-quality factor.[200]

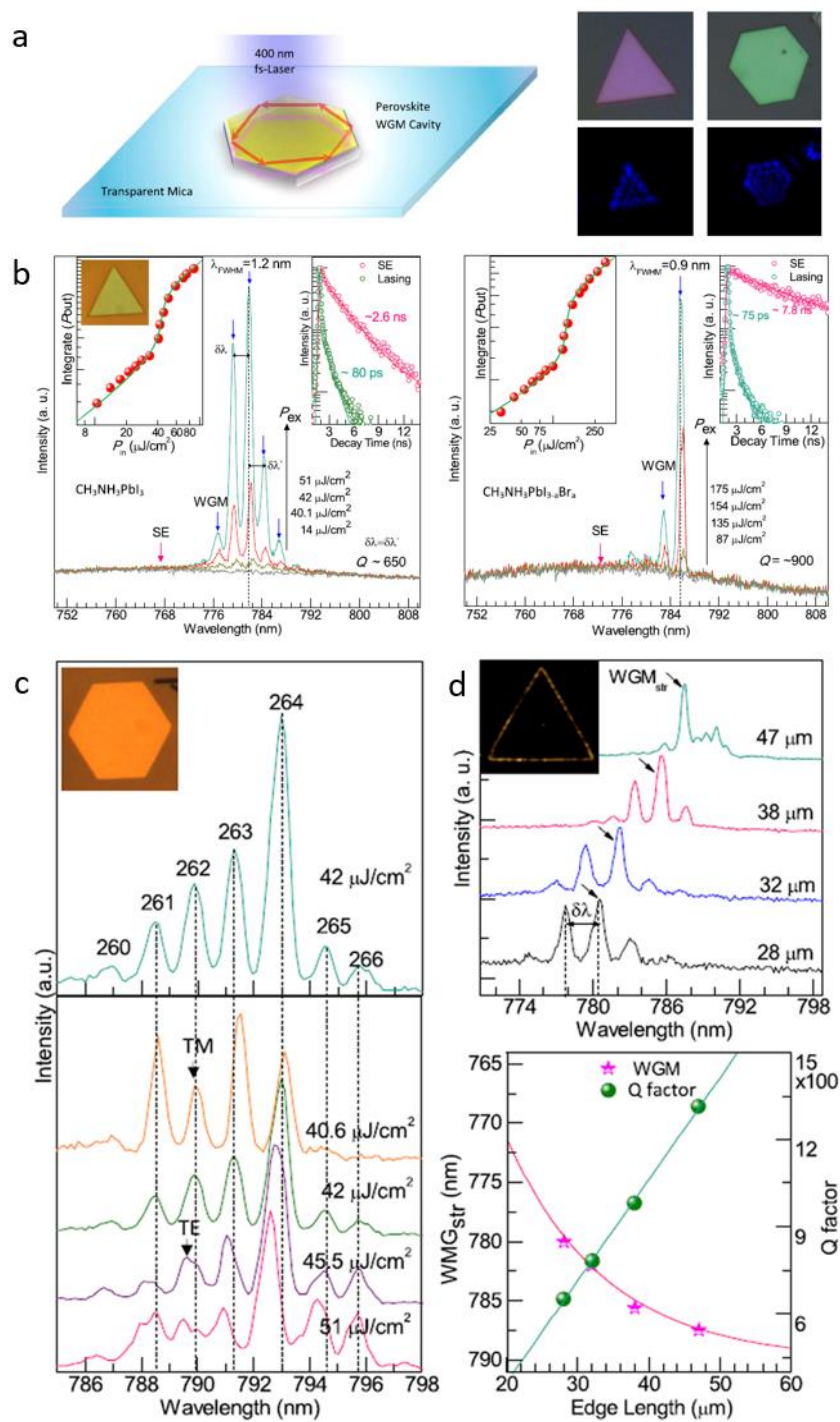

**Figure 38.** Lasing characterizations of perovskites whispering-gallery-mode nanocavities. (a) Schematic of optical setup and Far-field optical image of the $CH_3NH_3PbI_3$ nanoplatelets under the illumination of white light (upper panel) and incidence laser (bottom panel). b) The evolution from spontaneous emission (SE, ~768 nm, peak center indicated by red arrow) to lasing (whispering-gallery modes indicated by blue arrows) and parallel steady-states SE, lasing and time-resolved photoluminescence measurement on in a typical $CH_3NH_3PbI_3$ triangular nanoplatelet. c) Lasing spectra of hexagonal $CH_3NH_3PbI_3$ nanoplatelets and lasing mode evaluation as pumping fluence (bottom panel). d) The wavelength of lasing modes (pink star dots) and Q-factor (dark yellow dots) as a function of the triangular cavity edge length. Reprinted with permission from Ref.[202], copyright 2014, American Chemical Society.

# 5. Conclusions and outlook

Exploring low-dimensional perovskites materials allows an alternative strategy towards the development of stable optoelectronic, photovoltaic and sensing devices with improved stability and performance. In general, low-dimensional perovskites can be confined quantum dots, colloidal nanoplatelets (NPls), nanosheets (NSs), and quasi-2D structures. The main difference between the two latter types is the type of insulating layers, which can be spacers or ligands [34]. More specifically, spacers separate the inorganic layer or lead halide octahedra of the perovskite, and thus, determining the number of layers (n) of the quasi-2D perovskites. Rather differently, in case of ligands, surface passivation occurs that determines the lateral dimensions and thickness of the formed NPls and NSs.

Despite great progress in the synthesis protocols for metal halide perovskite 2D morphologies, it is still challenging to achieve monodisperse 2D perovskite materials with a single PL peak and stable in their colloidal solutions. Degradation and fusion of the NPls and NSs were also observed after the cleaning and purification processes following the synthesis which are crucial steps for the removal of the residual chemicals that were not reacted and also a thermal/moisture-induced perovskite nanocrystal agglomeration was observed in nanocrystal films used in devices. [40,73] This sensitivity of the NPls/NSs affects also the performance of the devices using these materials (LEDs and photodetectors) and improvement of the 2D materials stability is demanded. In the quest to improve the stability, metal halide perovskite nanocrystals/2D material heterostructures have been proposed exhibiting better stability against moisture, humidity, and high temperature stresses due to the suppression of the agglomeration by the 2D crosslinking. [73] These heterostructures combine the properties of the perovskite nanomaterials with their unique properties of the 2D materials and the large lateral dimensions. The large number of different perovskite nanocrystals and together with the plethora of 2D materials (graphene-based materials, hexagonal boron nitride or Transition Metal Dichalcogenides) can result to heterostructures with designed functionalities. [41] Additionally, new physics and synergetic effects can emerge from the coupling between the two different materials and new or improved functionalities have been arisen due to the interfacial phenomena. The films of the perovskite nanocrystals/2D materials heterostructures provides not only an effective channel for carrier transport, as witnessed by much improved conductivity but also shows significantly better stability.

In addition, one of the drawbacks of the synthesis protocols described extensively in this review for both types of 2D perovskite-based structures is the low yield. The development of large-scale synthesis procedures which will be cheap and easy is still a real challenge.

Long-term stability at ambient conditions or more harsh environments such as high temperature, direct irradiation, light and humidity have to be carefully addressed when we are interested to use the 2D materials in applications. The careful choice of a protective ligand or a shell proposed as an effective way to improve the stability of the perovskite nanocrystals but the effect of them on device performance is something that has to be studied as the conductivity and the PL is suppressed. [204,204] The encapsulation of the 2D perovskite-based materials in a matrix or a different material could be another way, but the sensitivity of these materials makes this more difficult.[204] Another notable future perspective worthy to be explored would be the development of composite inorganic oxide glasses upon encapsulating low-dimensional perovskite nanocrystals like NSs and NPls. The advantages of incorporating typical three-dimensional (3D) perovskite nanocrystals within transparent glasses towards improving stability while reducing toxicity have been demonstrated already. [205] The latter is of great importance if we consider that the toxic lead metal remains the main candidate for the majority of perovskite optoelectronic and photovoltaic devices. Along similar lines, polymers have been also employed as suitable hosts. [206] Moreover, the glass encapsulation approach offers an important tool for controlling the growth and morphology of the perovskite nanocrystals within the glass matrix by means of ultrafast pulsed and continuous wave laser sources. In particular, the formation of highly luminescent perovskite domains and micro-patterns within the glass matrices is achieved, paving the way towards advanced optoelectronic applications. [205] Also, it has been shown that the embedment of transition metal dichalcogenides 2D materials within silver-containing phosphate glasses results to the robust room-temperature B-exciton emission, as induced by the plasmon interactions between the randomly placed silver nanoparticles and the embedded few layers of the 2D material. [207] Thus, based on the above, it would be a great scientific achievement to incorporate 2D PNCs within glass matrices for three main reasons. First, the already improved stability of the 2D PNCs will be further enhanced. Second, upon using laser sources the crystalline morphology can be modified towards the development of advanced 2D and 2D-3D perovskite nanocrystal configurations. Finally, the induced interactions between glass components and the perovskite NSs and NPls are expected to induced interesting photoluminescence features that could be exploited towards next-generation optoelectronic and photonic devices.

Furthermore, the toxicity of the perovskite-based 2D materials in both NSs/NPls or heterostructured architectures due to the lead remain a major issue for using them in practical applications. To address the toxicity issues, significant efforts have been made to replace $Pb^{2+}$ with other environmentally friendly metal cations such as $Sn^{2+}$, $Bi^{3+}$, $Cu^{2+}$, $Sb^{3+}$, and $Ge^{2+}$. Last years the development of the lead-free perovskite nanocrystals is under investigation and new protocols and materials were emerged but the protocols on lead free NSs/NPls is still limited. [113,115] The number of the reports on the lead-free perovskite nanocrystals/2D materials heterostructures is even less. [65] Beside this, the performances of them in optoelectronic or photocatalysis are still unsatisfactory compared with their Pb-based 2D materials.[41] Sn-based perovskites have appropriate bandgaps for optoelectronic applications, but they display very low stability due to the oxidation of $Sn^{2+}$ to $Sn^{4+}$ under ambient conditions while the Ge- and Bi-based perovskites which are stable possess a large bandgap, which limit the light absorption in long-wavelength range.[208] Therefore, it is high desirable to design new 2D Pb-free perovskite materials with appropriate bandgap and large absorption coefficient and high stability.

In many applications of the perovskite-based 2D materials, compact and smooth films of high quality free of pinholes and cracks is needed. This is real challenge as many factors during this fabrication remain unexplored and have to be controlled. Moreover, the removal of the capping ligands or treatment of the nanocrystal surface is a necessity in order to fabricate such films with enhanced electrical properties.[209] For such purposes, various methods for this treatment have been proposed, but many times are insufficient which result in the release of nanocrystals from the surface or cause their undesired growth of the nanocrystals. These processes are more trivial in the case of the sensitive 2D materials with only a few atomic layers thickness. These affect also the stability of the devices in which are utilized such materials. The development of new efficient strategies for the effective treatment without affecting their primary structural or morphological features is a requirement.

In the tremendously emerging field of LEDs, scientists have identified other key factors that appear to be critical for enhancing the external quantum efficiency of perovskite-based LED devices. Namely, such factors include radiative recombination in the emitter component, charge injection features, electron-hole coupling effects, and number of generated photons emitted out of the device. [34] In addition, the humidity and thermal stability of the perovskite nanocrystals has been enhanced with their conjugation with the 2D materials displaying better performance and higher color purity compared to the single perovskite nanocrystals in LEDs.[51,52] Within the last five years or so, there has been great progress on putting together the pieces towards optimizing efficiency of LEDs. The great achievements on the design and fabrication of modern LED devices, arise from the incredible progress on developing new synthesis routes that allow the facile control of dimensionality, size, and shape of stable 2D perovskite nanocrystals or by achieving strong coupling between the perovskite nanocrystals and the 2D materials in the heterostructures. Apart from the evolution of the synthesis methods, post-fabrication treatments and advanced device architectures play also an important role. Based on this, it is fair to say that the scientific community has resolved the aforementioned scientific challenges to a great extent. The future appears even more promising towards the design of LEDs with supreme efficiency throughout the visible spectral range.

According to the application of the 2D perovskite materials as photocatalysts, the long-term catalytic cycling and operational stability has to be improved. [210] To enhance the resistance to degradation, core−shell structures (e.g., oxide capping layers) will help to limit contact between the metal halide perovskites and any polar solvent. No core shell structure including NPl as core is reported to date. 2D heterostructures found to be effective for the same reasons, increase the stability of the photocatalysts and they offer increased catalytic sites and sufficient charge separation compared to the pure perovskite nanocrystals and the 2D materials that are characterized from strong radiative recombination and insufficient stability. Interfacial engineering and strong coupling between the two materials will enhance further the photocatalytic performance.

In terms of perovskite solar cells based on low-dimensional perovskites, it becomes apparent that in order to achieve comparable power conversion efficiencies (PCEs) with the corresponding devices of typical 3D perovskites, the employment of quasi-2D perovskite films becomes a necessity. Nevertheless, as highlighted by Liu et al. [155], the formation of quasi-2D perovskites may be achieved in a controllable manner, by means of stacking together perovskite NSs. Thus, a great perspective would be to search for more strategies in order to transform the extremely stable perovskite NSs to advanced quasi-2D domains, that appear to be drastically more efficient in energy conversion devices (Section 4.4).

The transformation of the disadvantage of the metal halide perovskite nanocrystals to be moisture sensitive to advantage lead Ran et al to design a humidity detector based on $CH_3NH_3PbI_{3-x}Cl_x$ nanosheet arrays. [38] Also, the fact that the perovskite materials which are characterized by a high charge mobility can give an electrical readout as a result of variations in the environmental gas concentration result to gas sensors of ultra-low gas concentrations. [211,212] These materials can be an alternative to oxide-based sensing elements demonstrated enhanced room temperature gas sensing ability without using an external triggering such as temperature or UV irradiation in order to operate. The use of heterostructures in gas sensors can become a promising alternative to other gas sensitive materials, due to the protective character of 2D materials resulting from its high hydrophobicity. The hydrophobic properties of the graphene can protect the perovskite counterpart, enabling their use for ambient monitoring. However a proof of concept to use them for breath analysis for the detection of health-related biomarkers was reported.[192]

In summary, we believe that the 2D perovskites materials in the form of NSs/NPls or the nanocrystals/ 2D material heterostructures are a new family of structures with excellent tunability, highly dynamic structural features, and exciting optoelectronic properties.[213] The single-phase 2D perovskites need control of their thickness and homogeneity while the heterostructures need strong coupling between the two materials and interfacial engineering in order to achieve high performance in optoelectronic, photovoltaic sensing or photocatalytic devices. Progress on the materials design is essential and elucidation of the synergetic effects and interfacial phenomena and good understanding of the underlying mechanisms are demanded as well.

## Acknowledgements

This research was funded by H2020 InComEss Project-Innovative polymer-based Composite systems for high-efficient Energy scavenging and storage (Project ID: 862597) and by the European Union's Horizon 2020 framework programme for research and innovation under the NFFA-Europe-Pilot project (grant agreement no. 101007417).

IK would like to acknowledge support from the TheSmartMat project, "Laser-assisted development of composite thermochromic materials for energy smart and safe buildings", ΕΣΠΑ 2014-2020, RIS3Crete, G.A. KPHP1-0032623.

AK would like to acknowledge FLAG-ERA Joint Transnational Call 2019 for transnational research projects in synergy with the two FET Flagships Graphene Flagship & Human Brain Project - ERA-NETS 2019b (PeroGaS: MIS 5070514).

# References

1. C. Otero-Martínez et al., "Colloidal Metal-Halide Perovskite Nanoplatelets: Thickness-Controlled Synthesis, Properties, and Application in Light-Emitting Diodes," Advanced Materials **34**(10), 2107105 (2022) [doi:10.1002/adma.202107105].

2. A. Kostopoulou et al., "Perovskite nanocrystals for energy conversion and storage," Nanophotonics **8**(10), 1607–1640, De Gruyter (2019) [doi:10.1515/nanoph-2019-0119].

3. A. Kostopoulou, E. Kymakis, and E. Stratakis, "Perovskite nanostructures for photovoltaic and energy storage devices," J. Mater. Chem. A **6**(21), 9765–9798, The Royal Society of Chemistry (2018) [doi:10.1039/C8TA01964A].

4. L. C. Schmidt et al., "Nontemplate Synthesis of CH3NH3PbBr3 Perovskite Nanoparticles," J. Am. Chem. Soc. **136**(3), 850–853, American Chemical Society (2014) [doi:10.1021/ja4109209].

5. M. V. Kovalenko, L. Protesescu, and M. I. Bodnarchuk, "Properties and potential optoelectronic applications of lead halide perovskite nanocrystals," Science **358**(6364), 745–750, American Association for the Advancement of Science (2017) [doi:10.1126/science.aam7093].

6. N. Mondal, A. De, and A. Samanta, "Achieving Near-Unity Photoluminescence Efficiency for Blue-Violet-Emitting Perovskite Nanocrystals," ACS Energy Lett. **4**(1), 32–39, American Chemical Society (2019) [doi:10.1021/acsenergylett.8b01909].

7. G. Nedelcu et al., "Fast Anion-Exchange in Highly Luminescent Nanocrystals of Cesium Lead Halide Perovskites (CsPbX3, X = Cl, Br, I)," Nano Lett. **15**(8), 5635–5640, American Chemical Society (2015) [doi:10.1021/acs.nanolett.5b02404].

8. J. A. Sichert et al., "Quantum Size Effect in Organometal Halide Perovskite Nanoplatelets," Nano Lett. **15**(10), 6521–6527, American Chemical Society (2015) [doi:10.1021/acs.nanolett.5b02985].

9. V. A. Hintermayr et al., "Tuning the Optical Properties of Perovskite Nanoplatelets through Composition and Thickness by Ligand-Assisted Exfoliation," Advanced Materials **28**(43), 9478–9485 (2016) [doi:10.1002/adma.201602897].

10. L. Polavarapu et al., "Advances in Quantum-Confined Perovskite Nanocrystals for Optoelectronics," Advanced Energy Materials **7**(16), 1700267 (2017) [doi:10.1002/aenm.201700267].

11. Y. Bekenstein et al., "Highly Luminescent Colloidal Nanoplates of Perovskite Cesium Lead Halide and Their Oriented Assemblies," J. Am. Chem. Soc. **137**(51), 16008–16011, American Chemical Society (2015) [doi:10.1021/jacs.5b11199].

12. Q. A. Akkerman et al., "Solution Synthesis Approach to Colloidal Cesium Lead Halide Perovskite Nanoplatelets with Monolayer-Level Thickness Control," J. Am. Chem. Soc. **138**(3), 1010–1016, American Chemical Society (2016) [doi:10.1021/jacs.5b12124].

13. Y. Wang et al., "Two-Dimensional van der Waals Epitaxy Kinetics in a Three-Dimensional Perovskite Halide," Crystal Growth & Design **15**(10), 4741–4749, American Chemical Society (2015) [doi:10.1021/acs.cgd.5b00949].

14. W. Niu et al., "Exfoliation of self-assembled 2D organic-inorganic perovskite semiconductors," Appl. Phys. Lett. **104**(17), 171111, American Institute of Physics (2014) [doi:10.1063/1.4874846].

15. L. Dou et al., "Atomically thin two-dimensional organic-inorganic hybrid perovskites," Science **349**(6255), 1518–1521 (2015) [doi:10.1126/science.aac7660].

16. M. A. Uddin et al., "Growth of Highly Stable and Luminescent CsPbX3 (X = Cl, Br, and I) Nanoplates via Ligand Mediated Anion Exchange of CsPbCl3 Nanocubes with AlX3," Chem. Mater. **32**(12), 5217–5225, American Chemical Society (2020) [doi:10.1021/acs.chemmater.0c01325].

17. A. Kostopoulou et al., "Laser-Induced Morphological and Structural Changes of Cesium Lead Bromide Nanocrystals," 4, Nanomaterials **12**(4), 703, Multidisciplinary Digital Publishing Institute (2022) [doi:10.3390/nano12040703].

18.     Y. Tong et al., "Highly Luminescent Cesium Lead Halide Perovskite Nanocrystals with Tunable Composition and Thickness by Ultrasonication," Angewandte Chemie International Edition **55**(44), 13887–13892 (2016) [doi:10.1002/anie.201605909].

19.     G. Almeida et al., "Role of Acid–Base Equilibria in the Size, Shape, and Phase Control of Cesium Lead Bromide Nanocrystals," ACS Nano **12**(2), 1704–1711, American Chemical Society (2018) [doi:10.1021/acsnano.7b08357].

20.     A. Chakrabarty et al., "Precursor-Mediated Synthesis of Shape-Controlled Colloidal CsPbBr3 Perovskite Nanocrystals and Their Nanofiber-Directed Self-Assembly," Chem. Mater. **32**(2), 721–733, American Chemical Society (2020) [doi:10.1021/acs.chemmater.9b03700].

21.     J. Shamsi et al., "Colloidal Synthesis of Quantum Confined Single Crystal CsPbBr3 Nanosheets with Lateral Size Control up to the Micrometer Range," J. Am. Chem. Soc. **138**(23), 7240–7243, American Chemical Society (2016) [doi:10.1021/jacs.6b03166].

22.     Y. Wu et al., "In Situ Passivation of PbBr64– Octahedra toward Blue Luminescent CsPbBr3 Nanoplatelets with Near 100% Absolute Quantum Yield," ACS Energy Lett. **3**(9), 2030–2037, American Chemical Society (2018) [doi:10.1021/acsenergylett.8b01025].

23.     O. Vybornyi, S. Yakunin, and M. V. Kovalenko, "Polar-solvent-free colloidal synthesis of highly luminescent alkylammonium lead halide perovskite nanocrystals," Nanoscale **8**(12), 6278–6283, The Royal Society of Chemistry (2016) [doi:10.1039/C5NR06890H].

24.     L. Protesescu et al., "Dismantling the 'Red Wall' of Colloidal Perovskites: Highly Luminescent Formamidinium and Formamidinium–Cesium Lead Iodide Nanocrystals," ACS Nano **11**(3), 3119–3134, American Chemical Society (2017) [doi:10.1021/acsnano.7b00116].

25.     G. H. Ahmed et al., "Pyridine-Induced Dimensionality Change in Hybrid Perovskite Nanocrystals," Chem. Mater. **29**(10), 4393–4400, American Chemical Society (2017) [doi:10.1021/acs.chemmater.7b00872].

26.     I. Levchuk et al., "Brightly Luminescent and Color-Tunable Formamidinium Lead Halide Perovskite FAPbX3 (X = Cl, Br, I) Colloidal Nanocrystals," Nano Lett. **17**(5), 2765–2770, American Chemical Society (2017) [doi:10.1021/acs.nanolett.6b04781].

27.     I. Lignos et al., "Unveiling the Shape Evolution and Halide-Ion-Segregation in Blue-Emitting Formamidinium Lead Halide Perovskite Nanocrystals Using an Automated Microfluidic Platform," Nano Lett. **18**(2), 1246–1252, American Chemical Society (2018) [doi:10.1021/acs.nanolett.7b04838].

28.     J. Shamsi et al., "Metal Halide Perovskite Nanocrystals: Synthesis, Post-Synthesis Modifications, and Their Optical Properties," Chem. Rev. **119**(5), 3296–3348, American Chemical Society (2019) [doi:10.1021/acs.chemrev.8b00644].

29.     H. Huang et al., "Spontaneous Crystallization of Perovskite Nanocrystals in Nonpolar Organic Solvents: A Versatile Approach for their Shape-Controlled Synthesis," Angewandte Chemie International Edition **58**(46), 16558–16562 (2019) [doi:10.1002/anie.201906862].

30.     Y. Ding et al., "Tin-assisted growth of all-inorganic perovskite nanoplatelets with controllable morphologies and complementary emissions," CrystEngComm **21**(14), 2388–2397, The Royal Society of Chemistry (2019) [doi:10.1039/C9CE00027E].

31.     W. Zhai et al., "Solvothermal Synthesis of Ultrathin Cesium Lead Halide Perovskite Nanoplatelets with Tunable Lateral Sizes and Their Reversible Transformation into Cs4PbBr6 Nanocrystals," Chemistry of Materials, American Chemical Society (2018) [doi:10.1021/acs.chemmater.8b00612].

32.     L. Niu et al., "Controlled Growth and Reliable Thickness-Dependent Properties of Organic–Inorganic Perovskite Platelet Crystal," Advanced Functional Materials **26**(29), 5263–5270 (2016) [doi:10.1002/adfm.201601392].

33.     A. Mandal et al., "Photodetectors with High Responsivity by Thickness Tunable Mixed Halide Perovskite Nanosheets," ACS Appl. Mater. Interfaces **13**(36), 43104–43114, American Chemical Society (2021) [doi:10.1021/acsami.1c13452].

34.     C. Otero-Martínez et al., "Colloidal Metal-Halide Perovskite Nanoplatelets: Thickness-Controlled Synthesis, Properties, and Application in Light-Emitting Diodes," Advanced Materials **34**(10), 2107105 (2022) [doi:10.1002/adma.202107105].

35.     F. Bai et al., "Lead-free, air-stable ultrathin Cs3Bi2I9 perovskite nanosheets for solar cells," Solar Energy Materials and Solar Cells **184**, 15–21 (2018) [doi:10.1016/j.solmat.2018.04.032].

36.     Y. Dai and H. Tüysüz, "Rapid Acidic Media Growth of Cs3Bi2Br9 Halide Perovskite Platelets for Photocatalytic Toluene Oxidation," Solar RRL **5**(7), 2100265 (2021) [doi:10.1002/solr.202100265].

37.     L.-Y. Wu et al., "Two-Dimensional Metal Halide Perovskite Nanosheets for Efficient Photocatalytic CO2 Reduction," Solar RRL **5**(8), 2100263 (2021) [doi:10.1002/solr.202100263].

38.     K. Ren et al., "Turning a disadvantage into an advantage: synthesizing high-quality organometallic halide perovskite nanosheet arrays for humidity sensors," J. Mater. Chem. C **5**(10), 2504–2508, The Royal Society of Chemistry (2017) [doi:10.1039/C6TC05165K].

39.     C. Rodà et al., "O2 as a molecular probe for nonradiative surface defects in CsPbBr3 perovskite nanostructures and single crystals," Nanoscale **11**(16), 7613–7623, The Royal Society of Chemistry (2019) [doi:10.1039/C9NR01133A].

40.     Y. Bekenstein et al., "Highly Luminescent Colloidal Nanoplates of Perovskite Cesium Lead Halide and Their Oriented Assemblies," J. Am. Chem. Soc. **137**(51), 16008–16011, American Chemical Society (2015) [doi:10.1021/jacs.5b11199].

41.     Z. Zhang et al., "Metal Halide Perovskite/2D Material Heterostructures: Syntheses and Applications," Small Methods **5**(4), 2000937 (2021) [doi:10.1002/smtd.202000937].

42.     D.-H. Kwak et al., "High performance hybrid graphene–CsPbBr3–xIx perovskite nanocrystal photodetector," RSC Adv. **6**(69), 65252–65256, The Royal Society of Chemistry (2016) [doi:10.1039/C6RA08699C].

43.     Y.-F. Xu et al., "A CsPbBr3 Perovskite Quantum Dot/Graphene Oxide Composite for Photocatalytic CO2 Reduction," J. Am. Chem. Soc. **139**(16), 5660–5663, American Chemical Society (2017) [doi:10.1021/jacs.7b00489].

44.     X. Tang et al., "CsPbBr3/Reduced Graphene Oxide nanocomposites and their enhanced photoelectric detection application," Sensors and Actuators B: Chemical **245**, 435–440 (2017) [doi:10.1016/j.snb.2017.01.168].

45.     J.-S. Chen et al., "0D–2D and 1D–2D Semiconductor Hybrids Composed of All Inorganic Perovskite Nanocrystals and Single-Layer Graphene with Improved Light Harvesting," Particle & Particle Systems Characterization **35**(2), 1700310 (2018) [doi:10.1002/ppsc.201700310].

46.     X.-X. Guo et al., "Engineering a CsPbBr3-based nanocomposite for efficient photocatalytic CO2 reduction: improved charge separation concomitant with increased activity sites," RSC Adv. **9**(59), 34342–34348, The Royal Society of Chemistry (2019) [doi:10.1039/C9RA07236E].

47.     H. Huang et al., "In situ growth of all-inorganic perovskite nanocrystals on black phosphorus nanosheets," Chem. Commun. **54**(19), 2365–2368, The Royal Society of Chemistry (2018) [doi:10.1039/C8CC00029H].

48.     S. Muduli et al., "Photoluminescence Quenching in Self-Assembled CsPbBr3 Quantum Dots on Few-Layer Black Phosphorus Sheets," Angewandte Chemie International Edition **57**(26), 7682–7686 (2018) [doi:10.1002/anie.201712608].

49.     M. Ou et al., "Amino-Assisted Anchoring of CsPbBr3 Perovskite Quantum Dots on Porous g-C3N4 for Enhanced Photocatalytic CO2 Reduction," Angewandte Chemie International Edition **57**(41), 13570–13574 (2018) [doi:10.1002/anie.201808930].

50.     Q. Zhang et al., "In situ growth of α-CsPbI3 perovskite nanocrystals on the surface of reduced graphene oxide with enhanced stability and carrier transport quality," J. Mater. Chem. C **7**(22), 6795–6804, The Royal Society of Chemistry (2019) [doi:10.1039/C9TC01012B].

51.     L. Qiu et al., "One-pot in situ synthesis of CsPbX3@h-BN (X = Cl, Br, I) nanosheet composites with superior thermal stability for white LEDs," J. Mater. Chem. C **7**(14), 4038–4042, The Royal Society of Chemistry (2019) [doi:10.1039/C9TC00505F].

52.     Y. Li et al., "Room-Temperature Synthesis of Two-Dimensional Hexagonal Boron Nitride Nanosheet-Stabilized CsPbBr3 Perovskite Quantum Dots," ACS Appl. Mater. Interfaces **11**(8), 8242–8249, American Chemical Society (2019) [doi:10.1021/acsami.8b20400].

53.     Y. Pu et al., "Enhancing effects of reduced graphene oxide on photoluminescence of CsPbBr3 perovskite quantum dots," J. Mater. Chem. C **8**(22), 7447–7453, The Royal Society of Chemistry (2020) [doi:10.1039/D0TC01069C].

54.     Md. S. Hassan et al., "Enhanced Photocurrent Owing to Shuttling of Charge Carriers across 4-Aminothiophenol-Functionalized MoSe2–CsPbBr3 Nanohybrids," ACS Appl. Mater. Interfaces **12**(6), 7317–7325, American Chemical Society (2020) [doi:10.1021/acsami.9b20050].

55.     M. Peng et al., "All-Inorganic CsPbBr3 Perovskite Nanocrystals/2D Non-Layered Cadmium Sulfide Selenide for High-Performance Photodetectors by Energy Band Alignment Engineering," Advanced Functional Materials **31**(42), 2105051 (2021) [doi:10.1002/adfm.202105051].

56.     H. Bian et al., "2D-C3N4 encapsulated perovskite nanocrystals for efficient photo-assisted thermocatalytic CO2 reduction," Chem. Sci. **13**(5), 1335–1341, The Royal Society of Chemistry (2022) [doi:10.1039/D1SC06131C].

57.     H. Wu et al., "Interfacial Charge Behavior Modulation in Perovskite Quantum Dot-Monolayer MoS2 0D-2D Mixed-Dimensional van der Waals Heterostructures," Advanced Functional Materials **28**(34), 1802015 (2018) [doi:10.1002/adfm.201802015].

58.     H. Wu et al., "All-Inorganic Perovskite Quantum Dot-Monolayer MoS2 Mixed-Dimensional van der Waals Heterostructure for Ultrasensitive Photodetector," Advanced Science **5**(12), 1801219 (2018) [doi:10.1002/advs.201801219].

59.     F. Wang et al., "A noble-metal-free MoS2 nanosheet-coupled MAPbI3 photocatalyst for efficient and stable visible-light-driven hydrogen evolution," Chem. Commun. **56**(22), 3281–3284, The Royal Society of Chemistry (2020) [doi:10.1039/D0CC00095G].

60.     A. Kostopoulou et al., "Laser-Assisted Fabrication for Metal Halide Perovskite-2D Nanoconjugates: Control on the Nanocrystal Density and Morphology," 4, Nanomaterials **10**(4), 747, Multidisciplinary Digital Publishing Institute (2020) [doi:10.3390/nano10040747].

61.     Y.-C. Pu et al., "Methylamine lead bromide perovskite/protonated graphitic carbon nitride nanocomposites: interfacial charge carrier dynamics and photocatalysis," J. Mater. Chem. A **5**(48), 25438–25449, The Royal Society of Chemistry (2017) [doi:10.1039/C7TA08190A].

62.     Y. Wu et al., "Composite of CH3NH3PbI3 with Reduced Graphene Oxide as a Highly Efficient and Stable Visible-Light Photocatalyst for Hydrogen Evolution in Aqueous HI Solution," Advanced Materials **30**(7), 1704342 (2018) [doi:10.1002/adma.201704342].

63.     J. Casanova-Cháfer et al., "Gas Sensing Properties of Perovskite Decorated Graphene at Room Temperature," 20, Sensors **19**(20), 4563, Multidisciplinary Digital Publishing Institute (2019) [doi:10.3390/s19204563].

64.     Q. Wang et al., "Graphene oxide wrapped CH3NH3PbBr3 perovskite quantum dots hybrid for photoelectrochemical CO2 reduction in organic solvents," Applied Surface Science **465**, 607–613 (2019) [doi:10.1016/j.apsusc.2018.09.215].

65.     T. Wang et al., "Lead-free double perovskite Cs2AgBiBr6/RGO composite for efficient visible light photocatalytic H2 evolution," Applied Catalysis B: Environmental **268**, 118399 (2020) [doi:10.1016/j.apcatb.2019.118399].

66.     H. Lee et al., "Halide Perovskite Nanocrystal-Enabled Stabilization of Transition Metal Dichalcogenide Nanosheets," Small **18**(6), 2106035 (2022) [doi:10.1002/smll.202106035].

67.     X.-D. Wang et al., "In Situ Construction of a Cs2SnI6 Perovskite Nanocrystal/SnS2 Nanosheet Heterojunction with Boosted Interfacial Charge Transfer," J. Am. Chem. Soc. **141**(34), 13434–13441, American Chemical Society (2019) [doi:10.1021/jacs.9b04482].

68.     X. Wang et al., "CsPbBr3 perovskite nanocrystals anchoring on monolayer MoS2 nanosheets for efficient photocatalytic CO2 reduction," Chemical Engineering Journal **416**, 128077 (2021) [doi:10.1016/j.cej.2020.128077].

69.     S. Park et al., "Photocatalytic hydrogen generation from hydriodic acid using methylammonium lead iodide in dynamic equilibrium with aqueous solution," 1, Nat Energy **2**(1), 1–8, Nature Publishing Group (2016) [doi:10.1038/nenergy.2016.185].

70.     Y. Jiang et al., "In Situ Synthesis of Lead-Free Halide Perovskite Cs2AgBiBr6 Supported on Nitrogen-Doped Carbon for Efficient Hydrogen Evolution in Aqueous HBr Solution," ACS Applied Materials & Interfaces, American Chemical Society (2021) [doi:10.1021/acsami.0c21588].

71.     J. Casanova-Chafer et al., "The role of anions and cations in the gas sensing mechanisms of graphene decorated with lead halide perovskite nanocrystals," Chem. Commun. **56**(63), 8956–8959, The Royal Society of Chemistry (2020) [doi:10.1039/D0CC02984J].

72.     Y. Wang et al., "Hybrid Graphene–Perovskite Phototransistors with Ultrahigh Responsivity and Gain," Advanced Optical Materials **3**(10), 1389–1396 (2015) [doi:10.1002/adom.201500150].

73.     Q. Wang et al., "μ-Graphene Crosslinked CsPbI3 Quantum Dots for High Efficiency Solar Cells with Much Improved Stability," Advanced Energy Materials **8**(22), 1800007 (2018) [doi:10.1002/aenm.201800007].

74.     L. Polavarapu et al., "Advances in Quantum-Confined Perovskite Nanocrystals for Optoelectronics," Advanced Energy Materials **7**(16), 1700267 (2017) [doi:10.1002/aenm.201700267].

75.     A. Dey et al., "State of the Art and Prospects for Halide Perovskite Nanocrystals," ACS Nano **15**(7), 10775–10981, American Chemical Society (2021) [doi:10.1021/acsnano.0c08903].

76.     M. Imran et al., "Benzoyl Halides as Alternative Precursors for the Colloidal Synthesis of Lead-Based Halide Perovskite Nanocrystals," J. Am. Chem. Soc. **140**(7), 2656–2664, American Chemical Society (2018) [doi:10.1021/jacs.7b13477].

77.     J. A. Sichert et al., "Quantum Size Effect in Organometal Halide Perovskite Nanoplatelets," Nano Lett. **15**(10), 6521–6527, American Chemical Society (2015) [doi:10.1021/acs.nanolett.5b02985].

78.     Z. Yuan et al., "Highly luminescent nanoscale quasi-2D layered lead bromide perovskites with tunable emissions," Chem. Commun. **52**(20), 3887–3890, The Royal Society of Chemistry (2016) [doi:10.1039/C5CC09762B].

79.     M. C. Weidman et al., "Highly Tunable Colloidal Perovskite Nanoplatelets through Variable Cation, Metal, and Halide Composition," ACS Nano **10**(8), 7830–7839, American Chemical Society (2016) [doi:10.1021/acsnano.6b03496].

80.     J. Cho et al., "Ligand-Mediated Modulation of Layer Thicknesses of Perovskite Methylammonium Lead Bromide Nanoplatelets," Chem. Mater. **28**(19), 6909–6916, American Chemical Society (2016) [doi:10.1021/acs.chemmater.6b02241].

81.     S. Bhaumik et al., "Highly stable, luminescent core–shell type methylammonium–octylammonium lead bromide layered perovskite nanoparticles," Chem. Commun. **52**(44), 7118–7121, The Royal Society of Chemistry (2016) [doi:10.1039/C6CC01056C].

82.     S. Kumar et al., "Efficient Blue Electroluminescence Using Quantum-Confined Two-Dimensional Perovskites," ACS Nano **10**(10), 9720–9729, American Chemical Society (2016) [doi:10.1021/acsnano.6b05775].

83.     I. Levchuk et al., "Ligand-assisted thickness tailoring of highly luminescent colloidal CH3NH3PbX3 (X = Br and I) perovskite nanoplatelets," Chem. Commun. **53**(1), 244–247, The Royal Society of Chemistry (2016) [doi:10.1039/C6CC09266G].

84.     C. C. Stoumpos et al., "Ruddlesden–Popper Hybrid Lead Iodide Perovskite 2D Homologous Semiconductors," Chem. Mater. **28**(8), 2852–2867, American Chemical Society (2016) [doi:10.1021/acs.chemmater.6b00847].

85.     J.-C. Blancon et al., "Extremely efficient internal exciton dissociation through edge states in layered 2D perovskites," Science **355**(6331), 1288–1292, American Association for the Advancement of Science (2017) [doi:10.1126/science.aal4211].

86. C. J. Dahlman et al., "Structural Evolution of Layered Hybrid Lead Iodide Perovskites in Colloidal Dispersions," ACS Nano **14**(9), 11294–11308, American Chemical Society (2020) [doi:10.1021/acsnano.0c03219].

87. M. P. Hautzinger et al., "Band Edge Tuning of Two-Dimensional Ruddlesden–Popper Perovskites by A Cation Size Revealed through Nanoplates," ACS Energy Lett. **5**(5), 1430–1437, American Chemical Society (2020) [doi:10.1021/acsenergylett.0c00450].

88. D. N. Minh et al., "Room-Temperature Synthesis of Widely Tunable Formamidinium Lead Halide Perovskite Nanocrystals," Chem. Mater. **29**(13), 5713–5719, American Chemical Society (2017) [doi:10.1021/acs.chemmater.7b01705].

89. D. Yu et al., "Room-Temperature Ion-Exchange-Mediated Self-Assembly toward Formamidinium Perovskite Nanoplates with Finely Tunable, Ultrapure Green Emissions for Achieving Rec. 2020 Displays," Advanced Functional Materials **28**(19), 1800248 (2018) [doi:10.1002/adfm.201800248].

90. H. Fang et al., "Few-layer formamidinium lead bromide nanoplatelets for ultrapure-green and high-efficiency light-emitting diodes," Nano Res. **12**(1), 171–176 (2019) [doi:10.1007/s12274-018-2197-3].

91. S. Peng et al., "Effective Surface Ligand-Concentration Tuning of Deep-Blue Luminescent FAPbBr3 Nanoplatelets with Enhanced Stability and Charge Transport," ACS Applied Materials & Interfaces, American Chemical Society (2020) [doi:10.1021/acsami.0c08552].

92. V. A. Hintermayr et al., "Tuning the Optical Properties of Perovskite Nanoplatelets through Composition and Thickness by Ligand-Assisted Exfoliation," Advanced Materials **28**(43), 9478–9485 (2016) [doi:10.1002/adma.201602897].

93. X. Yang et al., "Patterned Perovskites for Optoelectronic Applications," Small Methods **2**(10), 1800110 (2018) [doi:10.1002/smtd.201800110].

94. S.-Y. Liang et al., "High-Resolution Patterning of 2D Perovskite Films through Femtosecond Laser Direct Writing," Advanced Functional Materials **32**(38), 0224957 (2022) [doi:10.1002/adfm.202204957].

95. X. Li et al., "CsPbX3 Quantum Dots for Lighting and Displays: Room-Temperature Synthesis, Photoluminescence Superiorities, Underlying Origins and White Light-Emitting Diodes," Advanced Functional Materials **26**(15), 2435–2445 (2016) [doi:10.1002/adfm.201600109].

96. L. Protesescu et al., "Nanocrystals of Cesium Lead Halide Perovskites (CsPbX3, X = Cl, Br, and I): Novel Optoelectronic Materials Showing Bright Emission with Wide Color Gamut," Nano Lett. **15**(6), 3692–3696, American Chemical Society (2015) [doi:10.1021/nl5048779].

97. S. Sun et al., "Ligand-Mediated Synthesis of Shape-Controlled Cesium Lead Halide Perovskite Nanocrystals via Reprecipitation Process at Room Temperature," ACS Nano **10**(3), 3648–3657, American Chemical Society (2016) [doi:10.1021/acsnano.5b08193].

98. K.-H. Wang et al., "Large-Scale Synthesis of Highly Luminescent Perovskite-Related CsPb2Br5 Nanoplatelets and Their Fast Anion Exchange," Angewandte Chemie International Edition **55**(29), 8328–8332 (2016) [doi:10.1002/anie.201602787].

99. A. Pan et al., "Insight into the Ligand-Mediated Synthesis of Colloidal CsPbBr3 Perovskite Nanocrystals: The Role of Organic Acid, Base, and Cesium Precursors," ACS Nano **10**(8), 7943–7954, American Chemical Society (2016) [doi:10.1021/acsnano.6b03863].

100. X. Sheng et al., "Polarized Optoelectronics of CsPbX3 (X = Cl, Br, I) Perovskite Nanoplates with Tunable Size and Thickness," Advanced Functional Materials **28**(19), 1800283 (2018) [doi:10.1002/adfm.201800283].

101. F. Bertolotti et al., "Crystal Structure, Morphology, and Surface Termination of Cyan-Emissive, Six-Monolayers-Thick CsPbBr3 Nanoplatelets from X-ray Total Scattering," ACS Nano **13**(12), 14294–14307, American Chemical Society (2019) [doi:10.1021/acsnano.9b07626].

102. J. Shamsi et al., "Stable Hexylphosphonate-Capped Blue-Emitting Quantum-Confined CsPbBr3 Nanoplatelets," ACS Energy Lett. **5**(6), 1900–1907, American Chemical Society (2020) [doi:10.1021/acsenergylett.0c00935].

103.     Q. Zeng et al., "Revealing the Aging Effect of Metal-Oleate Precursors on the Preparation of Highly Luminescent CsPbBr3 Nanoplatelets," J. Phys. Chem. Lett. **12**(10), 2668–2675, American Chemical Society (2021) [doi:10.1021/acs.jpclett.1c00300].

104.     L. G. Bonato et al., "Revealing the Role of Tin(IV) Halides in the Anisotropic Growth of CsPbX3 Perovskite Nanoplates," Angewandte Chemie International Edition **59**(28), 11501–11509 (2020) [doi:10.1002/anie.202002641].

105.     J. Zhao et al., "Amino Acid-Mediated Synthesis of CsPbBr3 Perovskite Nanoplatelets with Tunable Thickness and Optical Properties," Chem. Mater. **30**(19), 6737–6743, American Chemical Society (2018) [doi:10.1021/acs.chemmater.8b02396].

106.     X. Zhang et al., "Water-Assisted Size and Shape Control of CsPbBr3 Perovskite Nanocrystals," Angewandte Chemie International Edition **57**(13), 3337–3342 (2018) [doi:10.1002/anie.201710869].

107.     B. J. Bohn et al., "Boosting Tunable Blue Luminescence of Halide Perovskite Nanoplatelets through Postsynthetic Surface Trap Repair," Nano Lett. **18**(8), 5231–5238, American Chemical Society (2018) [doi:10.1021/acs.nanolett.8b02190].

108.     D. Chen et al., "Ultrathin CsPbX3 (X = Cl, Br, I) nanoplatelets: solvothermal synthesis and optical spectroscopic properties," Dalton Trans. **47**(29), 9845–9849, The Royal Society of Chemistry (2018) [doi:10.1039/C8DT01720D].

109.     A. Kostopoulou et al., "Laser-Induced Morphological and Structural Changes of Cesium Lead Bromide Nanocrystals," 4, Nanomaterials **12**(4), 703, Multidisciplinary Digital Publishing Institute (2022) [doi:10.3390/nano12040703].

110.     Q. Pan et al., "Microwave-assisted synthesis of high-quality 'all-inorganic' CsPbX3 (X = Cl, Br, I) perovskite nanocrystals and their application in light emitting diodes," J. Mater. Chem. C **5**(42), 10947–10954, The Royal Society of Chemistry (2017) [doi:10.1039/C7TC03774K].

111.     A. Wang et al., "Controlled Synthesis of Lead-Free and Stable Perovskite Derivative Cs2SnI6 Nanocrystals via a Facile Hot-Injection Process," Chem. Mater. **28**(22), 8132–8140, American Chemical Society (2016) [doi:10.1021/acs.chemmater.6b01329].

112.     A. B. Wong et al., "Strongly Quantum Confined Colloidal Cesium Tin Iodide Perovskite Nanoplates: Lessons for Reducing Defect Density and Improving Stability," Nano Lett. **18**(3), 2060–2066, American Chemical Society (2018) [doi:10.1021/acs.nanolett.8b00077].

113.     L. Lian et al., "Colloidal synthesis of lead-free all-inorganic cesium bismuth bromide perovskite nanoplatelets," CrystEngComm **20**(46), 7473–7478, The Royal Society of Chemistry (2018) [doi:10.1039/C8CE01060A].

114.     J. Huang et al., "Ultrathin lead-free double perovskite cesium silver bismuth bromide nanosheets," Nano Res. **14**(11), 4079–4086 (2021) [doi:10.1007/s12274-021-3343-x].

115.     Z. Liu et al., "Synthesis of Lead-Free Cs2AgBiX6 (X = Cl, Br, I) Double Perovskite Nanoplatelets and Their Application in CO2 Photocatalytic Reduction," Nano Lett. **21**(4), 1620–1627, American Chemical Society (2021) [doi:10.1021/acs.nanolett.0c04148].

116.     L. Lv et al., "Generalized colloidal synthesis of high-quality, two-dimensional cesium lead halide perovskite nanosheets and their applications in photodetectors," Nanoscale **8**(28), 13589–13596, The Royal Society of Chemistry (2016) [doi:10.1039/C6NR03428D].

117.     Y. Zhang et al., "Metal Halide Perovskite Nanosheet for X-ray High-Resolution Scintillation Imaging Screens," ACS Nano **13**(2), 2520–2525, American Chemical Society (2019) [doi:10.1021/acsnano.8b09484].

118.     R. Sun et al., "In situ preparation of two-dimensional ytterbium ions doped all-inorganic perovskite nanosheets for high-performance visual dual-bands photodetectors," Nano Energy **93**, 106815 (2022) [doi:10.1016/j.nanoen.2021.106815].

119.     E. Shi et al., "Two-dimensional halide perovskite lateral epitaxial heterostructures," 7805, Nature **580**(7805), 614–620, Nature Publishing Group (2020) [doi:10.1038/s41586-020-2219-7].

120.     D. Pan et al., "Deterministic fabrication of arbitrary vertical heterostructures of two-dimensional Ruddlesden–Popper halide perovskites," 2, Nat. Nanotechnol. **16**(2), 159–165, Nature Publishing Group (2021) [doi:10.1038/s41565-020-00802-2].

121.     J. Zhang et al., "Establishing charge-transfer excitons in 2D perovskite heterostructures," 1, Nat Commun **11**(1), 2618, Nature Publishing Group (2020) [doi:10.1038/s41467-020-16415-1].

122.     S. Parveen, K. K. Paul, and P. K. Giri, "Precise Tuning of the Thickness and Optical Properties of Highly Stable 2D Organometal Halide Perovskite Nanosheets through a Solvothermal Process and Their Applications as a White LED and a Fast Photodetector," ACS Applied Materials & Interfaces, American Chemical Society (2020) [doi:10.1021/acsami.9b20896].

123.     Y. Gao et al., "Copper-doping defect-lowered perovskite nanosheets for deep-blue light-emitting diodes," Journal of Colloid and Interface Science **607**, 1796–1804 (2022) [doi:10.1016/j.jcis.2021.09.061].

124.     Y. Ling et al., "Bright Light-Emitting Diodes Based on Organometal Halide Perovskite Nanoplatelets," Advanced Materials **28**(2), 305–311 (2016) [doi:10.1002/adma.201503954].

125.     J. Si et al., "Efficient and High-Color-Purity Light-Emitting Diodes Based on In Situ Grown Films of CsPbX3 (X = Br, I) Nanoplates with Controlled Thicknesses," ACS Publications, 24 October 2017, [doi:10.1021/acsnano.7b05191] (accessed 22 July 2022).

126.     S. Kumar et al., "Ultrapure Green Light-Emitting Diodes Using Two-Dimensional Formamidinium Perovskites: Achieving Recommendation 2020 Color Coordinates," Nano Lett. **17**(9), 5277–5284, American Chemical Society (2017) [doi:10.1021/acs.nanolett.7b01544].

127.     D. Yang et al., "Large-scale synthesis of ultrathin cesium lead bromide perovskite nanoplates with precisely tunable dimensions and their application in blue light-emitting diodes," Nano Energy **47**, 235–242 (2018) [doi:10.1016/j.nanoen.2018.03.019].

128.     B. J. Bohn et al., "Boosting Tunable Blue Luminescence of Halide Perovskite Nanoplatelets through Postsynthetic Surface Trap Repair," Nano Lett. **18**(8), 5231–5238, American Chemical Society (2018) [doi:10.1021/acs.nanolett.8b02190].

129.     Y. Wu et al., "In Situ Passivation of PbBr64– Octahedra toward Blue Luminescent CsPbBr3 Nanoplatelets with Near 100% Absolute Quantum Yield," ACS Energy Lett. **3**(9), 2030–2037, American Chemical Society (2018) [doi:10.1021/acsenergylett.8b01025].

130.     R. L. Z. Hoye et al., "Identifying and Reducing Interfacial Losses to Enhance Color-Pure Electroluminescence in Blue-Emitting Perovskite Nanoplatelet Light-Emitting Diodes," ACS Energy Lett. **4**(5), 1181–1188, American Chemical Society (2019) [doi:10.1021/acsenergylett.9b00571].

131.     W. Yin et al., "Multidentate Ligand Polyethylenimine Enables Bright Color-Saturated Blue Light-Emitting Diodes Based on CsPbBr3 Nanoplatelets," ACS Energy Lett. **6**(2), 477–484, American Chemical Society (2021) [doi:10.1021/acsenergylett.0c02651].

132.     H. Lin et al., "Stable and Efficient Blue-Emitting CsPbBr3 Nanoplatelets with Potassium Bromide Surface Passivation," Small **17**(43), 2101359 (2021) [doi:10.1002/smll.202101359].

133.     P. Tonkaev et al., "Acceleration of radiative recombination in quasi-2D perovskite films on hyperbolic metamaterials," Appl. Phys. Lett. **118**(9), 091104, American Institute of Physics (2021) [doi:10.1063/5.0042557].

134.     P. Guo et al., "Hyperbolic Dispersion Arising from Anisotropic Excitons in Two-Dimensional Perovskites," Phys. Rev. Lett. **121**(12), 127401, American Physical Society (2018) [doi:10.1103/PhysRevLett.121.127401].

135.     S. Heo et al., "Dimensionally Engineered Perovskite Heterostructure for Photovoltaic and Optoelectronic Applications," Advanced Energy Materials **9**(45), 1902470 (2019) [doi:10.1002/aenm.201902470].

136.     N. Jiang et al., "2D/3D Heterojunction perovskite light-emitting diodes with tunable ultrapure blue emissions," Nano Energy **97**, 107181 (2022) [doi:10.1016/j.nanoen.2022.107181].

137.     J. Liu et al., "Two-Dimensional CH3NH3PbI3 Perovskite: Synthesis and Optoelectronic Application," ACS Nano **10**(3), 3536–3542, American Chemical Society (2016) [doi:10.1021/acsnano.5b07791].

138.    X. Hu et al., "High-Performance Flexible Broadband Photodetector Based on Organolead Halide Perovskite," Advanced Functional Materials **24**(46), 7373–7380 (2014) [doi:10.1002/adfm.201402020].

139.    G. Wang et al., "Wafer-scale growth of large arrays of perovskite microplate crystals for functional electronics and optoelectronics," Science Advances **1**(9), e1500613, American Association for the Advancement of Science (2015) [doi:10.1126/sciadv.1500613].

140.    X. Qin et al., "Perovskite Photodetectors based on CH3NH3PbI3 Single Crystals," Chemistry – An Asian Journal **11**(19), 2675–2679 (2016) [doi:10.1002/asia.201600430].

141.    X. Liu et al., "Low-Voltage Photodetectors with High Responsivity Based on Solution-Processed Micrometer-Scale All-Inorganic Perovskite Nanoplatelets," Small **13**(25), 1700364 (2017) [doi:10.1002/smll.201700364].

142.    Z. Yang et al., "Engineering the Exciton Dissociation in Quantum-Confined 2D CsPbBr3 Nanosheet Films," Advanced Functional Materials **28**(14), 1705908 (2018) [doi:10.1002/adfm.201705908].

143.    J. Song et al., "Monolayer and Few-Layer All-Inorganic Perovskites as a New Family of Two-Dimensional Semiconductors for Printable Optoelectronic Devices," Advanced Materials **28**(24), 4861–4869 (2016) [doi:10.1002/adma.201600225].

144.    X. Li et al., "Constructing Fast Carrier Tracks into Flexible Perovskite Photodetectors To Greatly Improve Responsivity," ACS Nano **11**(2), 2015–2023, American Chemical Society (2017) [doi:10.1021/acsnano.6b08194].

145.    B. Xin et al., "Self-Patterned CsPbBr3 Nanocrystals for High-Performance Optoelectronics," ACS Applied Materials & Interfaces, American Chemical Society (2019) [doi:10.1021/acsami.8b17249].

146.    Y. Shen et al., "Two-dimensional CsPbBr$\less$sub$\greater$3$\less$/sub$\greater$/PCBM heterojunctions for sensitive, fast and flexible photodetectors boosted by charge transfer," Nanotechnology **29**(8), 085201, IOP Publishing (2018) [doi:10.1088/1361-6528/aaa456].

147.    Z. Yang et al., "Flexible all-inorganic photoconductor detectors based on perovskite/hole-conducting layer heterostructures," J. Mater. Chem. C **6**(25), 6739–6746, The Royal Society of Chemistry (2018) [doi:10.1039/C8TC02093K].

148.    M. He et al., "Chemical decoration of CH3NH3PbI3 perovskites with graphene oxides for photodetector applications," Chem. Commun. **51**(47), 9659–9661, The Royal Society of Chemistry (2015) [doi:10.1039/C5CC02282G].

149.    X. Song et al., "Boosting Two-Dimensional MoS2/CsPbBr3 Photodetectors via Enhanced Light Absorbance and Interfacial Carrier Separation," ACS Appl. Mater. Interfaces **10**(3), 2801–2809, American Chemical Society (2018) [doi:10.1021/acsami.7b14745].

150.    H. Wu et al., "Ligand Engineering for Improved All-Inorganic Perovskite Quantum Dot-MoS2 Monolayer Mixed Dimensional van der Waals Phototransistor," Small Methods **3**(7), 1900117 (2019) [doi:10.1002/smtd.201900117].

151.    "High performance photodetector based on 2D CH3NH3PbI3 perovskite nanosheets - IOPscience," <https://iopscience.iop.org/article/10.1088/1361-6463/aa5623/meta> (accessed 11 July 2022).

152.    A. Mardani, F. Kazemi, and B. Kaboudin, "Photo-tunable oxidation of toluene and its derivatives catalyzed by TBATB," Journal of Photochemistry and Photobiology A: Chemistry **414**, 113301 (2021) [doi:10.1016/j.jphotochem.2021.113301].

153.    P. Qiu et al., "Fabricating Surface-Functionalized CsPbBr3/Cs4PbBr6 Nanosheets for Visible-Light Photocatalytic Oxidation of Styrene," Frontiers in Chemistry **8** (2020).

154.    R. Li et al., "Few-layer black phosphorus-on-MAPbI3 for superb visible-light photocatalytic hydrogen evolution from HI splitting," Applied Catalysis B: Environmental **259**, 118075 (2019) [doi:10.1016/j.apcatb.2019.118075].

155.    Y. Liu et al., "Self-Assembly of Two-Dimensional Perovskite Nanosheet Building Blocks into Ordered Ruddlesden–Popper Perovskite Phase," J. Am. Chem. Soc. **141**(33), 13028–13032, American Chemical Society (2019) [doi:10.1021/jacs.9b06889].

156.    J. Yan et al., "Recent progress in 2D/quasi-2D layered metal halide perovskites for solar cells," J. Mater. Chem. A **6**(24), 11063–11077, The Royal Society of Chemistry (2018) [doi:10.1039/C8TA02288G].

157.    C. Lan et al., "Two-dimensional perovskite materials: From synthesis to energy-related applications," Materials Today Energy **11**, 61–82 (2019) [doi:10.1016/j.mtener.2018.10.008].

158.    P. Liu et al., "High-Quality Ruddlesden–Popper Perovskite Film Formation for High-Performance Perovskite Solar Cells," Advanced Materials **33**(10), 2002582 (2021) [doi:10.1002/adma.202002582].

159.    E.-B. Kim et al., "A review on two-dimensional (2D) and 2D-3D multidimensional perovskite solar cells: Perovskites structures, stability, and photovoltaic performances," Journal of Photochemistry and Photobiology C: Photochemistry Reviews **48**, 100405 (2021) [doi:10.1016/j.jphotochemrev.2021.100405].

160.    X. Zhao, T. Liu, and Y.-L. Loo, "Advancing 2D Perovskites for Efficient and Stable Solar Cells: Challenges and Opportunities," Advanced Materials **34**(3), 2105849 (2022) [doi:10.1002/adma.202105849].

161.    E. Elahi et al., "A review on two-dimensional (2D) perovskite material-based solar cells to enhance the power conversion efficiency," Dalton Trans. **51**(3), 797–816, The Royal Society of Chemistry (2022) [doi:10.1039/D1DT02991F].

162.    S. Miao et al., "2D Material and Perovskite Heterostructure for Optoelectronic Applications," 12, Nanomaterials **12**(12), 2100, Multidisciplinary Digital Publishing Institute (2022) [doi:10.3390/nano12122100].

163.    I. C. Smith et al., "A Layered Hybrid Perovskite Solar-Cell Absorber with Enhanced Moisture Stability," Angewandte Chemie International Edition **53**(42), 11232–11235 (2014) [doi:10.1002/anie.201406466].

164.    L. N. Quan et al., "Ligand-Stabilized Reduced-Dimensionality Perovskites," J. Am. Chem. Soc. **138**(8), 2649–2655, American Chemical Society (2016) [doi:10.1021/jacs.5b11740].

165.    X. Zhang et al., "Stable high efficiency two-dimensional perovskite solar cells via cesium doping," Energy Environ. Sci. **10**(10), 2095–2102, The Royal Society of Chemistry (2017) [doi:10.1039/C7EE01145H].

166.    X. Zhang et al., "Orientation Regulation of Phenylethylammonium Cation Based 2D Perovskite Solar Cell with Efficiency Higher Than 11%," Advanced Energy Materials **8**(14), 1702498 (2018) [doi:10.1002/aenm.201702498].

167.    R. Yang et al., "Oriented Quasi-2D Perovskites for High Performance Optoelectronic Devices," Advanced Materials **30**(51), 1804771 (2018) [doi:10.1002/adma.201804771].

168.    K. T. Cho et al., "Water-Repellent Low-Dimensional Fluorous Perovskite as Interfacial Coating for 20% Efficient Solar Cells," Nano Lett. **18**(9), 5467–5474, American Chemical Society (2018) [doi:10.1021/acs.nanolett.8b01863].

169.    M. Abuhelaiqa et al., "Mixed cation 2D perovskite: a novel approach for enhanced perovskite solar cell stability," Sustainable Energy Fuels **6**(10), 2471–2477, The Royal Society of Chemistry (2022) [doi:10.1039/D1SE01721G].

170.    Y.-W. Jang et al., "Intact 2D/3D halide junction perovskite solar cells via solid-phase in-plane growth," 1, Nat Energy **6**(1), 63–71, Nature Publishing Group (2021) [doi:10.1038/s41560-020-00749-7].

171.    Y. Huang et al., "Stable Layered 2D Perovskite Solar Cells with an Efficiency of over 19% via Multifunctional Interfacial Engineering," J. Am. Chem. Soc. **143**(10), 3911–3917, American Chemical Society (2021) [doi:10.1021/jacs.0c13087].

172.    H. Ren et al., "Efficient and stable Ruddlesden–Popper perovskite solar cell with tailored interlayer molecular interaction," 3, Nat. Photonics **14**(3), 154–163, Nature Publishing Group (2020) [doi:10.1038/s41566-019-0572-6].

173.    C. Liang et al., "Two-dimensional Ruddlesden–Popper layered perovskite solar cells based on phase-pure thin films," 1, Nat Energy **6**(1), 38–45, Nature Publishing Group (2021) [doi:10.1038/s41560-020-00721-5].

174.    K. Li et al., "High Efficiency Perovskite Solar Cells Employing Quasi-2D Ruddlesden-Popper/Dion-Jacobson Heterojunctions," Advanced Functional Materials **32**(21), 2200024 (2022) [doi:10.1002/adfm.202200024].

175.    W. Li et al., "Light-activated interlayer contraction in two-dimensional perovskites for high-efficiency solar cells," 1, Nat. Nanotechnol. **17**(1), 45–52, Nature Publishing Group (2022) [doi:10.1038/s41565-021-01010-2].

176.    J. Liang et al., "A finely regulated quantum well structure in quasi-2D Ruddlesden–Popper perovskite solar cells with efficiency exceeding 20%," Energy Environ. Sci. **15**(1), 296–310, The Royal Society of Chemistry (2022) [doi:10.1039/D1EE01695D].

177.    M. Shao et al., "Over 21% Efficiency Stable 2D Perovskite Solar Cells," Advanced Materials **34**(1), 2107211 (2022) [doi:10.1002/adma.202107211].

178.    M. Hadadian et al., "Enhancing Efficiency of Perovskite Solar Cells via N-doped Graphene: Crystal Modification and Surface Passivation," Advanced Materials **28**(39), 8681–8686 (2016) [doi:10.1002/adma.201602785].

179.    E. Serpetzoglou et al., "Improved Carrier Transport in Perovskite Solar Cells Probed by Femtosecond Transient Absorption Spectroscopy," ACS Appl. Mater. Interfaces **9**(50), 43910–43919, American Chemical Society (2017) [doi:10.1021/acsami.7b15195].

180.    I. Konidakis et al., "Improved Charge Carrier Dynamics of $CH_3NH_3PbI_3$ Perovskite Films Synthesized by Means of Laser-Assisted Crystallization," ACS Appl. Energy Mater. **1**(9), 5101–5111, American Chemical Society (2018) [doi:10.1021/acsaem.8b01152].

181.    L.-L. Jiang et al., "Passivated Perovskite Crystallization via g-C3N4 for High-Performance Solar Cells," Advanced Functional Materials **28**(7), 1705875 (2018) [doi:10.1002/adfm.201705875].

182.    Z. Guo et al., "High Electrical Conductivity 2D MXene Serves as Additive of Perovskite for Efficient Solar Cells," Small **14**(47), 1802738 (2018) [doi:10.1002/smll.201802738].

183.    Z. Wu et al., "Efficient planar heterojunction perovskite solar cells employing graphene oxide as hole conductor," Nanoscale **6**(18), 10505–10510, The Royal Society of Chemistry (2014) [doi:10.1039/C4NR03181D].

184.    J.-S. Yeo et al., "Highly efficient and stable planar perovskite solar cells with reduced graphene oxide nanosheets as electrode interlayer," Nano Energy **12**, 96–104 (2015) [doi:10.1016/j.nanoen.2014.12.022].

185.    Q. Zhao et al., "High efficiency perovskite quantum dot solar cells with charge separating heterostructure," 1, Nat Commun **10**(1), 2842, Nature Publishing Group (2019) [doi:10.1038/s41467-019-10856-z].

186.    Y. Hu et al., "Hybrid Perovskite/Perovskite Heterojunction Solar Cells," ACS Nano **10**(6), 5999–6007, American Chemical Society (2016) [doi:10.1021/acsnano.6b01535].

187.    C. Travan and A. Bergmann, "NO2 and NH3 Sensing Characteristics of Inkjet Printing Graphene Gas Sensors," 15, Sensors **19**(15), 3379, Multidisciplinary Digital Publishing Institute (2019) [doi:10.3390/s19153379].

188.    M. Rodner et al., "Graphene Decorated with Iron Oxide Nanoparticles for Highly Sensitive Interaction with Volatile Organic Compounds," 4, Sensors **19**(4), 918, Multidisciplinary Digital Publishing Institute (2019) [doi:10.3390/s19040918].

189.    O. Ovsianytskyi et al., "Highly sensitive chemiresistive H2S gas sensor based on graphene decorated with Ag nanoparticles and charged impurities," Sensors and Actuators B: Chemical **257**, 278–285 (2018) [doi:10.1016/j.snb.2017.10.128].

190.     Y. Ren et al., "Detection of sulfur dioxide gas with graphene field effect transistor," Appl. Phys. Lett. **100**(16), 163114, American Institute of Physics (2012) [doi:10.1063/1.4704803].

191.     D. Panda et al., "Selective detection of carbon monoxide (CO) gas by reduced graphene oxide (rGO) at room temperature," RSC Adv. **6**(53), 47337–47348, The Royal Society of Chemistry (2016) [doi:10.1039/C6RA06058G].

192.     J. Casanova-Chafer et al., "Perovskite@Graphene Nanohybrids for Breath Analysis: A Proof-of-Concept," 8, Chemosensors **9**(8), 215, Multidisciplinary Digital Publishing Institute (2021) [doi:10.3390/chemosensors9080215].

193.     J. Casanova-Chafer et al., "Unraveling the Gas-Sensing Mechanisms of Lead-Free Perovskites Supported on Graphene," ACS Sens., American Chemical Society (2022) [doi:10.1021/acssensors.2c01581].

194.     S. Kondo et al., "Photoluminescence and stimulated emission from microcrystalline $\mathrm{Cs}\mathrm{Pb}\mathrm{Cl}_{3}$ films prepared by amorphous-to-crystalline transformation," Phys. Rev. B **70**(20), 205322, American Physical Society (2004) [doi:10.1103/PhysRevB.70.205322].

195.     A. L. Alvarado-Leaños et al., "Lasing in Two-Dimensional Tin Perovskites," ACS Nano **16**(12), 20671–20679, American Chemical Society (2022) [doi:10.1021/acsnano.2c07705].

196.     C. M. Raghavan et al., "Low-Threshold Lasing from 2D Homologous Organic–Inorganic Hybrid Ruddlesden–Popper Perovskite Single Crystals," Nano Lett. **18**(5), 3221–3228, American Chemical Society (2018) [doi:10.1021/acs.nanolett.8b00990].

197.     H. Zhang et al., "2D Ruddlesden–Popper Perovskites Microring Laser Array," Advanced Materials **30**(15), 1706186 (2018) [doi:10.1002/adma.201706186].

198.     C. Qin et al., "Stable room-temperature continuous-wave lasing in quasi-2D perovskite films," 7823, Nature **585**(7823), 53–57, Nature Publishing Group (2020) [doi:10.1038/s41586-020-2621-1].

199.     Y. Li et al., "Lasing from Laminated Quasi-2D/3D Perovskite Planar Heterostructures," Advanced Functional Materials **32**(27), 2200772 (2022) [doi:10.1002/adfm.202200772].

200.     Z. Zheng et al., "Space-Confined Synthesis of 2D All-Inorganic CsPbI3 Perovskite Nanosheets for Multiphoton-Pumped Lasing," Advanced Optical Materials **6**(22), 1800879 (2018) [doi:10.1002/adom.201800879].

201.     Z. Hu et al., "Advances in metal halide perovskite lasers: synthetic strategies, morphology control, and lasing emission," AP **3**(3), 034002, SPIE (2021) [doi:10.1117/1.AP.3.3.034002].

202.     Q. Zhang et al., "Room-Temperature Near-Infrared High-Q Perovskite Whispering-Gallery Planar Nanolasers," Nano Lett. **14**(10), 5995–6001, American Chemical Society (2014) [doi:10.1021/nl503057g].

203.     Q. Zhang et al., "High-Quality Whispering-Gallery-Mode Lasing from Cesium Lead Halide Perovskite Nanoplatelets," Advanced Functional Materials **26**(34), 6238–6245 (2016) [doi:10.1002/adfm.201601690].

204.     C. Zhang et al., "Core/Shell Metal Halide Perovskite Nanocrystals for Optoelectronic Applications," Advanced Functional Materials **31**(19), 2100438 (2021) [doi:10.1002/adfm.202100438].

205.     I. Konidakis, A. Karagiannaki, and E. Stratakis, "Advanced composite glasses with metallic, perovskite, and two-dimensional nanocrystals for optoelectronic and photonic applications," Nanoscale **14**(8), 2966–2989, The Royal Society of Chemistry (2022) [doi:10.1039/D1NR07711B].

206.     S. N. Raja et al., "Encapsulation of Perovskite Nanocrystals into Macroscale Polymer Matrices: Enhanced Stability and Polarization," ACS Appl. Mater. Interfaces **8**(51), 35523–35533, American Chemical Society (2016) [doi:10.1021/acsami.6b09443].

207.     A. S. Sarkar et al., "Robust B-exciton emission at room temperature in few-layers of MoS2:Ag nanoheterojunctions embedded into a glass matrix," 1, Sci Rep **10**(1), 15697, Nature Publishing Group (2020) [doi:10.1038/s41598-020-72899-3].

208.     Y. Tang, X. Cao, and Q. Chi, *Two-Dimensional Halide Perovskites for Emerging New-Generation Photodetectors*, in Two-dimensional Materials for Photodetector, IntechOpen (2017) [doi:10.5772/intechopen.71032].

209.     F. Chen et al., "Long-term optical and morphological stability of CsPbBr3 nanocrystal-based films," Materials Research Bulletin **134**, 111107 (2021) [doi:10.1016/j.materresbull.2020.111107].

210.     H. Huang et al., "Solar-Driven Metal Halide Perovskite Photocatalysis: Design, Stability, and Performance," ACS Energy Lett. **5**(4), 1107–1123, American Chemical Society (2020) [doi:10.1021/acsenergylett.0c00058].

211.     A. Argyrou et al., "Highly sensitive ozone and hydrogen sensors based on perovskite microcrystals directly grown on electrodes," Journal of Materiomics (2021) [doi:10.1016/j.jmat.2021.07.002].

212.     K. Brintakis et al., "Ligand-free all-inorganic metal halide nanocubes for fast, ultra-sensitive and self-powered ozone sensors," Nanoscale Adv. **1**(7), 2699–2706, RSC (2019) [doi:10.1039/C9NA00219G].

213.     E. Shi et al., "Two-dimensional halide perovskite nanomaterials and heterostructures," Chem. Soc. Rev. **47**(16), 6046–6072, The Royal Society of Chemistry (2018) [doi:10.1039/C7CS00886D].